\documentclass[aps,pra,amsmath,twocolumn,amssymb,footinbib,superscriptaddress]{revtex4-1}

\usepackage{amssymb,amsfonts,amsmath}

\usepackage{color}
\usepackage{float}
\usepackage[apple mac]{inputenc}
\usepackage[english]{babel}
\usepackage{times}
\usepackage{latexsym}
\usepackage{fancyhdr}
\usepackage{verbatim}
\usepackage{graphicx}
\usepackage{epsf}
\usepackage{subfigure}
\usepackage{bm}
\usepackage{epstopdf}

\usepackage[colorlinks=true , citecolor=blue,urlcolor=blue]{hyperref}

\definecolor{Nathanblue}{rgb}{0.,0.24,0.51}
\newcommand{\blue}{\color{Nathanblue}}

\def\be{\begin{equation}}
\def\ee{\end{equation}}
	
\def\bs#1{\mathbf{#1}}

\begin{document}

\title{{\blue Loading Ultracold Gases in Topological Floquet Bands: \\ Current and Center-of-Mass Responses}}

\author{Alexandre Dauphin}
\email{alexandre.dauphin@icfo.es}
\affiliation{ICFO-Institut de Ciencies Fotoniques, The Barcelona Institute of Science and Technology, 08860 Castelldefels (Barcelona), Spain}

\author{Duc-Thanh Tran}
\email{ducttran@ulb.ac.be}
\affiliation{Center for Nonlinear Phenomena and Complex Systems, Universit\'e Libre de Bruxelles, CP 231, Campus Plaine, B-1050 Brussels, Belgium}

\author{Maciej Lewenstein}
\affiliation{ICFO-Institut de Ciencies Fotoniques, The Barcelona Institute of Science and Technology, 08860 Castelldefels (Barcelona), Spain}
\affiliation{ICREA, Passeig de Lluis Companys 23, E-08010 Barcelona, Spain}

\author{Nathan Goldman}
\email{ngoldman@ulb.ac.be}
\affiliation{Center for Nonlinear Phenomena and Complex Systems, Universit\'e Libre de Bruxelles, CP 231, Campus Plaine, B-1050 Brussels, Belgium}

\begin{abstract}

Topological band structures can be designed by subjecting lattice systems to time-periodic modulations, as was recently demonstrated in cold atoms and photonic crystals. However, changing the topological nature of Floquet Bloch bands from trivial to non-trivial, by progressively launching the time-modulation, is necessarily accompanied with gap-closing processes:~this has important consequences for the loading of particles into a target Floquet band with non-trivial topology, and hence, on the subsequent measurements. In this work, we analyse how such loading sequences can be optimized in view of probing the topology of Floquet bands through transport measurements. In particular, we demonstrate the robustness of center-of-mass responses, as compared to current responses, which present important irregularities due to an interplay between the micro-motion of the drive and inter-band interference effects. The results presented in this work illustrate how probing the center-of-mass displacement of atomic clouds offers a reliable method to detect the topology of Floquet bands, after realistic loading sequences.

\end{abstract}
\maketitle

\section{Introduction}

Shaking lattices periodically in time can be used as a powerful tool to control the tunneling matrix elements of engineered quantum systems~\cite{Grossmann:1991}, such as ultracold atoms trapped in optical lattices~\cite{eckardt2005a,eckardt2007}. This idea was implemented successfully, ten years ago, in pioneering experiments~\cite{lignier2007,Kierig:2008,sias2008}, which eventually led to the control of the Mott insulator--superfluid phase transition through external shaking~\cite{zenesini2009}; see~\cite{Morsch_2010,Eckardt:2016Review} for reviews. It was then realized~\cite{Eckardt_2010} that this method could be further exploited to generate artificial magnetic fields and frustrated magnetism, when applied to unusual lattice geometries, such as triangular lattices; this was demonstrated in a series of experiments, which explored classical frustrated magnetism~\cite{struck_2011}, tunable gauge potentials \cite{struck_2012}, and Ising-XY models \cite{struck_2013}, based on shaken optical lattices. Other recent developments studied how various forms of lattice shaking could produce staggered-vortex superfluids~\cite{Lim:2008}, uniform magnetic fields~\cite{Kolovsky:2011,Bermudez:2011,creffield_2014,creffield_2016}, non-Abelian gauge potentials~\cite{hauke_2012,kosior_2014}, spin-orbit coupling~\cite{struck_2014},  topological insulators~\cite{hauke_2012,baur_2014,Zheng:2014,nascimbene2015}, frustration in Fermi-Bose systems~\cite{sacha_2012}, lattices of sub-wavelength spacing~\cite{nascimbene2015}, helical 1D channels~\cite{Budich:2016}, and synthetic dimensions~\cite{Price:2016}. Moreover, the use of time-periodic pulse sequences was also proposed as a promising alternative to generate artificial magnetic fields~\cite{Sorensen:2005} and spin-orbit coupling~\cite{Xu:2013,Anderson:2013,goldman_2014} in cold-atom setups. Importantly, the experimental implementation of well-designed time-periodic modulations in ultracold gases recently allowed for the realization of emblematic models of topological quantum matter~\cite{Goldman:2016Review}, such as the Hofstadter~\cite{Miyake_2013,Aidelsburger_2013,Aidelsburger_2014,Kennedy:2015NP,Tai:2016} and Haldane~\cite{Jotzu_2014} models. 

In fact, the general strategy according to which a desired lattice model can be engineered by subjecting a quantum system to a proper time-periodic modulation has been applied to a wide range of physical platforms~\cite{oka_09,kitagawa2010,Calvo:2011,kitagawa11,lindner2011,cayssol2013,goldman_2014,bukov_2015,Goldman:2016Review,Eckardt:2016Review,Lu:2014Review}. This powerful method, which is generally termed \emph{Floquet engineering} in the contemporary condensed-matter literature, can be formulated in simple terms: Consider a lattice system described by a Hamiltonian $\hat H_0$, whose band structure is denoted $E_0 (\bs k)$, where $\bs k$ is the quasi-momentum; if one subjects this system to a high-frequency time-modulation $\hat V (t)$, its long-time dynamics can be accurately captured by an effective Hamiltonian $\hat H_{\text{eff}}$, which results from an interplay between the static Hamiltonian $\hat H_0$ and the modulation $\hat V (t)$. Interestingly, the spectrum $E_{\text{eff}} (\bs k)$ associated with the effective Hamiltonian $\hat H_{\text{eff}}$, the so-called Floquet bands, can show interesting features not present in the initial spectrum $E_0 (\bs k)$, such as non-trivial topological properties~\cite{oka_09,kitagawa2010,Rudner:2013,goldman_2014,Carpentier:2015,Nathan:2015}. 

Specific schemes leading to Floquet bands of non-trivial topological nature have been identified in a wide range of systems, including irradiated materials~\cite{oka_09,Calvo:2011,kitagawa11,lindner2011,cayssol2013,torres_2014}, ultracold gases in optical lattices~\cite{Aidelsburger_2014,Jotzu_2014,Flaschner_2016}, quantum walks~\cite{Kitagawa_2012,Cardano_2016,Cardano2016a,groh_2016}, photonic crystals~\cite{Rechtsman_2013,mukherjee_2016,Maczewsky:2016,Lu:2014Review}, and mechanical systems~\cite{Salerno:2016A,Huber:2016}. Such experiments demonstrated quantum-Hall responses~\cite{Aidelsburger_2014,Jotzu_2014} and dynamical phase transitions~\cite{Flaschner_2016} in cold atoms, and topological edge modes in photonics~\cite{Kitagawa_2012,Rechtsman_2013,mukherjee_2016,Maczewsky:2016}.  

In this context, the problem of how to prepare the system in the ground-state of a Floquet band structure appears as a recurrent and crucial question~\cite{eckardt2007,eckardt2009,zenesini2009,Baur:2013,goldman_2014,Aidelsburger_2014,nascimbene2015,Seetharam:2015,Kennedy:2015NP,Ho_2016,novicenko_2016}. Intuitively, one would propose to initially prepare the system in the ground-state associated with the static Hamiltonian $\hat H_0$, and then slowly ramp up the drive; indeed, this protocol would gently deform the bands $E_0 (\bs k)\!\rightarrow E_{\text{eff}} (\bs k)$, in view of reaching the desired Floquet state at the end of the ramp.  However, in general, the initial and final (target) states cannot be adiabatically connected, meaning that this preparation scheme is vitiated by an error, which is generically associated with inter-band (Landau-Zener) processes. Importantly, this issue necessarily arises when loading particles into topological Floquet bands, when starting from a topologically-trivial state, since a topological phase transition necessarily involves a non-adiabatic (gap closing) event~\cite{Hasan_2010}. Therefore, a cold-atom experiment that aims to study the topological properties of a specific Floquet band will necessarily present a finite fraction of its atomic cloud populating other bands, independently of heating processes associated with the drive~\cite{lacki_2012,lacki_2013,lazarides_2014,dalessio_2014,Bukov:2015,bilitewski_2015,Choudhury:2015,strater_2016,Bukov:2016,celi_2016,lellouch_2016}; see Ref.~\cite{Aidelsburger_2014} for measurements of such band-population dynamics. A natural question then arises:~What are the transport properties of such partially-filled topological Floquet bands in response to an applied force, and how do these compare to the standard topological responses (e.g.~the quantized Hall conductivity) associated with completely filled bands, as found in conventional static systems?

Recently, a series of works have investigated a connected problem, namely, the fate of quantum systems as one dynamically modifies the topological nature of their underlying band structure~\cite{caio_2015,dalessio_2015,caio_2016,hu_2016,wilson_2016,dehghani_2016,nurunal_2016,wangc_2016}; see Ref.~\cite{Flaschner_2016} for a recent experiment. These studies highlighted the subtle differences between three distinct notions, which become crucial in this out-of-equilibrium framework: (a) the topology of the evolving many-body state, (b) the topology of the underlying (instantaneous) single-particle band structure, and (c) the topological character of observable effects (e.g.~robust transport properties). For the sake of concreteness, let us illustrate these concepts on a striking example~\cite{caio_2015,dalessio_2015,caio_2016}: Consider an initial state $\vert \psi \rangle$, which corresponds to a completely filled band $E_{\text{trivial}} (\bs k)$ of trivial topology (i.e.~zero Chern number); one then performs a sudden change in the system parameters, such that the state $\vert \psi (t) \rangle$ evolves according to another Hamiltonian, which is associated with a non-trivial band structure $E_{\text{non-trivial}} (\bs k)$  characterized by non-zero Chern numbers. Since the time-evolution is assumed to be unitary, the Chern number related to the evolving state $\vert \psi (t) \rangle$ is necessarily constant, and hence, this quantity cannot be exploited to signal a change in the underlying bands' topology~\cite{caio_2015,caio_2016}. However, this change does manifest itself in a variety of observable phenomena, such as chiral edge currents~\cite{caio_2015,caio_2016,dehghani_2016}, robust bulk responses~\cite{hu_2016,wilson_2016,nurunal_2016} and vortex patterns in momentum distributions~\cite{Flaschner_2016,wangc_2016}. Interestingly, quantization of transport coefficients~\cite{nurunal_2016} and winding numbers~\cite{wangc_2016} can still be identified in this out-of-equilibrium context.

\subsection{Scope of the paper}

In this work, we propose to extend these previous studies by investigating a series of effects that are particularly relevant to the realization and observation of Floquet topological matter using ultracold atoms in optical lattices~\cite{Goldman:2016Review,Eckardt:2016Review}. First of all, we highlight the subtle issue of state preparation in schemes involving resonant time-modulations~\cite{goldman_2015,bukov_2015,Eckardt:2016Review}, and the related necessity to introduce a detuning in the state-preparation protocol to optimize the loading of atoms in a target Floquet band~\cite{Aidelsburger_2014}. Then, we study the behavior of the current density in response to an external force, applied at the end of the loading process, and analyse its strong irregularities~\cite{hu_2016}, which are due to inevitable Landau-Zener transitions during the state preparation. We then demonstrate how these irregularities in the current are in fact smoothened when probing another observable, namely, the centre-of-mass (COM) displacement of the atomic cloud in response to an applied force. Importantly, this shows how visualizing COM motion~\cite{dauphin_2013,Aidelsburger_2014,Price:2016} can indeed be used to extract the topology of the underlying (Floquet) bands, under a realistic state-preparation scheme.  Besides, we analyse the effects of the micro-motion associated with the time-periodic drive on currents and COM displacements. This study emphasizes the presence of three different characteristic time scales in the problem: (a) the long-time dynamics over which the topology of Floquet bands is measured (e.g.~through COM displacements), (b) the period of oscillations associated with the micro-motion due to the time-periodic drive, (c) the period of oscillations associated with inter-band interferences, due to the partial filling of the Floquet bands at the end of the loading sequence. We discuss how these different time scales affect current and COM responses.

\subsection{Outline}

The rest of the paper is organized as follows. In Section~\ref{sec:section1}, we discuss several notions related to the general framework of Floquet-engineered systems, and emphasize important subtleties associated with the state preparation and the topological characterization in these systems. We then establish equations of motion that capture the transport properties of partially-filled bands, which are relevant to describe the dynamics after the loading into topological Floquet bands. We then illustrate these concepts and results in Section~\ref{sec:model}, where we analyze an original two-band model leading to non-trivial Floquet bands of large flatness ratio. This Section aims to identify and highlight the diverse effects that affect the dynamics and transport properties of prepared Floquet states, such as the micro-motion due to the drive, inter-band interference effects, and corrections to the lowest-order effective-Hamiltonian approach. In particular, we numerically estimate how the Chern numbers extracted from the dynamics, and which characterize the topology of the underlying Floquet bands, depend on the duration of an optimized loading sequence. We present concluding remarks in Section~\ref{sec:conclusion}.


\section{General framework and notions}
\label{sec:section1}

In this first Section, we discuss general notions of Floquet engineering, emphasizing the issue of state preparation in the context of Floquet bands with non-trivial topology (Sections~\ref{sec:sec:floqueteng} and~\ref{sec:sec:prepstate}), with a special emphasis on the case of resonant modulations. This introductory part is also meant to define the notations used throughout the rest of the paper. We then introduce equations of motion to describe the transport properties of a generic prepared state, which includes the effects due to incomplete band populations, in Section~\ref{sec:sec:gscom}.  

\subsection{Floquet engineering and the effective Hamiltonian: A brief summary}
\label{sec:sec:floqueteng}


\subsubsection{General notions}

In this work, we analyse the transport properties of non-interacting particles moving on a lattice, whose band structure is modified by a time-periodic modulation. We are thus interested in systems that are described by a time-dependent Hamiltonian of the form~\cite{rahav_03_pra,goldman_2014,eckardt_15,mikami_15}
\begin{align}
\hat{H}(t)&=\hat{H}_0+\hat{V}(t), \label{eq:eqgen}\\
&=\hat{H}_0+\sum_{j=1}^{\infty}\hat{H}^{(j)} e^{ij \Omega t}+\hat{H}^{(-j)} e^{-ij \Omega t}, \notag
\end{align}
where $\hat{H}_0$ denotes a static Hamiltonian, and where $\hat{V}(t)$ is a time-periodic modulation of period $T=2\pi/\Omega$. Since the system is time periodic, the time-evolution operator can be written in the general form
\begin{equation}
\hat{U}(t,t_0)=e^{-i\hat{\mathcal{K}}(t)} e^{-i\hat{\mathcal{H}}_\text{eff}(t-t_0)}e^{i\hat{\mathcal{K}}(t_0)} ,
\label{eq:evolution}
\end{equation}
where $\hat{\mathcal{H}}_\text{eff}$ is a (static) effective Hamiltonian and where $\hat{\mathcal{K}}(t)$ is a time-periodic ``kick'' operator, which has a zero time-average over one period~\cite{rahav_03_pra,goldman_2014}. On the one hand, the effective Hamiltonian describes the long-time dynamics of the system.  On the other hand, the kick operator describes the micromotion undergone by the system within each period of the drive. In the following, we set $\hbar\!=\!1$ except otherwise stated.

In the high-frequency limit, namely when $ \Omega$ is much larger compared to all other characteristic energy scales in the problem (e.g.~the hopping amplitude in the lattice framework), both operators can be developed in a perturbative series in powers of $1/\Omega$ \cite{goldman_2014,eckardt_15,mikami_15}:
\begin{align}
\hat{\mathcal{H}}_\text{eff} &=\hat{H}_0+\frac{1}{\Omega}\sum_{j=1}^{\infty}\frac{1}{j}[\hat{H}^{(j)},\hat{H}^{(-j)}]+\mathcal{O}(1/\Omega^2), \label{eq:heffpkick}\\
\hat{\mathcal{K}}(t) &=\frac{1}{i\, \Omega}\sum_{j=1}^\infty\frac{1}{j }\left(\hat{H}^{(j)}e^{ij\Omega t}-\hat{H}^{(-j)}e^{-ij\Omega t}\right)+\mathcal{O}(1/\Omega^2). \notag
\end{align}
where the components $\hat{H}^{(j)}$ were introduced in Eq.~\eqref{eq:eqgen}. Importantly, the band structure associated with $\hat{\mathcal{H}}_\text{eff}$ can strongly differ from that related to the static Hamiltonian $\hat{H}_0$: this is the essence of Floquet engineering, which can thus be exploited to create non-trivial topological band structures by tailoring the time-modulation $\hat V (t)$. 

We recall that the effects of the micro-motion, as captured by the operator $\hat{\mathcal{K}}(t)$, can be put aside by probing the dynamics stroboscopically at times $t_{N}\!=\!T \times N$ where $N\!\in\!\mathbb{N}$ (we set $t_0\!=\!0$). Indeed, the form of the corresponding time-evolution operator [Eq.~\eqref{eq:evolution}]
\begin{equation}
\hat{U}(t_{N})=e^{-i t_N \hat{\mathcal{H}}_\text{F}}, \qquad \hat{\mathcal{H}}_\text{F}= e^{-i\hat{\mathcal{K}}(0)}\hat{\mathcal{H}}_\text{eff}e^{i\hat{\mathcal{K}}(0)} ,
\label{eq:strobo}
\end{equation}
indicates that the stroboscopical dynamics is completely governed by the band structure associated with the effective Hamiltonian $\hat{\mathcal{H}}_\text{eff}$; we recall that these so-called Floquet bands (or quasi-energy bands) are defined modulo $\Omega$, since $\hat{U}(t_{N})\!=\!e^{-i \frac{2 \pi N}{\Omega} \hat{\mathcal{H}}_\text{F}}$.

\subsubsection{Resonant time-modulations}

An important family of periodically-driven quantum systems is based on \emph{resonant} time-modulations~\cite{goldman_2015,bukov_2015,Eckardt:2016Review}; this corresponds to situations where the drive frequency $\Omega$ is taken to be resonant with some energy separation that is intrinsic to the static system. Such schemes were shown to provide a powerful and versatile tool for the design of topological Floquet bands in cold atoms~\cite{hauke_2012,baur_2014,goldman_2015}. In the following, we will be interested in such configurations, and we will focus on the case of two-site superlattices, with inequivalent lattice sites denoted $A$ and $B$. The corresponding time-dependent Hamiltonian will thus be taken to be of the form
\begin{align}
&\hat H(t) = \hat H_0 + \hat V (t), \qquad \hat V (t+2 \pi/ \Omega)=\hat V (t),  \\
&\hat{H}_0=\hat{T}+(\Omega+\delta)\sum_{B \text{ sites}}\hat c^\dagger_B \hat c_B,\label{ham_res}
\end{align}
where $\hat{T}$ describes hopping on the superlattice, and where the second term in \eqref{ham_res} describes an offset of amplitude $\Omega$ between $A$ and $B$ sites; we also introduced a detuning $\delta\ll \Omega$, which will play an important role below, when discussing the loading of atoms into Floquet bands. In the schemes described by Eq.~\eqref{ham_res}, the energy offset (and hence, the drive frequency $\Omega$) is generally chosen to be much larger than all tunneling amplitudes associated with the hopping term $\hat{T}$, which means that tunneling is suppressed in the absence of the periodic drive. The latter then restores tunneling through resonant processes, so that tunneling amplitudes can then be controlled by tuning the modulation strength. Furthermore, the phase of the drive can be used to induce artificial gauge fields in the effective lattice system, opening the possibility to generate (Floquet) bands with non-trivial topological features, such as non-zero Chern numbers \cite{hauke_2012,baur_2014,Aidelsburger_2014,goldman_2015}. 

As a technical remark, let us note that although the time-dependent Hamiltonian \eqref{ham_res} has an apparent diverging term proportional to $\Omega$ when considering the infinite-frequency limit $\Omega\!\rightarrow\infty$, one may still write the perturbative series for the effective Hamiltonian, Eq~\eqref{eq:heffpkick}, in a moving frame~\cite{goldman_2014,goldman_2015,bukov_2015}.

\subsection{Topology of Floquet bands and state preparation}
\label{sec:sec:prepstate}


\subsubsection{Topology in the high-frequency regime}

In static systems, Bloch bands present two fundamental properties, which are local in quasi-momentum space: their band velocity $\partial_{\bs k} E (\bs k)$ and their Berry curvature $\mathcal{F} (\bs k)$. Considering two-dimensional lattices, the Berry curvature in a given band $E (\bs k)$ reads~\cite{xiao_2010}
\be
\mathcal{F} (\bs k) = i \left ( \langle \partial_{k_x} u \vert \partial_{k_y} u \rangle -  \langle \partial_{k_y} u \vert \partial_{k_x} u \rangle \right ),
\ee
where $\vert u (\bs k) \rangle$ denotes the Bloch states in the band. The Berry curvature can be integrated over the first Brillouin zone (BZ) to give the topologically-invariant Chern number of the band, $$\nu \!=\! (1/2 \pi) \int_{\text{BZ}} \mathcal{F}(\bs k) \, \text d^2k.$$ A striking manifestation of this topological invariant is found in the integer quantum Hall effect, where the Hall conductivity of a completely filled band with Chern number $\nu$ is quantized according to the TKNN formula: $\sigma_H\!=\!(e^2/h) \nu$, where $e$ denotes the electron charge~\cite{xiao_2010,Hasan_2010}. 

As in the case of static systems, the Floquet (quasi-energy) bands $E_{\text{eff}} (\bs k)$ associated with the effective Hamiltonian~\eqref{eq:heffpkick}-\eqref{eq:strobo} can also display geometrical properties~\cite{kitagawa2010}, as captured by the Berry curvature $\mathcal{F}_{\text{eff}} (\bs k)$; in this case, 
\be
\mathcal{F}_{\text{eff}} (\bs k) = i \left ( \langle \partial_{k_x} u_{\text{eff}} \vert \partial_{k_y} u_{\text{eff}} \rangle -  \langle \partial_{k_y} u_{\text{eff}} \vert \partial_{k_x} u_{\text{eff}} \rangle \right ),
\ee
where $\vert u_{\text{eff}} (\bs k) \rangle$  refers to the Bloch states associated with the effective band $E_{\text{eff}} (\bs k)$, namely, the eigenstates of the effective Hamiltonian (or Floquet states). Similarly, in the high-frequency limit of the drive, the topology of a Floquet band can be characterized by the Chern number $\nu_{\text{eff}} \!=\! (1/2 \pi) \int_{\text{BZ}} \mathcal{F}_{\text{eff}} \, \text d^2k$; see Refs.~\cite{kitagawa2010,Rudner:2013,Nathan:2015}. Hence, in this regime of the drive, the topological characterization of Floquet systems is strictly equivalent to that of static systems. This indicates that completely filling a Floquet band with non-zero Chern number $\nu_{\text{eff}} \!\ne\!0$ should give rise to topological responses reminiscent of the quantum-Hall effect. 

One should point out that, away from the high-frequency limit of the drive, the micro-motion can strongly affect the topological characterization presented above~\cite{kitagawa2010,Rudner:2013,Carpentier:2015,Nathan:2015,mukherjee_2016,Maczewsky:2016}, which is entirely based on the properties of the effective Hamiltonian only. However, such a situation will not be addressed in this work, which focuses on the high-frequency regime of Floquet systems. 

\subsubsection{State preparation: General considerations}\label{Section_preparation}

\begin{figure}[h]
\begin{center}
	\includegraphics[scale=0.9]{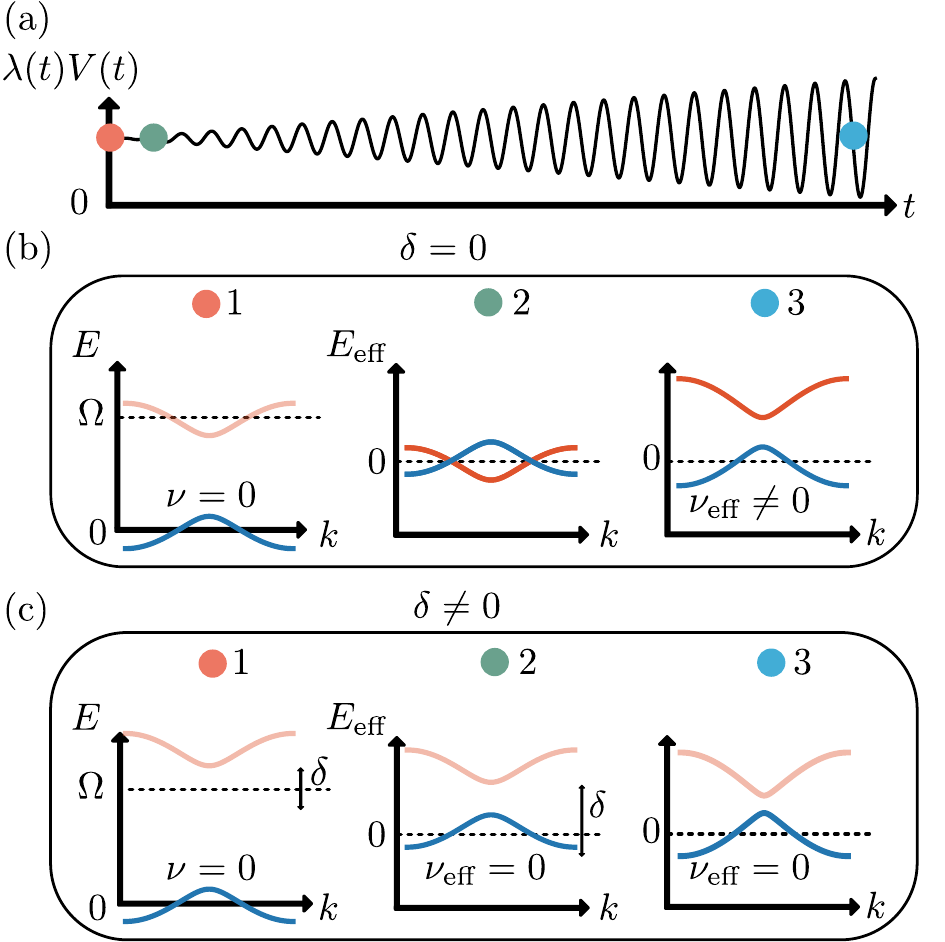}
	\caption{The loading into Floquet bands: Turning on the time-modulation $V(t)$. (a) Sketch of the ramp considered to progressively increase the strength of the time-modulation, $\lambda (t)V(t)$. (b) Illustration of the corresponding (instantaneous) band structure and population: (1) Initially, the lowest band associated with the static Hamiltonian $\hat H_0$ is completely filled; the system is topologically trivial, as dictated by the Chern number $\nu\!=\!0$ of the populated band; the two bands are separated by a gap of order $\Omega$ due to the offset between $A$ and $B$ sites [Eq.~\eqref{ham_res}]; (2) As soon as the time-modulation is turned on, the band structure is described in terms of instantaneous Floquet bands, $E_{\text{eff}}$, which are defined modulo $\Omega$; in this picture, the bands strongly overlap, thus leading to severe band repopulation during the early stage of the ramp [$\lambda (t)\!\approx\! 0$]; (3) At the end of the ramp, the Floquet bands are topologically non-trivial, as marked by non-zero Chern numbers $\nu_{\text{eff}}\!\ne\!0$; here, both Floquet bands (with opposite Chern numbers) are largely populated, indicating the inefficiency of this scheme to prepare the system in the lowest Floquet band only. (c) Same protocol as for (b), except that a detuning $\delta$ has been introduced [Eq.~\eqref{ham_res}] so as to avoid any overlap between the Floquet bands during the entire ramp $\lambda(t)V(t)$; at the end of the sequence, only the lowest band is occupied; in this case, the populated Floquet band has a trivial topology $\nu_{\text{eff}}\!=\!0$, due to the large detuning $\delta$. A controlled transition, from trivial to non-trivial Floquet bands, can then be induced by removing the detuning; see Fig.~\ref{fig:figsketch}.}
\label{fig:detuning}
\end{center}
\end{figure}

A crucial issue therefore concerns the possibility of realizing a state that completely fills a desired Floquet band of non-trivial topology ($\nu_{\text{eff}} \!\ne\!0$). As discussed in the introductory section, such a state preparation is spoiled by the topological transition that necessarily occurs when deforming the initial (trivial) bands  into the desired Floquet bands with non-zero Chern number. Formally, the state-preparation scenario is described by a time-dependent Hamiltonian of the form
\be
\hat H(t) = \hat H_0 + \lambda (t) \hat V (t),\label{prep_scheme}
\ee
where the operators have the same properties as in Eq.~\eqref{ham_res}, and where we have introduced the ramp function $\lambda (t)$, which is zero at time $t\!=\!t_0$, and which then smoothly reaches its final value $\lambda (\tau)\!=\!1$ after some duration $\tau$; see Fig.~\ref{fig:detuning} (a). If the ramp duration is taken to be much longer than the driving period $T$, there is a separation of time scales, and one can then analyze the system during the ramp in terms of an \emph{instantaneous} effective Hamiltonian $\hat{\mathcal{H}}_\text{eff}(t)$, which is now evaluated locally in time (i.e.~using Eq.~\eqref{eq:heffpkick} for fixed values of $\lambda$); see also Ref.~\cite{novicenko_2016} for a more rigorous treatment of slowly-varying time-modulating systems. In this simplest picture, the instantaneous Floquet bands $E_{\text{eff}} (\bs k; \lambda (t))$, and hence the corresponding instantaneous Chern numbers $\nu_{\text{eff}} (\lambda (t))$, can be evaluated dynamically during the ramp; note that one eventually recovers the Chern number of the desired Floquet band $\nu_{\text{eff}}$ for times $t \!\ge\! \tau$.

We now describe a specific situation of interest, where one starts with a trivial static system, described by $\hat H_0$, and then ramps up the drive $\hat V (t)$ to generate Floquet bands with non-trivial topology ($\nu_{\text{eff}} \!\ne\!0$). For the sake of simplicity, we consider the case of two-band models, noting that generalizations to multi-band systems are straightforward. In this case, the two instantaneous Floquet bands necessarily undergo a gap-closing event at some critical time $t^* \!<\! \tau$, accompanied with a change in the instantaneous Chern numbers 
\be
\nu_{\text{eff}} (t < t^*)\!=\!0\!\longrightarrow\!\nu_{\text{eff}}  (t > t^*)\!\ne\!0.
\ee 

Importantly, if the initial state corresponds to a completely filled trivial band ($\nu\!=\!0$), the Chern number associated with the evolving (many-body) state $\vert \psi (t) \rangle$, which is defined by
\be
\nu_{\text{state}} (t) = \frac{i}{2 \pi} \int \text d^2k \left ( \langle \partial_{k_x} \psi \vert \partial_{k_y} \psi \rangle -  \langle \partial_{k_y} \psi \vert \partial_{k_x} \psi \rangle \right ),\label{Chern_state}
\ee 
necessarily remains zero at all times~\cite{dalessio_2015,caio_2015,caio_2016}. However, the invariance of $\nu_{\text{state}} (t)$ under unitary time-evolution is not incompatible with a change in the  Chern numbers of the instantaneous Floquet bands defined above $\nu_{\text{eff}} (t)\!\ne\!\nu_{\text{state}} (t)$. In particular, the triviality of the evolving state [$\nu_{\text{state}} (t) \!=\!0$] does not preclude the observation of the underlying Floquet bands topology. Indeed, by minimizing the impact of the gap closing event (using well-designed ramps~\cite{Ho_2016}), one can optimize the loading of particles into a given topological Floquet band, and observe clear manifestations of its topology. Surprisingly, as we discuss below, the gap closing can strongly affect certain observables, even in the limit of very long ramps; see also Ref.~\cite{hu_2016}.

\begin{figure}[h]
\begin{center}
	\includegraphics[scale=1]{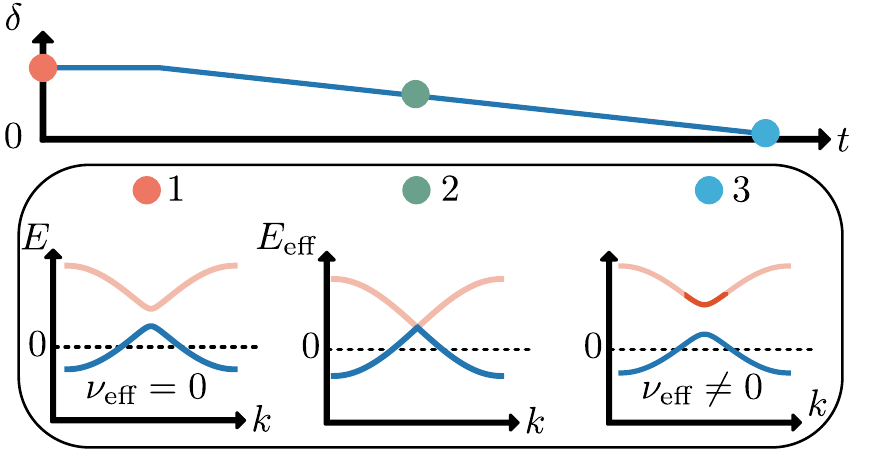}
\end{center}
\caption{The second step of the loading sequence. As the detuning is progressively decreased, the Floquet bands introduced in Fig.~\ref{fig:detuning}(c) undergo a controlled topological phase transition, leading to a limited excited fraction in the upper band. At the end of the sequence, most of the particles occupy the lowest Floquet band, with non-zero Chern number $\nu_{\text{eff}}\!\ne\!0$. This second step of the loading protocol can be optimized by increasing the duration of the ramp $\delta (t)$, i.e.~by minimizing Landau-Zener transitions during the gap closing event; see also Eq.~\eqref{eq:lz}.}
\label{fig:figsketch}
\end{figure}

\subsubsection{State preparation in the case of resonant modulations: Introducing a two-step loading sequence}\label{Section_preparation}

Before discussing this aspect further, let us emphasize another interesting issue, which specifically appears when the drive frequency $\Omega$ is chosen to be (quasi-)resonant with energy offsets inherent to the static Hamiltonian $\hat H_0$, as in Eq.~\eqref{ham_res}. In this configuration, the energy bands are initially separated by a gap of order $\Omega$ in the absence of the drive ($\lambda\!=\!0$). Let us now discuss the band-loading scheme, Eq.~\eqref{prep_scheme}, assuming that our initial state consists of the lowest-energy band being completely filled. Importantly, as soon as the time-periodic modulation is turned on ($\lambda\!\ne\!0$), our theoretical description replaces these static bands $E (\bs k)$ by instantaneous Floquet (quasi-energy) bands $E_{\text{eff}} (\bs k; \lambda (t))$; as illustrated in Fig.~\ref{fig:detuning} (b), although the initial bands were gapped, the instantaneous Floquet bands are gapless at the early stage of the ramp due to the fact that these are defined modulo $\Omega$. As the drive amplitude is increased, particles then distribute themselves between the two Floquet bands, through strong and uncontrollable inter-band transitions. Although the final Floquet bands are well isolated and topologically non-trivial at the end of the ramp, the system will be far from constituting a topological insulating state, due to the large population of particles in both Floquet bands; Fig.~\ref{fig:detuning} (b).

One way to avoid this issue is to introduce a detuning $\delta$, as in Eq.~\eqref{ham_res}, so as to avoid any gap closing event during the ramp performed on the drive amplitude $\lambda (t)$; see Fig.~\ref{fig:detuning} (c). Indeed, the detuning can be chosen such that a single Floquet band remains completely populated during the entire duration of the ramp; note that in this case, both Floquet bands then remain topologically trivial over the entire ramp. However, one can then perform a second ramp on the detuning $\delta\!\rightarrow\delta(t)$, so as to induce a well-controlled topological phase transition. The advantage of this protocol relies on that the related gap closing event typically occurs at a few singular points in the Brillouin zone, see Fig.~\ref{fig:figsketch}. Hence, the particle transfer to the other Floquet band only takes place through Landau-Zener transitions within these singular band-touching regions, which can be limited by using ramps of long duration. 

We point out that the introduction of a detuning was also considered in the loading process of Ref.~\cite{Aidelsburger_2014}; however, in that work, this effect was used to control the number of (Floquet) bands during the launch of the drive, and not to inhibit inter-band transitions.

\subsection{Equations of motion for partially-filled bands: current vs center-of-mass drift}
\label{sec:sec:gscom}
 
In the previous Section, we discussed how the loading of particles into topological Floquet bands is necessarily associated with a gap-closing event, hence leading to partially-filled bands at the end of the ramp.  In this Section, we investigate how this partial filling of the bands influence the transport equations in the presence of an external force (added after the loading process). In particular, the calculations presented below highlight the drastically different behaviors expected for current and center-of-mass observables in the case of partially-filled bands.

In this Section, we disregard any effect related to the micro-motion due to the drive. In this sense, the results presented below are general: they can be used to describe transport in static systems with partially-filled energy bands, but also, the stroboscopic time-evolution of periodically-driven quantum systems with partially-filled Floquet bands. The notations used below are thus chosen to be as general as possible.

\subsubsection{Equations of motion}

Let us consider a general two-band system, with bands denoted $E_{1,2}(\bs k)$. We are interested in the time-evolution of states $| \psi_\mathbf{k}\rangle$, defined at  quasi-momentum $\mathbf{k}$, which are given by a linear combination of the Bloch states $|u_1(\mathbf{k})\rangle$ and $|u_2(\mathbf{k})\rangle$ associated with the two bands:
\begin{equation}
| \psi_\mathbf{k}\rangle=\alpha(\mathbf{k})|u_1(\mathbf{k})\rangle+\beta(\mathbf{k})|u_2(\mathbf{k})\rangle .\label{state_alpha_beta}
\end{equation}
In the following, we assume that the coefficients satisfy $\vert \alpha(\mathbf{k}) \vert ^2+\vert \beta(\mathbf{k}) \vert ^2\!=\!1$, meaning that the particle density in the system is given by $n\!=\!1/A_{\text{cell}}$, where $A_{\text{cell}}$ denotes the area of the unit cell. We note that this indeed corresponds to the situation treated in this work: one starts from a completely filled band ($n\!=\!1/A_{\text{cell}}$), then the particles  simply distribute themselves between the two Floquet bands during the loading process, while keeping the density fixed.

In the presence of an external force $\mathbf{F}$, the dynamics of the state $| \psi_\mathbf{k}\rangle$ is found to be governed by the following equations of motions (see Appendix~\ref{sec:semiclassic}):
\begin{align}
& \mathbf{k}=\mathbf{k}_0+\mathbf{F} \, t, \label{k-time} \\
& \mathbf{v}(\mathbf{k})=\mathbf{v}_\text{band}(\mathbf{k})+\mathbf{v}_\mathcal{F}(\mathbf{k})+\mathbf{v}_\text{inter}(\mathbf{k}),\label{velo_three}
\end{align}
where $\mathbf{v}(\mathbf{k})$ denotes the average velocity in the state $| \psi_\mathbf{k}\rangle$. The two first contributions to the velocity are intuitive, as they simply correspond to the usual band and anomalous velocities~\cite{xiao_2010} weighted by the band populations, namely
\begin{align}
&\mathbf{v}_\text{band} (\bs k)=\vert \alpha (\bs k) \vert^2 \partial_\mathbf{k}E_{1}+\vert \beta (\bs k) \vert^2 \partial_\mathbf{k}E_{2},
\label{eq:band}\\
&\mathbf{v}_\mathcal{F}(\bs k)= - \, \mathbf{F}\times\mathbf{1}_z (\vert \alpha(\bs k) \vert^2 \mathcal{F}^1+\vert \beta(\bs k) \vert^2 \mathcal{F}^2),
\label{eq:berry}
\end{align}
where $\mathcal{F}^{1,2}$ denote the Berry curvature of the two bands, and where the populations $\alpha(\mathbf{k})$ and $\beta(\mathbf{k})$ implicitly depend on time through Eq.~\eqref{k-time}. In the following, we suppose that the applied force is sufficiently weak so as to neglect inter-band transitions, which would lead to an extra time-dependence of these coefficients. The third contribution in Eq.~\eqref{velo_three} corresponds to an inter-band-interference effect, which takes the explicit form (see Appendix~\ref{sec:semiclassic})
\begin{align}
\mathbf{v}_\text{inter}\!=\!2\text{Re}\Big \{& \alpha^*\beta\exp{\left [-i\int^t_{t_0}\text{d}t \, (E_2-E_1) \right ]}\!\exp{\left [i(\gamma_1-\gamma_2)\right ]} \notag \\
&\times\Big [(E_1-E_2)\langle\partial_\mathbf{k}u_1|u_2\rangle \notag \\
&\quad +i\,\mathbf{F}\cdot\dfrac{\langle\partial_\mathbf{k}u_1|u_2\rangle}{E_1-E_2}(\partial_\mathbf{k}E_2-\partial_\mathbf{k}E_1)\Big ] \Big \},
\label{eq:vinterband}
\end{align}
where $\gamma_i(t)\!=\!i\int_{t_0}^t\text{d}t'\langle u_i(t')|\partial_tu_i(t')\rangle$ is associated with the usual geometric phase of the band $i$~\cite{xiao_2010}, and where $t_0$ denotes the time at which the force is applied. We point out that this inter-band contribution does not correspond to a transfer of particles between the bands, but rather, it describes an interference effect between particles occupying the two bands. Interestingly, the factor  $\exp{[-i\int \text{d}t (E_2-E_1)]}$ in Eq.~\eqref{eq:vinterband} can lead to strong oscillations in the inter-band velocity $\mathbf{v}_\text{inter}(t)$. The period of these oscillations can be roughly related to the quantity $2\pi/(E_2-E_1)$, which is bounded by $2\pi/\Delta_{\text{gap}}$, where $\Delta_{\text{gap}}$ denotes the energy gap. Hence, in transport experiments operating in the linear regime (i.e.~$a F\!\ll\!\Delta_{\text{gap}}$, where $a$ is the lattice spacing and $F$ the strength of the applied force), these oscillations are much faster than typical Bloch oscillations, $T_{\text{Bloch}}\!=\!2 \pi/(aF)\!\gg\!2\pi/\Delta_{\text{gap}}$, which sets a natural time scale for realistic observation times. 

We point out that the inter-band contribution to the velocity $\mathbf{v}_\text{inter}$ vanishes when integrated over one period of its oscillation, under the reasonable assumption that all the other factors in Eq.~\eqref{eq:vinterband} are constant over each period. Consequently, isolating the effects of the anomalous velocity $\mathbf{v}_\mathcal{F}$ and band velocity $\mathbf{v}_\text{band}$ would require, in general, to perform a time-average of the dynamics over long observation times $t\!\gtrsim\!T_{\text{Bloch}}\!\gg\!2\pi/\Delta_{\text{gap}}$. We remind that the band velocity $\mathbf{v}_\text{band}$ can be annihilated by considering uniform filling of the bands~\cite{dauphin_2013,Aidelsburger_2014,price_2016}, or using model-dependent symmetry properties~\cite{dauphin_2013}, or by performing two successive transport experiments using opposite forces~\cite{Price:2012}.  

\subsubsection{Currents and center-of-mass observables}

We now discuss the transport properties of the system introduced in Section~\ref{Section_preparation}. Since the loading sequence starts with a completely filled band, the current density after the ramp is given by the contribution of the many states~\eqref{state_alpha_beta} corresponding to all quasi-momenta $\bs{k}$ within the entire BZ; the explicit form of the coefficients $(\alpha (\bs k),\beta (\bs k))$ is determined by the model under consideration and the duration of the second ramp [Fig.~\ref{fig:figsketch}]. Using Eq.~\eqref{velo_three}, the total current density is then given by the general expression
\begin{equation}
\mathbf{j}(t)=\dfrac{1}{(2\pi)^2}\int_\text{BZ}\text d^2k \,\left (\mathbf{v}_\text{band}(\mathbf{k},t)+\mathbf{v}_\mathcal{F}(\mathbf{k},t)+\mathbf{v}_\text{inter}(\mathbf{k},t) \right) ,\label{current_density}
\end{equation}
where the three time-dependent contributions to the velocity are given in Eqs.~\eqref{eq:band}-\eqref{eq:vinterband}. As the particle density is constant and given by $n\!=\!1/A_{\text{cell}}$, the center-of-mass (COM) velocity is then simply given by~\cite{dauphin_2013,price_2016}
\be
\mathbf{v}_\text{CM}(t)\!=\!A_{\text{cell}}\;\mathbf{j}(t).\label{COM_vel}
\ee 
Hence, we observe that both the current density and the COM velocity are typically rapidly-oscillating quantities due to the inter-band contribution $\mathbf{v}_\text{inter}(\mathbf{k},t)$.

In contrast, the center-of-mass drift given by
\begin{align}
\bs x_\text{CM}&=\bs x_0+\int_{t_0}^t \text d t' \mathbf{v}_\text{CM}(t'),
\label{eq:xcm}\\
&\approx \bs x_0+ \dfrac{A_{\text{cell}}}{(2\pi)^2} \int_{t_0}^t \text d t' \int_\text{BZ}\text d^2k \left (\mathbf{v}_\text{band}(\mathbf{k},t)+\mathbf{v}_\mathcal{F}(\mathbf{k},t) \right ), \notag
\end{align}
constitutes a more stable observable, as the time-integration in Eq.~\eqref{eq:xcm}, when performed over reasonable observation times $t\!\gg\!2\pi/\Delta_{\text{gap}}$, annihilates the contribution from the oscillating inter-band velocity. 

Finally, we discuss the relevant case where the bands are filled in a uniform manner after the ramp, i.e.~when the coefficients $\alpha$ and $\beta$ are constant numbers. In this case, the band-velocity contribution to the COM dynamics \eqref{eq:xcm} vanishes. Furthermore, the contribution due to the anomalous velocity is then directly proportional to the Chern numbers of the individual (Floquet) bands, $\nu_1$ and $\nu_2$. In the long-time limit, where the inter-band velocity contribution vanishes, the COM displacement  \eqref{eq:xcm} then yields
\begin{equation}
\begin{split}
\bs x_\text{CM}&=\bs x_0+ t\dfrac{A_{\text{cell}}}{(2\pi)^2}\, \int_\text{BZ} \text d^2k \, \mathbf{v}_\mathcal{F} (\bs k)\\
&=\bs x_0 - \frac{tA_{\text{cell}}}{2\pi}\,  \mathbf{F}\times\mathbf{1}_z \left (\vert \alpha \vert^2 \nu_1+\vert \beta \vert^2 \nu_2 \right).
\end{split}
\end{equation}
Hence, in the case of uniformly-filled (Floquet) bands, one recovers the COM behavior that was observed in the experiment of Ref.~\cite{Aidelsburger_2014}, where the uniformity of the band-populations was confirmed through band-mapping measurements. One should emphasize that in the general case, the population of the Floquet bands after the ramp (and hence, during the transport experiment) are non-uniform~\cite{Jotzu_2014}, in which case the contribution of the band velocity to transport should generally be taken into account.

\subsubsection{Hierarchy of oscillation frequencies}

We conclude this Section by emphasizing the existence of three distinct frequencies that potentially appear in the responses of Floquet systems: the drive frequency $\Omega$, the frequency related to the inter-band velocity $\omega_{\text{int}}\!\sim\!\Delta_{\text{gap}}$, and the Bloch oscillation frequency $\omega_{\text{B}}\!=\!a F$. In usual situations, we have the hierarchy
\be
\Omega\gg\omega_{\text{int}} \gg\omega_{\text{B}}.
\ee
Hence, one generally recovers the behavior of standard static systems (as captured by the single-band semi-classical equations of motion~\cite{xiao_2010}), whenever one applies a low-pass filter at the frequency $\omega_{\text{int}}$ to the response signal; see Fig.~\ref{fig:vcmfull} in Section~\ref{section_chern_measurement} for an illustration.


\section{The model}
\label{sec:model}
In this section, we illustrate the general results discussed in Section \ref{sec:section1} by solving numerically the dynamics of a specific two-band model, which realizes Floquet bands with non-zero Chern numbers. We first introduce an original time-dependent Hamiltonian, well suited for cold-atom implementations, and we derive the related effective Hamiltonian. Then, we study the loading into a specific Floquet band of non-trivial topology, starting from a trivial configuration, and analyze the corresponding transport properties. The latter are systematically compared to the theoretical predictions of Section~\ref{sec:sec:gscom}. 

\subsection{Time-dependent Hamiltonian and effective flat-band model}
\label{sec:hamiltonian}

\begin{figure}[h]
\begin{center}
	\includegraphics[scale=1]{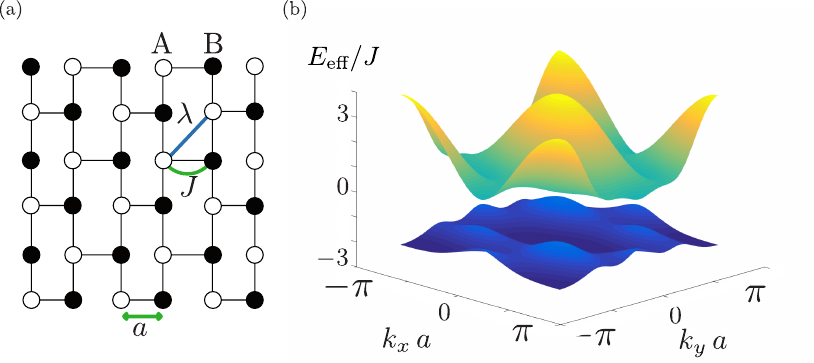}
\end{center}
	\caption{ (a) Sketch of the brickwall lattice and its NN and NNN links (with hopping parameters $J$ and $\lambda$, respectively). (b) Energy spectrum associated with effective Hamiltonian in Eq.~\eqref{eq:heff}, in the absence of detuning  $\delta\!=\!0$. The other system parameters satisfy $J^\text{eff}_{ij}\!=\!J$, $\lambda^\text{eff}_{ij}\!=\!0.3J$ and $\mathbf{q}=(6,2)/a$. The lowest quasi-flat band is separated from the higher band by a gap $\Delta_{\text{gap}}\!=\!1.73J$, and it is topologically non-trivial, with Chern number $\nu_{\text{eff}}\!=\!1$.}
\label{fig:lattespec}
\end{figure} 

In this Section, we introduce a novel scheme that realizes a two-band Hamiltonian leading to non-trivial Floquet bands, with the advantage that one of the corresponding topological bands can be made almost perfectly flat. Inspired by Ref.~\cite{baur_2014}, we combine a 2-site superlattice with a resonant time-modulation so as to realize the Chern-insulator model of Refs.~\cite{Alba_2011,goldman_2013,Anisimovas:2014}, which indeed features non-trivial flat bands in certain parameters regimes~\cite{goldman_2013}.

Specifically, we consider non-interacting particles on a time-modulated brickwall lattice~\cite{Jotzu_2014}, of lattice spacing $a$; see Fig.~\ref{fig:lattespec}(a). Considering a tight-binding approach, the second-quantized Hamiltonian is taken to be of the form
\begin{equation}
\hat{H}(t)=\hat{H}_0+\hat{V}(t),
\label{eq:hamiltonientemporel}
\end{equation}
where the static part is given by
\begin{equation}
\hat{H}_0=-\sum_{\langle i,j\rangle}J_{ij}\hat{c}^\dagger_i\hat{c}_j-\sum_{\langle\langle i,j\rangle\rangle}\lambda_{ij}\hat{c}^\dagger_i\hat{c}_{j}+\Delta\sum_{\text{$B$ sites }}\hat{c}^\dagger_i\hat{c}_i.
\label{eq:hstatic}
\end{equation}
This Hamiltonian includes nearest-neighbor (NN) and next-to-nearest neighbor (NNN) hopping terms, which are characterized by the tunneling matrix elements $J_{ij}$ and $\lambda_{ij}$, respectively. The static Hamiltonian also includes a large energy offset $\Delta\!\gg\! (J_{ij}, \lambda_{ij})$ between $A$ and $B$ sites, which inhibits bare NN tunneling. The time-dependent part of the Hamiltonian $\hat{H}(t)$ consists in an onsite modulation that acts on all the sites, with frequency $\Omega$,
\begin{equation}
\hat{V}(t)=K\sum_i\cos{\left(\Omega t+\mathbf{q}\cdot\mathbf{r}_i\right)}\hat{c}^\dagger_i \hat{c}_i,\label{drive_def}
\end{equation}
where $\mathbf{r}_i$ denotes the site positions. In an actual cold-atom implementation, the quantity $\mathbf{q}$ would be directly related to the wave vectors associated with the lasers generating the lattice modulation~\cite{baur_2014,goldman_2015}. Considering the case of a (nearly) resonant modulation $\Delta=\Omega+\delta$, and considering the high-frequency limit $\Omega\gg(\delta, J_{ij}, \lambda_{ij})$, the effective Hamiltonian reads \cite{goldman_2015}
\begin{align}
\hat{\mathcal{H}}_\text{eff}=&-\sum_{\langle i,j\rangle} iJ^\text{eff}_{ij}e^{i\, \mathbf{q}\cdot(\mathbf{r}_i+\mathbf{r}_j)/2}\hat{c}^\dagger_i\hat{c}_j-\sum_{\langle\langle i,j\rangle\rangle}\lambda^\text{eff}_{ij}\hat{c}^\dagger_i\hat{c}_{j}\notag \\
&+\delta\sum_{\text{$B$ sites}}\hat{c}^\dagger_i\hat{c}_i+\mathcal{O}\left(\dfrac{1}{\Omega}\right),
\label{eq:hameffzero}
\end{align}
where the effective tunneling matrix elements are given by (see Appendix~\ref{app:derivham})
 \begin{equation}
 \begin{cases}
 J^\text{eff}_{ij}=(-1)^\alpha J_{ij}\times\mathcal{J}_1\left[\dfrac{2K}{\Omega}\sin{\left(\mathbf{q}\cdot\dfrac{\mathbf{r}_j-\mathbf{r}_i}{2}\right)}\right],\\
 \lambda^\text{eff}_{ij}=\lambda_{ij}\times\mathcal{J}_0\left[\dfrac{2K}{\Omega}\sin{\left(\mathbf{q}\cdot\dfrac{\mathbf{r}_j-\mathbf{r}_{i}}{2}\right)} \right].
 \end{cases}
 \label{eq:renormalization}
 \end{equation}
Here  $\mathcal{J}_{0,1}(x)$ denote the Bessel functions of the first kind, and the sign function $(-1)^\alpha$ is positive for hopping from $A$ to $B$ sites and negative otherwise. Interestingly, the first line in Eq.~\eqref{eq:hameffzero} corresponds to the (static) Hamiltonian of Ref.~\cite{Alba_2011,goldman_2013,Anisimovas:2014}. Building on the results of Ref.~\cite{goldman_2013}, we propose to set the system parameters such as to realize the special values $J_{ij}^\text{eff}\!=\!J$, $\lambda_{ij}^\text{eff}\!=\!0.3J$ and $\mathbf{q}\!=\!(6,2)/a$, which realize a non-trivial topological band structure with a quasi-flat lowest band; see Fig.~\ref{fig:lattespec}(b) and Appendix~\ref{app:derivham}. Besides, in the following, we set the modulation strength to the value $K\!=\!2\Omega$. 

In the second line of Eq.~\eqref{eq:hameffzero}, the detuning  $\delta$ appears as an effective offset between $A$ and $B$ sites; according to Ref.~\cite{goldman_2013}, the offset $\delta$ can be tuned so as to generate topological phase transitions, from a topological configuration ($\delta\!=\!0$) to a trivial configuration ($\delta\!\gg\!0$).

\subsection{Preparation of the initial state using two successive ramps}
\label{sec:prepgs}

In this Section, we analyze the loading of particles into the lowest (Floquet) band associated with the effective Hamiltonian derived in the previous Section [Fig.~\ref{fig:lattespec}(b)]. Since we deal with a scheme based on resonant modulations, we study a loading protocol that consists of two successive ramps (i.e.~a ramp on the modulation strength $K(t)$ followed by a ramp on the detuning $\delta$), as introduced in Section~\ref{Section_preparation}. This analysis allows one to estimate the populations in the two Floquet bands at the end of the loading process, which will be crucial for our study of transport. In particular, we discuss the inter-band transfer that occurs during the inevitable gap-closing event, using the Landau-Zener formula.

\subsubsection{Starting with a trivial ground state}

Let us first describe the initial state that we use in our numerical calculations. We consider that particles are initially confined in a certain disc of radius $r$, in the absence of the modulation. The spectrum of the static Hamiltonian $\hat H_0$, Eq.~\eqref{eq:hstatic}, displays a large energy gap $\Delta_{\text{gap}}\!\sim\! \Delta$ due to the offset between $A$ and $B$ sites, and we assume that all the states corresponding to the lowest band are uniformly populated. This situation could either correspond to a Fermi gas where the Fermi energy is located within the gap, or a thermal Bose gas for which the temperature is large compared to the band-width of the lowest band but small compared to the band-gap. In our numerical simulations, we propose to describe the latter situation and consider an incoherent state of the form
\begin{equation}
\vert \psi_0 \rangle=\sum_{i \in \,\text{lowest band}} e^{i \theta_i}\vert \phi_i \rangle ,
\end{equation}
which we then project unto the disc of radius $r$; see also Ref.~\cite{price_2016}. Here, $ \vert \phi_i\rangle$ denotes the eigenstates of the static Hamiltonian $\hat{H}_0$ and $\theta_i$ are random phases. When averaged over about $50$ realizations, we verified that the initial state indeed uniformly populates the lowest band of the static Hamiltonian $\hat H_0$ with a 99.9\% probability. Besides, we also verified that the time-evolution of the initial state, to be discussed below, well converges when averaging the dynamics over $50$ realizations. 

We emphasize that this choice of the initial state is not crucial, as the results presented below can equally be obtained starting with a completely filled band of fermions. However, this choice does offer practical advantages in terms of numerical calculations.

\subsubsection{Ramping up the time-modulation}

\begin{figure}[h!]
\begin{center}
	\includegraphics[scale=1]{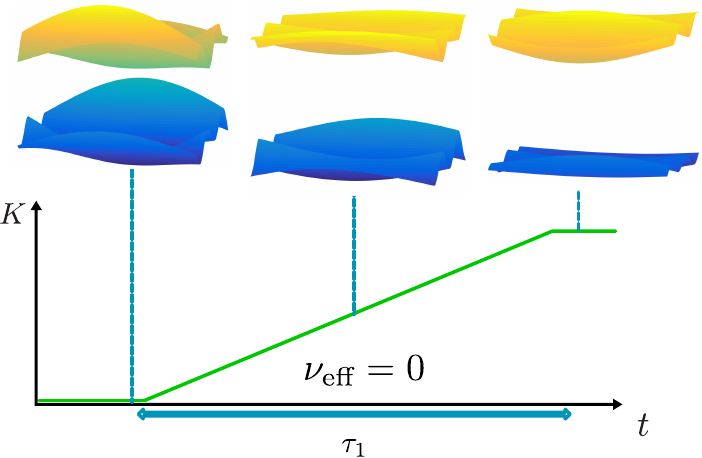}
	\caption{First step of the loading sequence: The modulation is slowly turned on by linearly increasing the amplitude of the modulation in the interval $[0,K]$. The corresponding instantaneous energy band $E_{\text{eff}}$ of the effective Hamiltonian in Eq.~\eqref{eq:hameffzero} are shown at different times during the ramp. Note that the detuning $\delta_\text{in}$ prevents any overlap between the bands during the entire duration of the ramp; see also Fig.~\ref{fig:detuning} (c). The topology of the bands is trivial throughout this first step of the loading sequence. The systems parameters are adjusted such that $J^\text{eff}_{ij}\!=\!J$, $\lambda^\text{eff}_{ij}\!=\!0.3J$, $\mathbf{q}\!=\!(6,2)/a$ and 
	$\delta_\text{in}\!=\!10J$, at the end of the ramp.}
\label{fig:ramp1}
\end{center}
\end{figure} 

We then smoothly turn on the drive, Eq.~\eqref{drive_def}, and choose a large initial detuning $\delta_{\text{in}}\!=\!\Delta-\Omega\!=\!10J$ so as to avoid any overlapping of the initial Floquet quasi-energy bands (see SubSec.~\ref{sec:sec:prepstate}). As shown in Fig.~\ref{fig:ramp1}, we indeed find that the instantaneous Floquet bands remain well separated during the entire ramp. In our calculations, the launch of the drive amplitude $K (t)$ is chosen to follow a linear ramp, which extends over 200 periods of the drive, $\tau_1=200\,T$; at the end of this ramp, $K$ reaches its final value $K\!=\!2 \Omega$. Due to the absence of inter-band transitions over this first ramp, the system is found to occupy the lowest (Floquet) band of the effective Hamiltonian \eqref{eq:hameffzero} with a 99.9\% probability; we remind that this lowest band is topologically trivial due to the large detuning $\delta_\text{in}$, so that the system is still in a trivial phase at this stage of the loading process.

\subsubsection{Removing the detuning and the gap-closing process}

\begin{figure}[h]
\begin{center}
	\includegraphics[scale=1]{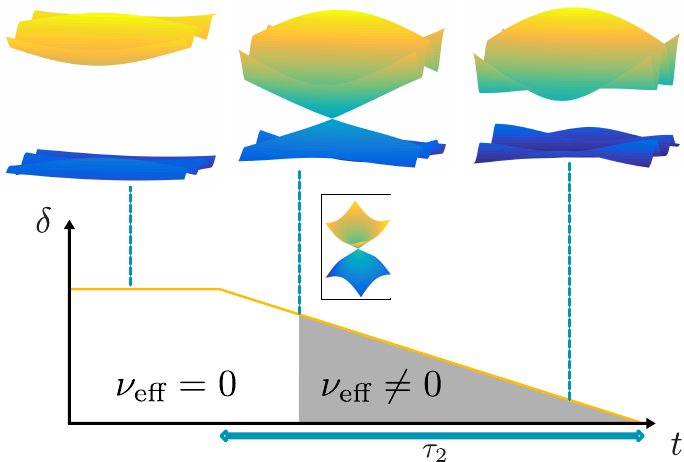}
	\caption{Second step of the loading sequence The detuning is slowly turned off through a linear ramp $\delta (t)$ of duration $\tau_2$. At some critical time, the instantaneous Floquet bands undergo a topological phase transition, as characterized by a change in the Chern number $\nu_{\text{eff}}\!\ne\!0$. The systems parameters are adjusted such that $J^\text{eff}_{ij}\!=\!J$, $\lambda^\text{eff}_{ij}\!=\!0.3J$ and $\mathbf{q}\!=\!(6,2)/a$.}
\label{fig:ramp2}
\end{center}
\end{figure} 

The next step consists in slowly removing the detuning $\delta$, so as to induce a controllable topological phase transition. This second ramp is illustrated together with the corresponding instantaneous Floquet bands in Fig.~\ref{fig:ramp2}. In particular, we find that the Chern numbers associated with these Floquet bands indeed become non-zero, after a critical time corresponding to the gap-closing event. 

\begin{figure}
\center
\includegraphics[scale=1]{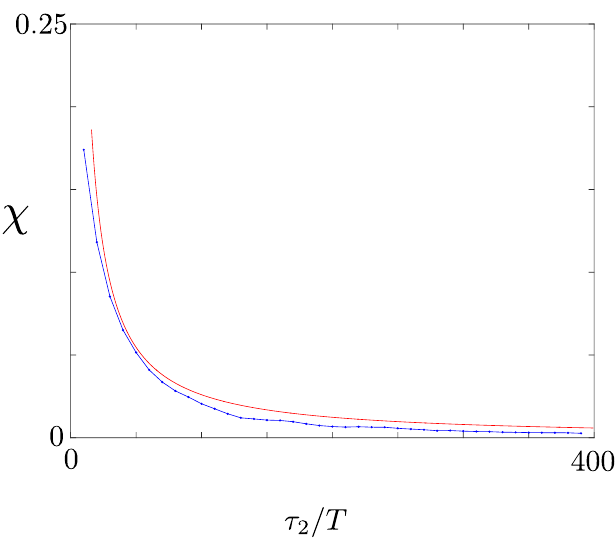}
\caption{Excited fraction $\chi$ as a function of the ramp duration $\tau_2$, for the system confined in a disk of radius $r\!=\!10a$. The numerical results (blue curve) is compared with the analytical Landau-Zener formula in Eq.~\eqref{eq:lz}, shown in red. }
\label{fig:lz}
\end{figure}

The duration of the second ramp $\tau_2$ allows one to control the Landau-Zener (inter-band) transitions that occur at the gap closing. In order to estimate the particle transfer from the lowest band to the excited band, we consider the momentum-representation of the effective Hamiltonian [see Eq.~\eqref{eq:heff}], and we include the time-dependence of the detuning $\delta$; using the standard Landau-Zener formula for linear ramps~\cite{Dziarmaga_2010}, one finds that the fraction of particles in the excited band is given by 
\begin{equation}
	\chi=\sum_\text{BZ}\exp{\left(-\dfrac{\pi\tau_2\vert g(\mathbf{k})\vert^2}{\delta_{\text{in}}}\right)},
	\label{eq:lz}
\end{equation}
where the model-dependent function $g(\mathbf{k})$ is defined in Appendix~\ref{app:derivham}, and where $\delta_{\text{in}}$ denotes the value of the detuning before performing the second ramp. The agreement between this analytical result and our numerical simulations, which are based on the full time-dependent Schr\"{o}dinger equation, is shown in Fig.~\ref{fig:lz}, where the excitation fraction $\chi$ is plotted as a function of the ramp duration $\tau_2$. These numerical and analytical results offer an instructive estimate for the ramp duration ($\tau_2\gtrsim 200 T$) that would be required in order to reach reasonably small excited fraction ($\chi\!\lesssim\!1\%$) in view of detecting the geometrical and topological properties of the loaded Floquet band.

\subsection{Dynamics of the atomic cloud and displacement of the center of mass}
\label{sec:com}

\begin{figure}[h]
\begin{center}
	\includegraphics[scale=1.4]{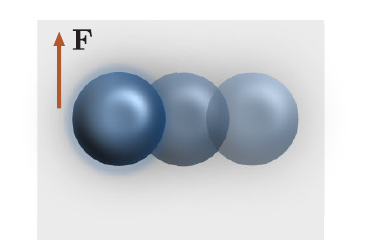}
\end{center}
	\caption{Sketch of the transport experiment. An atomic cloud is initially prepared and confined within a certain region (e.g.~a disk of radius $r$). After the preparation of the desired state, the confining potential is released and a force $\mathbf{F}$ is applied to the system. The center-of-mass performs a transverse drift whenever the occupied (Floquet) band is characterized by a non-zero Chern number~\cite{dauphin_2013,Aidelsburger_2014,price_2016}.}
\label{fig:prep_state}
\end{figure} 

In the previous Section, we analyzed the typical population that one would expect in a non-trivial Floquet band, using an optimized (two-ramp) loading sequence. In this Section, we build on these results and investigate the transport properties of the resulting system. As the Floquet bands are only partially filled after the loading process ($\chi\!\ne\!0$), the transport in response to an applied force is captured by the equations of motion discussed in Section~\ref{sec:sec:gscom}. It is the aim of this Section to validate these theoretical results through a numerical resolution of the full time-dependent problem.

In the numerical simulations presented in this Section, we consider the time-evolution of the state that was obtained after the two successive ramps (see previous Sections). The time-evolution is ruled by the time-dependent Hamiltonian in Eq.~\eqref{eq:hamiltonientemporel} to which we add a constant force aligned along the $y$ direction $\mathbf{F}\!=\! F_y \mathbf{1}_y$. In order to limit inter-band transitions during transport, we choose the value $F_y\!=\!0.1 J/a$, which is indeed reasonable when considering the large bulk gap $\Delta_{\text{gap}}\!\approx\! 1.7 J$ of the Floquet spectrum shown in Fig.~\ref{fig:lattespec}. The full dynamics is then described by the time-evolution operator
\begin{equation}
\hat{U}(t,t_0)\!=\!\hat{R}^\dagger(t)e^{-i\hat{\mathcal{K}}(t)} e^{-i\hat{\mathcal{H}}_\text{eff}(t-t_0)+\hat V_\text{force}}e^{i\hat{\mathcal{K}}(t_0)}\hat{R}(t_0)\text{,}
\label{eq:heff_force}
\end{equation}
where the operators associated with the micro-motion, $\hat{R}^\dagger(t)$ and $\hat{\mathcal{K}}(t)$, are explicitly given in Appendix~\ref{app:derivham}, and where the term associated with the force, $\hat V_\text{force}\!=\!-\sum_i \mathbf{F}\cdot{\mathbf{r}_i \, \hat c^\dagger_i \hat c_i}$, is simply added to the effective Hamiltonian (which is indeed valid in the high-frequency limit of the drive). Note that the force is set along the $y$ direction, so that a Hall drift is expected along the $x$ direction~\cite{dauphin_2013,price_2016}, due to the topological Floquet bands (see Fig.~\ref{fig:prep_state}).

In addition to the applied force, one also removes the confinement that was present during the loading process (see Section~\ref{sec:prepgs}). This removal of the confinement leads to a slight modification of the excited fraction $\chi$ calculated in the previous Section, due to the partial projection of the populated edge states unto the bulk states associated with the higher band; the modified fraction $\chi$, which will then define the relevant excited fraction during the transport experiment studied below, is shown in Fig.~\ref{fig:occupation} (a). In our calculations, the size of the system after releasing the confinement is chosen to be constituted of $80\!\times\!80$ sites, and we verified that the atomic cloud never touches the edges of the system over the considered simulation times. 

The state that one obtains numerically at the end of the loading process and after the removal of the confinement  will be denoted $\vert \psi_{\text{transp}}\rangle$. The following paragraphs aim to investigate the time-evolution of that relevant (``prepared") state, in response to an applied force.

We propose to organize our numerical investigations into a series of steps, so as to highlight the individual effects that come into play in the context of transport-experiments with Floquet bands. Indeed, the current and center-of-mass responses are affected by two main and independent effects: the rapid oscillations due to the partial filling of the Floquet bands (after the loading process), and oscillations due to the micro-motion associated with the time-modulated lattice. Our analysis below demonstrates the robustness of COM drifts with respect to these two spoiling effects, in contrast to the current which shows large oscillations. Additionally, we study how the transport coefficients expected for completely filled bands deviate from their quantized value, due to the incomplete filling of the Floquet bands. Eventually, we discuss the effects of higher-order terms in the effective Hamiltonian $\hat{\mathcal{H}}_\text{eff}$, which are often neglected in standard analysis of Floquet-engineered systems operating in the high-frequency limit. 

We point out that we focus our study on two physical observables: the center-of-mass velocity~\eqref{COM_vel} and the COM drift~\eqref{eq:xcm}. We remind that the former is directly related to the current density~\eqref{current_density}, which is measured in solid-state experiments, while the latter can be directly imaged in cold-atom experiments~\cite{Aidelsburger_2014}. 


\subsubsection{Dynamics under the effective Hamiltonian: Neglecting the micro-motion}
\label{sec:sec:sec:negmicro}

\begin{figure}[h]
\begin{center}
	\includegraphics[scale=1]{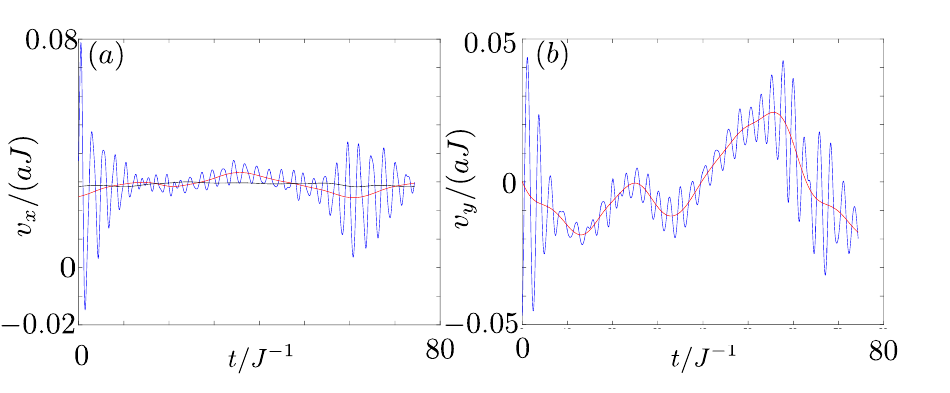}
	\caption{Velocity of the center of mass (blue line) along the $x$ and $y$ directions, as computed using the full equations of motion~\eqref{current_density}-\eqref{COM_vel}. Removing the contribution of the inter-band velocity term [Eq.~\eqref{eq:vinterband}] leads to a smoother behavior (red curve). The isolated contribution of the anomalous velocity [Eq.~\eqref{eq:berry}] is shown in black.}
\label{fig:bcurrent}	
\end{center}
\end{figure} 

In order to isolate the effects related to the partial filling of the Floquet bands, we first neglect all the effects due to the micro-motion and analyze the dynamics associated with the (static) effective Hamiltonian~\eqref{eq:hameffzero} in response to an applied force. To do so, we numerically solve the equations of motion presented in Section~\ref{sec:sec:gscom}, by including the Berry curvature $\mathcal{F}^{1,2}_{\text{eff}}(\bs k)$ and dispersions $E^{1,2}_{\text{eff}}(\bs k)$ of the corresponding Floquet bands. Besides, the coefficients $\alpha(\mathbf{k})$ and $\beta(\mathbf{k})$ that appear in Eqs.~\eqref{eq:band}-\eqref{eq:vinterband} are obtained by projecting the state $\vert \psi_{\text{transp}}\rangle$ unto the Bloch states of the effective Hamiltonian~\eqref{eq:hameffzero}. 

The $x$ and $y$ components of the COM velocity $\mathbf{v}_\text{CM}(t)$ are shown in Fig.~\ref{fig:bcurrent}, as a function of time (blue curve). This result illustrates the strong oscillations announced in Section~\ref{sec:sec:gscom}, whose period indeed coincides with the size of the bulk gap, $2\pi/\Delta_{\text{gap}}\!=\!3.63 J^{-1}$. In order to highlight the fact that these oscillations are indeed due to the inter-band velocity term~\eqref{eq:vinterband} only, we remove the contribution of this term by hand and show the corresponding result in Fig.~\ref{fig:bcurrent} (red curve): the rapid oscillations indeed vanish, but large oscillations remain; the latter are due to the band-velocity contribution, which is non-zero due to the non-uniform filling of the bands, and their period thus corresponds to the Bloch period $T_{\text{Bloch}}\!=\!2 \pi/(aF)$. This is confirmed by observing that all oscillations disappear when calculating the contribution of the anomalous velocity only (see black curve in Fig.~\ref{fig:bcurrent}). As a technical note, we point out that the band-velocity contribution is significantly stronger for the $y$ component of the COM velocity $\mathbf{v}_\text{CM}(t)$, which can be traced back to special symmetries in the model~\cite{dauphin_2013}. We also remind that the contribution of the band velocity could be removed in experiments using the schemes proposed in Ref.~\cite{Price:2012}.

Importantly, while the transverse COM velocity $v_\text{CM}^x(t)$ shown in Fig.~\ref{fig:bcurrent} strongly oscillates, we find that its time-average over a long observation time $\langle v_\text{CM}^x \rangle $ leads to a clear signature of the non-trivial topology of the mostly populated (Floquet) band. Indeed, one can compare this value with the prediction of the TKNN formula~\cite{thouless_1982,dauphin_2013,price_2016}
\be
v_\text{CM}^x= \frac{F_y A_{\text{cell}}}{2 \pi} \nu_1 ,\label{TKNN}
\ee
which corresponds to the ideal case where the lowest band, with Chern number $\nu_1$, is completely filled. Inserting the value $\langle v_\text{CM}^x \rangle $ extracted from our numerics, we find an approximate value for the Chern number of the populated Floquet band, $\nu_\text{approx}\!=\!0.92$. We note that the slight deviation from the real value $\nu_{\text{eff}}\!=\!1$ is mostly due to the non-ideal filling of the Floquet band. This indicates that time-averaging the COM velocity, or the current density, allows one to accurately measure the topology of the Floquet band, under realistic loading schemes. 

\begin{figure}[h]
\begin{center}
	\includegraphics[scale=1]{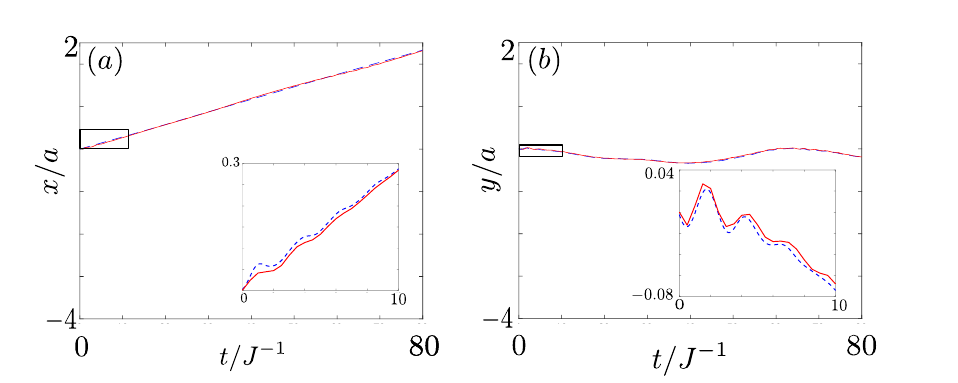}
	\caption{Center-of-mass displacement, for a state prepared using a ramp duration of $\tau_2\!=\!400T$, as computed using the equations of motion (blue dashed line) and using the lowest-order effective Hamiltonian~\eqref{eq:hameffzero} (red line).}
\label{fig:smheff}
\end{center}
\end{figure} 

Next, we compute the displacement of the COM, $\bs x_{\text{COM}} (t)$, and show the corresponding trajectories in Fig.~\ref{fig:smheff}. The almost perfectly linear trajectory along the transverse direction $x$ illustrates the robustness of the COM drift against the inter-band-velocity effects; a zoom into these curves shows the remnant of the related rapid oscillations, which survived upon integration over time (see insets in Fig.~\ref{fig:smheff}). This clearly illustrates the advantage of probing the COM drift instead of the current density (or COM velocity), when studying the transport properties of partially-filled bands (compare Fig.~\ref{fig:smheff} and Fig.~\ref{fig:bcurrent}). 

In order to validate the predictions of the equations of motion derived in Section~\ref{sec:sec:gscom}, we show the agreement between the resulting trajectories and those obtained through a full time-evolution of the  Schr\"{o}dinger equation, based on the effective Hamiltonian [Eq.~\eqref{eq:hameffzero}], in Fig.~\ref{fig:smheff}. These results show the validity of our equations of motion, in the regime of weak external forces.

\subsubsection{Including the micro-motion}
\label{sec:sec:sec:incmicro}
\begin{figure}[h]
\begin{center}
	\includegraphics[scale=1]{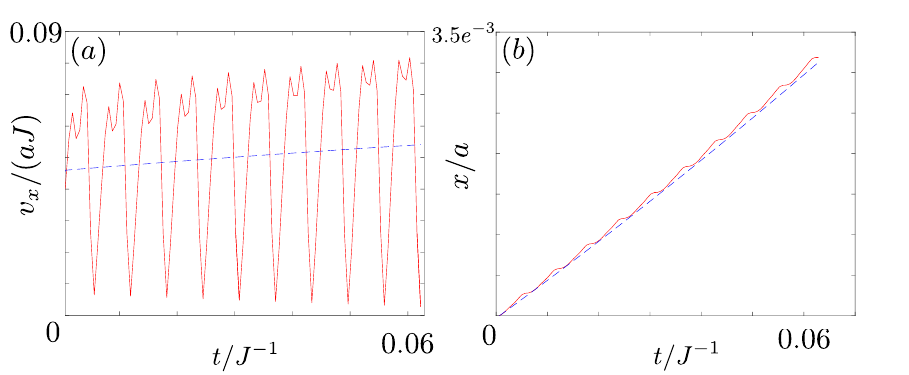}
	\caption{(a) Center-of-mass velocity along the $x$ direction, as obtained using the effective Hamiltonian only (dashed blue) and when including the effects due to the micro-motion (red curve), considering a drive frequency of $\Omega\!=\!1000J$.  (b) Center-of-mass displacement along the $x$ direction, as obtained using the effective Hamiltonian only (dashed blue) and  when including the micro-motion (red curve).}
	\label{fig:figmicro}
\end{center}
\end{figure} 

We now include the effects due to the micro-motion, considering the full time-evolution operator in Eq.~\eqref{eq:heff_force} (and the same prepared state $\vert \psi_{\text{transp}}$). The explicit expressions for the kick operator $\hat{\mathcal{K}} (t)$, calculated up to first order in $1/\Omega$, and the frame-transformation operator $\hat R (t)$, are given in Appendix~\ref{app:derivham}. Setting the drive frequency to the large value $\Omega\!=\!1000J$, we calculate the effects of the micro-motion on the COM transverse velocity $v^x_{\text{COM}} (t)$ and drift $x_{\text{COM}} (t)$. The results are shown in Fig.~\ref{fig:figmicro}, which compares the behavior of these observables in the presence (red curve) and absence (blue dotted curve) of micro-motion. We find that the COM velocity (and hence, the current density) strongly oscillates due to the micro-motion, with the expected period $T\!=\!2\pi/\Omega$. Importantly, the effects due to the micro-motion tend to vanish when integrated over time, due to the fact that the kick operator $\hat{\mathcal{K}} (t)$ has a zero average over a period of the drive~\cite{goldman_2014}. This is clearly reflected in the trajectory of the COM $x_{\text{COM}} (t)$, whose long-time dynamics is indeed shown to be most entirely governed by the effective Hamiltonian. This analysis further demonstrates the robustness of the COM drift, and hence, its relevance for the detection of Floquet bands' topology, as compared to current measurements.


\subsubsection{Higher-order corrections to the effective Hamiltonian}\label{section_higher}
\begin{figure}[h]
\begin{center}
	\includegraphics[scale=1]{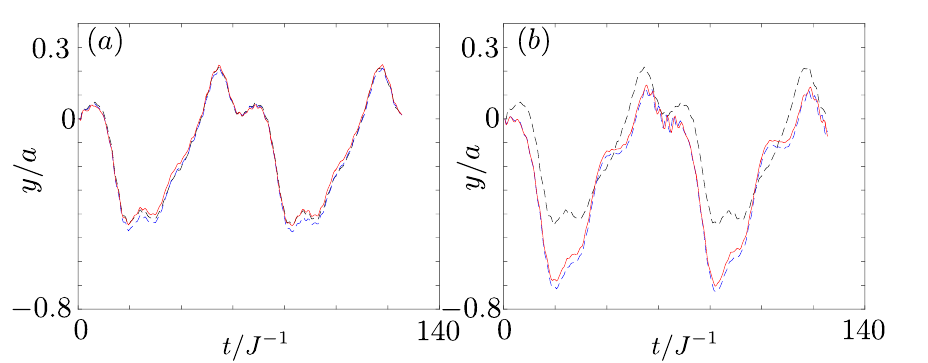}
	\caption{Center-of-mass displacement along the $y$ direction, in the presence of a force $\mathbf{F}\!=\!0.1J \mathbf{1}_y$, as calculated using three different Hamiltonian: the zeroth order effective Hamiltonian (dashed black), the effective hamiltonian with first order corrections (dashed blue) and the ``exact" effective hamiltonian, as computed numerically (red).  (a) Using a driving frequency $\Omega\!=\!1000J$. (b) Using $\Omega\!=\!100J$. Note the importance of considering the first-order corrections to the effective Hamiltonian in the latter case.}
\label{fig:figcompa}
\end{center}
\end{figure} 

The previous results built on the knowledge of the effective Hamiltonian in Eq.~\eqref{eq:hameffzero}, which indeed governs the long-time dynamics in the high-frequency limit $\Omega\!\rightarrow\!\infty$. In order to estimate the validity of this effective Hamiltonian when using finite frequencies, we compare the COM trajectories generated by the stroboscopic time-evolution operator~\eqref{eq:strobo}, using three different effective Hamiltonians: (a) the lowest-order effective Hamiltonian in Eq.~\eqref{eq:hameffzero}, (b) the same effective Hamiltonian but including first-order corrections [see Appendix~\ref{app:derivham}], and (c) the ``exact" effective Hamiltonian, as calculated numerically through the expression $\hat H_F\!=\!(i/T)\log \hat U (T)$, where the time-evolution operator over one period was numerically constructed using discrete time-steps $\delta_t\!\ll\!T$. The results are shown in Fig.~\ref{fig:figcompa}, which compares these trajectories along the $y$ direction, for two values of the drive frequency, $\Omega\!=\!1000J$ and $\Omega\!=\!100J$. We find that while the three curves perfectly match in the case of a very large frequency $\Omega\!=\!1000J$, the lowest-order effective Hamiltonian shows significant deviations in the case of lower frequencies ($\Omega\!=\!100J$). However, we note that the general (qualitative) behavior of the COM trajectory is captured by the lowest-order effective Hamiltonian in Eq.~\eqref{eq:hameffzero}, even for intermediate frequencies (here $\Omega\!=\!100J$). 

We point out that the first-order corrections considered here most entirely rectify any deviations between the predictions of the lowest-order effective Hamiltonian and the full-time (real) dynamics [Fig.~\ref{fig:figcompa} (b)], which helps understanding the origin of these deviations:~these are found to be caused by residual on-site potentials, which find their origin in the approximative nature of the resonance condition $\Delta\!=\!\Omega$, see also Refs.~\cite{Monika_thesis,bukov_2014} for more details. Finally, we note that, for the present model, these high-order effects are more pronounced for the motion taking place along the $y$ direction; in particular, we find that the Hall drift occurring along the $x$ direction is only slightly affected by these effects.

\subsubsection{Full time dynamics and the Chern number measurement}\label{section_chern_measurement}

We conclude this Section by presenting the dynamics of the prepared state $\vert \psi_{\text{transp}}\rangle$ using the complete time-evolution operator associated with the time-dependent Hamiltonian~\eqref{eq:hamiltonientemporel}, and including the force $\hat V_{\text{force}}$ within the static Hamiltonian $\hat H_0$. These calculations thus include all the effects associated with the micro-motion and all the corrections to the effective Hamiltonian in Eq.~\eqref{eq:hameffzero}; they also include residual inter-band transitions, not captured by the equations of motion of Section~\ref{sec:sec:gscom}. In practice, the full time-evolution operator is obtained by discretizing the dynamics into small time steps ($\delta_t\!=\!T/100$) and applying the Trotter formula to lowest order.

\begin{figure}[h]
\begin{center}
	\includegraphics[scale=1]{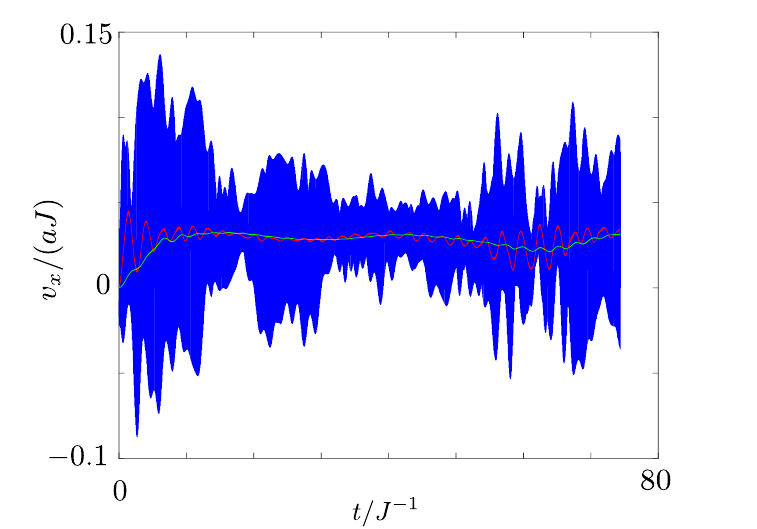}
	\caption{Center-of-mass velocity along the $x$ direction, as computed using the full time-dependent Hamiltonian (blue). The red curve shows the signal obtained by applying a low-pass filter at the micro-motion (driving) frequency, hence revealing the regular oscillations due to the inter-band velocity. A second low-pass filter can be applied to further remove these residual oscillations (see green curve); this reveals the contributions of the band and anomalous velocities.
\label{fig:vcmfull}}
\end{center}
\end{figure}

First, we show the transverse COM velocity $v^x_\text{CM}(t)$ resulting from these full-time-evolution calculations in Figure~\ref{fig:vcmfull}. As anticipated from our previous discussions (Section~\ref{sec:sec:sec:negmicro}), the COM velocity (and hence the current density), shows an irregular behavior (blue curve), stemming from an interplay between the micro-motion and oscillations due to the inter-band velocity~\eqref{eq:vinterband}. In order to further demonstrate these two individual effects, we apply a low-pass filter to this signal, at the driving frequency $\Omega$. The resulting (red) curve shows residual oscillations, reminiscent of those presented in Fig.~\ref{fig:bcurrent}, which can thus be attributed to the inter-band velocity contribution. Then, applying a second low-pass filter at the inter-band-velocity frequency $\omega_{\text{int}}$ is found to remove these residual oscillations: this second filter reveals the contributions of the anomalous and band velocities (compare with the red curve in Fig.~\ref{fig:bcurrent}). Altogether, these full-time-dynamics results show a qualitatively agreement with the individual behaviors discussed in Sections~\ref{sec:sec:sec:negmicro} and \ref{sec:sec:sec:incmicro}.

\begin{figure}[h]
\begin{center}
	\includegraphics[scale=1]{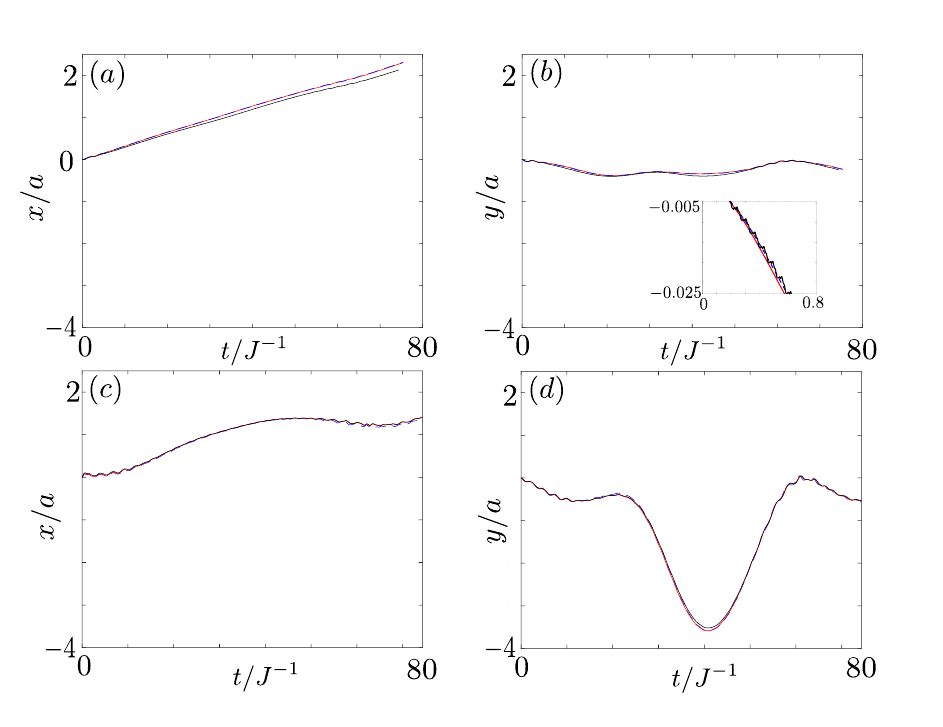}
	\caption{Center-of-mass displacement along the $x$ and $y$ direction, for a  state prepared using a ramp duration of (a)-(b) $\tau_2\!=\!400T$ and (c)-(d) $\tau_2=10T$. Here, the modulation frequency is set to the value $\omega\!=\!100J$. The red line corresponds to the stroboscopic evolution associated with the ``exact" (numerical) effective Hamiltonian; the blue dashed line corresponds to the calculation obtained by using the effective hamiltonian (including first order corrections); the dark line corresponds to the full time-dependent simulation. Note that the excited fraction corresponding to $\tau_2\!=\!400T$ is $\chi=3\%$, while it is $\chi=23\%$ for $\tau_2\!=\!10T$.}
\label{fig:dispcm}
\end{center}
\end{figure}

Next, we study the displacement of the COM drift $\bs{x}_\text{CM}(t)$ in Fig.~\ref{fig:dispcm}, for two different values of the ramp duration. Here, we compare the full-time dynamics with the one obtained through the analytical effective Hamiltonian in Eq.~\eqref{eq:hameffzero} (including first-order corrections), and also with the one corresponding to the numerical (``exact") effective Hamiltonian $\hat H_F$ introduced in Section~\ref{section_higher}. All curves show a good agreement, suggesting that the long-time dynamics are indeed well captured by the effective-Hamiltonian approach, considered in the previous Sections. A small inset in Fig.~\ref{fig:dispcm}(b)  highlights the micro-motion, not captured by the effective Hamiltonian; in agreement with the discussion of Section~\ref{sec:sec:sec:incmicro}, the oscillations due to the micro-motion are found to be negligible when probing the COM drift.

In the case of a long ramp duration ($\tau_2\!=\!400T$), one obtains a clear Hall drift along the $x$ direction, and residual Bloch oscillations along the $y$ direction; the latter oscillations are due to the finite excitation fraction in the upper Floquet band ($\chi=0.03$), and mainly appear in the motion along $y$ due to model-dependent symmetries. Applying the naive TKNN formula~\eqref{TKNN} to the Hall drift shown in Fig.~\ref{fig:dispcm}(a), i.e.~neglecting the contribution of the excited fraction to transport,
\be
x_\text{CM}(t)=x_\text{CM}(t=0) + t \frac{F_y A_{\text{cell}}}{2 \pi} \nu_1 ,\label{TKNN_2}
\ee
we find an approximate value for the Chern number of the populated Floquet band $\nu_\text{approx}\!=\!0.95$. This demonstrates that loading atoms into a Floquet band using a proper loading sequence, followed by a study of the cloud's COM drift, can indeed allow one to estimate the topology of the Floquet band in a satisfactory manner. 

\begin{figure}[h]
\begin{center}
	\includegraphics[scale=1]{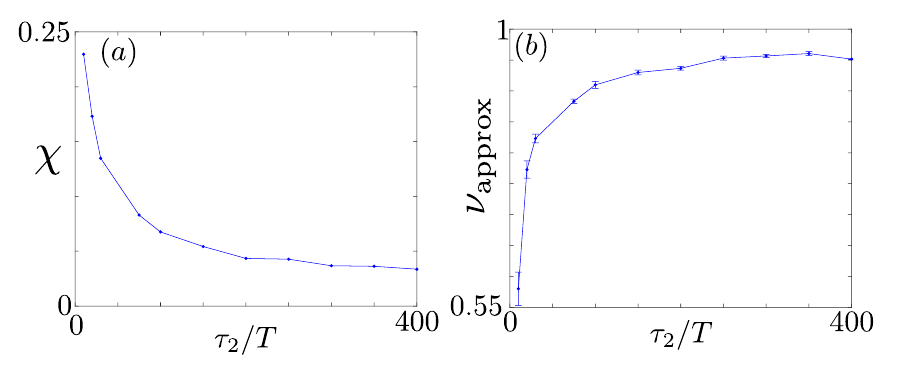}
\end{center}
\caption{(a) Excited fraction $\chi$ as a function of the ramp duration $\tau_2$, after removing the confinement used to prepare the state. (b)  Chern number extracted from the transverse COM displacement [Eq.~\eqref{TKNN_2}] as a function of the ramp duration $\tau_2$. The error bars reflects the quality of the linear regression. }
\label{fig:occupation}
\end{figure}

We also show in Fig.~\ref{fig:dispcm}(c)-(d) the COM displacement in the situation of a short ramp ($\tau_2\!=\!10T$). In this case, the excited fraction is important ($\chi\!=\!0.23$), and significant deviations to the ideal (completely-filled-band) behavior are observed (as manifested by large band-velocity contributions). Applying the naive TKNN formula~\eqref{TKNN_2} to a linear regression performed on the Hall drift shown in Fig.~\ref{fig:dispcm}(c) yields the non-zero, but unsatisfactory result, $\nu_\text{approx}\!=\!0.58$. In order to further illustrate the manner by which the Chern-number measurement based on Eq.~\eqref{TKNN_2} relies on the loading sequence, we show the extracted Chern number $\nu_\text{approx}$ as a function of the ramp duration $\tau_2$ in Fig.~\ref{fig:occupation}(b). From this plot, we estimate that the breakdown of the Chern-number measurement occurs around $\tau_2\!=\!100T$, where $\nu_\text{approx}\!=\!0.91$.

\section{Concluding remarks}
\label{sec:conclusion}

In this work, we investigated the transport properties of a non-interacting gas loaded into Floquet bands of non-trivial topology, setting the focus on the role of the loading sequence. Since the latter inevitably leads to inter-band processes, topological responses such as those revealed in current measurements typically show interference effects between the (potentially many) populated bands, leading to deviations from the standard quantum-Hall behavior. This work analyzed the fate of topological responses in this intriguing context, and demonstrated how robust signatures of topology can still be reached through COM observables. 

We provided a systematic analysis of the many effects that are specifically associated with the intrinsic time-dependence of Floquet-engineered setups, such as the effects of the micro-motion and corrections to the effective Hamiltonian that become significant away from the infinite-frequency limit. We also discussed the origin of strong irregularities and oscillations in current responses, which we attribute to the contribution of an inter-band-velocity term; the latter was shown to be present whenever the bands are partially populated, which necessarily occurs after the loading of particles in a topological Floquet band (starting from a trivial configuration) due to inevitable gap-closing and inter-band transfer processes. 

In our study, we have emphasized the importance of identifying an optimal loading sequence, so as to minimize inter-band transitions during the state-preparation process. As we showed, this is particularly crucial for Floquet systems based on resonant time-modulations, where quasi-energy (Floquet) bands typically overlap as soon as the drive is launched. We introduced a specific loading sequence, based on two successive ramps of the system parameters, to overcome such issues.

This work highlights how distinct observables can react very differently to undesired effects, such as those induced by strong micro-motion and inter-band effects. Here, we illustrated this fact by showing the regularity of the COM displacement, as compared to the rapidly-oscillating current density. This result can be simply explained through the equations of motion derived in Section~\ref{sec:sec:gscom}, where the regularity of the COM was shown to originate from the time-integration of the strongly-oscillating current density.

We also show how the topology of Floquet bands can indeed be probed after a loading sequence, independently of the fact that the Chern number associated with the evolving state [Eq.~\eqref{Chern_state}] is constant (and hence, trivial, when starting the loading sequence with a trivial static band structure). This highlights the fact that transport experiments are well suited to probe the geometry and topology of time-dependent (Floquet) systems, where transport is found to be governed by the topology of the underlying (instantaneous) bands. One should note, however, that this conclusion only strictly holds for Floquet systems that operate in the high-frequency regime: Away from this limit, the micro-motion becomes crucial in the description of the system's topology, and the understanding of transport in terms of instantaneous (Floquet) bands becomes more subtle~\cite{kitagawa2010,Rudner:2013,Carpentier:2015,Nathan:2015,mukherjee_2016,Maczewsky:2016}.

We emphasize that the results presented in this work are not limited to the loading and study of two-dimensional Floquet Chern bands, as these also apply to other types of Floquet topological bands, such as time-reversal-invariant ($Z_2$) topological bands~\cite{Carpentier:2015} and four-dimensional Floquet systems characterized by non-zero second Chern-numbers~\cite{Price:20154D,Price:2016}.

Finally, our study also sheds some light on the challenging problem of state-preparation in the context of (strongly) interacting Floquet systems~\cite{Grushin:2014,anisimovas2015}. In parallel to the issue of ``intrinsic" heating mechanisms, which arise due to an interplay between the drive and the interactions~\cite{lacki_2012,lacki_2013,lazarides_2014,dalessio_2014,Bukov:2015,bilitewski_2015,Choudhury:2015,strater_2016,Bukov:2016,celi_2016,lellouch_2016}, it appears crucial to identify realistic preparation schemes that avoid, or strongly reduce, the impact of (single-particle and many-body) level crossings~\cite{Ho_2016}. \\

A. D. and D.T. T. contributed equally to this work. We acknowledge insightful discussions with A. Bermudez, A. Celi, J.C. Budich, N. R. Cooper, P. Gaspard and P. Massignan. A.D. and M.L. acknowledge support from Adv.\ ERC grant OSYRIS, EU grant QUIC (H2020-FETPROACT-2014 No.\ 641122), EU STREP EQuaM, MINECO (Severo Ochoa grant SEV-2015-0522 and FOQUS FIS2013-46768), Generalitat de Catalunya (SGR 874), Fundaci\'o Privada Cellex, and CERCA Program/Generalitat de Catalunya.  D-T.T. is supported by the F.R.S.-F.N.R.S. (Belgium). N.G. is financed by the FRS-FNRS Belgium and by
the BSPO under PAI Project No. P7/18 DYGEST.

%

\appendix

\section{Equations of motion for partially-filled bands}
\label{sec:semiclassic}

In this Appendix, we calculate the mean velocity in a state given by a superposition of Bloch states
\begin{equation}
| \psi_\mathbf{k}\rangle=\alpha(\mathbf{k})|u_1(\mathbf{k})\rangle+\beta(\mathbf{k})|u_2(\mathbf{k})\rangle ,\label{state_alpha_beta_App}
\end{equation}
in the presence of an external force $\mathbf{F}$; see Section~\ref{sec:sec:gscom}. To do so, let us remind that the quasi-momentum $\mathbf{k}$ is implicitly time-dependent, through the expression $\mathbf{k}(t)\!=\!\mathbf{k}_0\!+\!\mathbf{F} \, t$. Besides, the adiabatic evolution within each band is well captured by the first-order-approximation formula~\cite{xiao_2010}
\begin{align}
|\phi_n(t)\rangle=&\exp{\left(\dfrac{-i}{\hbar}\int_{t_0}^t \text{d}t'E_n(t') \right)}\exp{(i\gamma_n(t))}\times \label{eq:phi} \\
&\times\Big[|u_n\rangle-i\hbar\sum_{n'\neq n}\dfrac{|u_{n'}\rangle\langle u_{n'}|\partial_tu_n\rangle}{E_n-E_{n'}}\Big], \, \quad n=1,2 \notag
\end{align}
where $|u_{1,2}(t)\rangle$ and $E_n(t)$ are the (instantaneous) Bloch states and dispersions at quasi-momentum $\mathbf{k}(t)$, and where $\gamma_n(t)$ is related to the Berry phase in the $n$th band. Hence, the time-evolution of the state in Eq.~\eqref{state_alpha_beta_App} can be approximated by
\begin{equation}
|\psi(t)\rangle=\alpha|\phi_1(t)\rangle+\beta|\phi_2(t)\rangle, \label{eq:psi}
\end{equation}
where the states $|\phi_{1,2}(t)\rangle$ follow the adiabatic motion dictated by Eq.~\eqref{eq:phi}. Evaluating the mean velocity in the time-evolving state~\eqref{eq:psi} then yields
\begin{align}
\dot{\mathbf{r}}(\mathbf{k})&=\langle\psi(t)|\partial_\mathbf{k}\hat{H}|\psi(t)\rangle \notag\\
&=\underbrace{|\alpha|^2\langle\phi_1|\partial_\mathbf{k}\hat{H}|\phi_1\rangle}_\text{I}+\underbrace{|\beta|^2\langle\phi_2|\partial_\mathbf{k}\hat{H}|\phi_2\rangle}_\text{II} \notag\\
&+ \underbrace{2\, \text{Re}\left(\alpha^*\beta \langle\phi_1|\partial_\mathbf{k}\hat{H}|\phi_2\rangle\right)}_\text{III},
\label{eq:semicla}
\end{align}
where $\hat{H}\!=\!\hat{H}(\bs k (t))$ is the underlying momentum-representation Hamiltonian.

The two first terms correspond to the weighted band velocity and anomalous velocity~\cite{xiao_2010}
\begin{align}
&\text{I}+\text{II}=\mathbf{v}_\text{band}+\mathbf{v}_\mathcal{F},\notag\\
&\mathbf{v}_\text{band}=\vert \alpha \vert^2 \partial_\mathbf{k}E_{1}+\vert \beta \vert^2 \partial_\mathbf{k}E_{2} \notag\\
&\mathbf{v}_\mathcal{F}= - \, \mathbf{F}\times\mathbf{1}_z (\vert \alpha \vert^2 \mathcal{F}^1+\vert \beta \vert^2 \mathcal{F}^2),\notag
\end{align}
where the Berry curvature $\mathcal F$ is defined in the main text. 

The third term in Eq.~\eqref{eq:semicla} comes from the interference between the  evolving states $\phi_{1,2} (t)$ associated with the two bands. At this order of the adiabatic approximation~\cite{xiao_2010}, one only keeps the first-order terms in $\dot{\mathbf{k}}$, which yields
\begin{align}
\text{III}=&2\text{Re}\Big[\alpha^*\beta\exp{\left(-\dfrac{i}{\hbar}\int^t_{t_0}\text{d}t\hspace{0.5 mm}E_2-E_1 \right)}\exp{\left(i(\gamma_1-\gamma_2)\right)}\times \notag\\
&\times\Big((E_1-E_2)\langle\partial_\mathbf{k}u_1|u_2\rangle+\notag\\
&+i\hbar\,\mathbf{F}\cdot\dfrac{\langle\partial_\mathbf{k}u_1|u_2\rangle}{E_1-E_2}(\partial_\mathbf{k}E_2-\partial_\mathbf{k}E_1)\Big)\Big],
\label{eq:interband}
\end{align}
which is the time-oscillating inter-band contribution, which we analyse in the main text.

\section{Derivation of the effective Hamiltonian and micro-motion operators}
\label{app:derivham}

\begin{figure}[h]
\center
\includegraphics[scale=1]{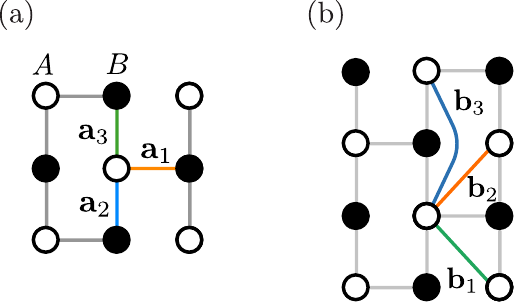}
\caption{The brickwall lattice: \text{(a)} Nearest-neighbor links. \text{(b)} Next-nearest-neighbor links.}
\label{fig:hopping}
\end{figure}

We start by defining the brickwall lattice, which is characterized by the nearest-neighbors vectors $\mathbf{a}_1=(1,0)$, $\mathbf{a}_2=(0,-1)$, $\mathbf{a}_3=(0,1)$ and next-nearest-neighbor vectors $\mathbf{b}_1=(1,-1)$, $\mathbf{b}_2=(1,1)$, $\mathbf{b}_3=(0,2)$; see Fig.~\ref{fig:hopping}.

The time-dependent Hamiltonian in Eq.~\eqref{eq:hamiltonientemporel} takes the form
\begin{align}
&\hat{H}(t)=\hat{H}_0+K\sum_i\cos{\left(\Omega t+\mathbf{q}\cdot\mathbf{r}_i\right)}\hat{c}^\dagger_i \hat{c}_i,\notag\\
&\hat{H}_0=-\sum_{\langle i,j\rangle}J_{ij}\hat{c}^\dagger_i\hat{c}_j-\sum_{\langle\langle i,j\rangle\rangle}\lambda_{ij}\hat{c}^\dagger_i\hat{c}_{j}+\Delta\sum_{\text{sites $B$ }}\hat{c}^\dagger_i\hat{c}_i,
\label{t_ham}
\end{align}
where $\Delta\!=\!\delta+\Omega$, and where the definition of each term is given in the main text.

First, assuming a strong modulation $K\!\sim\!\Omega$ and a weak detuning $\delta \ll \Omega$, we remove all divergent terms in Eq.~\eqref{t_ham} by performing the following gauge transformation
\begin{equation}
\begin{split}
 \hat{R}(t)=\exp\left(i  \int_0^t\hspace{1 mm}\text{dt} \hspace{1 mm}  \left[ K\sum_{i}\cos{\left(\Omega t+\mathbf{q}\cdot\mathbf{r}_i\right)} \hat{c}^\dagger_i \hat{c}_i  \right. \right. \\
\left. \left.  +\Omega\sum_{\text{B sites}} \hat{c}^\dagger_j\hat{c}_j \right]\right).
 \end{split}
 \end{equation}
In the new frame, the transformed Hamiltonian reads:
\begin{align}
&\hat{\mathcal{H}}(t)=(\hat{R}\hat{H}(t)\hat{R}^\dagger-i\hat{R}\partial_t\hat{R}^\dagger)\notag\\
&=\hat{R}\left(-\sum_{\langle i,j\rangle}J_{ij}\hat{c}^\dagger_i\hat{c}_j-\sum_{\langle\langle i,j\rangle\rangle}\lambda_{i,j}\hat{c}^\dagger_{i}\hat{c}_{j}+\delta\sum_{\text{$B$ sites}}\hat{c}^\dagger_i\hat{c}_i\right)\hat{R}^\dagger\notag\\
&=\hat{\mathcal{H}}_{(0)}+\sum_{n\neq0}\hat{\mathcal{H}}_{(n)}e^{i n\Omega t},\notag
\end{align}
where the Fourier components $\mathcal{H}_{(n)}$ are explicitly given by
\begin{equation}
\hat{\mathcal{H}}_{(n)}=-\sum_{\langle i,j\rangle}J_{ij}G(n,\mathbf{r}_j,\mathbf{r}_i)\hat{c}^\dagger_j\hat{c}_i-\sum_{\langle\langle ij\rangle\rangle}\lambda_{ij}F(n,\mathbf{r}_j,\mathbf{r}_i)\hat{c}^\dagger_j\hat{c}_i \label{Fourier_comp}
\end{equation}
where 
\begin{equation}
\begin{split}
F(n,\mathbf{r}_i,\mathbf{r}_j)=&\exp{\left(-in\left(\dfrac{\pi}{2}-\mathbf{q}\cdot\dfrac{\mathbf{r}_i+\mathbf{r}_j}{2}\right)\right)}\times\\
&\mathcal{J}_n\left(-\dfrac{2K}{\Omega}\sin{\left( \mathbf{q}\cdot\dfrac{\mathbf{r}_j-\mathbf{r}_i}{2} \right)}\right)
\end{split}
\end{equation}
and
\begin{equation}
G(n,\mathbf{r}_i,\mathbf{r}_j)=\begin{cases}
\underline{\text{for $\mathbf{r}_i$ on site A}}\\
\begin{split}
&\exp{\left(i(1-n)\left(\dfrac{\pi}{2}-\mathbf{q}\cdot\dfrac{\mathbf{r}_i+\mathbf{r}_j}{2}\right)\right)}\times\\
&\mathcal{J}_{n-1}\left(-\dfrac{2K}{\Omega}\sin{\left( \mathbf{q}\cdot\dfrac{\mathbf{r}_j-\mathbf{r}_i}{2} \right)}\right)
\end{split}\\ 
\underline{\text{for $\mathbf{r}_i$ on site B}}\\
\begin{split}
&\exp{\left(-i(1+n)\left(\dfrac{\pi}{2}-\mathbf{q}\cdot\dfrac{\mathbf{r}_i+\mathbf{r}_j}{2}\right)\right)}\times\\
&\mathcal{J}_{n+1}\left(-\dfrac{2K}{\Omega}\sin{\left( \mathbf{q}\cdot\dfrac{\mathbf{r}_j-\mathbf{r}_i}{2} \right)}\right)
\end{split}\\ 
\end{cases}
\end{equation}

Taking into account the change-of-frame transformations, the time-evolution operator in the initial frame can be partitioned in terms of the action of an effective hamiltonian (describing long-time dynamics) and the action of a kick-operator, which describes the micromotion within each period \cite{goldman_2013,goldman_2015}:
\begin{equation}
\hat{U}(t,t_0)=\hat{R}^\dagger(t)e^{-i\hat{\mathcal{K}}(t)}e^{-i\hat{\mathcal{H}}_\text{eff}(t-t_0)}e^{i\hat{\mathcal{K}}(t_0)}\hat{R}(t_0).
 \end{equation}
Up to the first order of $1/\Omega$, the effective Hamiltonian  is given by
\begin{equation}
\hat{\mathcal{H}}_{\text{eff}}=\hat{\mathcal{H}}_{(0)}+\dfrac{1}{\Omega}\sum_{n=1}^{\infty}\dfrac{1}{n}[\hat{\mathcal{H}}_{(n)},\hat{\mathcal{H}}_{(-n)}]+\mathcal{O}(1/\Omega^2),
\label{eq:effectif}
\end{equation}
and the micromotion operator is given by,
\begin{equation}
\hat{\mathcal{K}}(t)=\frac{1}{i\, \Omega}\sum_{n=1}^\infty\frac{1}{j }\left(\hat{\mathcal{H}}_{(n)}e^{in\omega t}-\hat{\mathcal{H}}_{(-n)}e^{-in\omega t}\right)+\mathcal{O}(1/\Omega^2).
\end{equation}
where the Fourier components are given in Eq.~\eqref{Fourier_comp}.

Interestingly, the zeroth-order effective Hamiltonian realises the model analyzed in Ref.~\cite{goldman_2013}, namely\begin{equation}
\begin{split}
\hat{\mathcal{H}}_\text{eff}=&-\sum_{\langle i,j\rangle} iJ^\text{eff}_{ij}e^{i\, \mathbf{q}\cdot(\mathbf{r}_i+\mathbf{r}_j)/2}\hat{c}^\dagger_i\hat{c}_j-\sum_{\langle\langle i,j\rangle\rangle}\lambda^\text{eff}_{ij}\hat{c}^\dagger_i\hat{c}_{j}\\
&+\delta\sum_{\text{$B$ sites}}\hat{c}^\dagger_i\hat{c}_i+\mathcal{O}(1/\Omega),
\label{eq:eqeff}
\end{split}
\end{equation}
where the effective tunneling matrix elements are given by
 \begin{equation}
 \begin{cases}
 J^\text{eff}_{ij}=(-1)^\alpha J_{ij}\times\mathcal{J}_1\Big[\dfrac{2K}{\Omega}\sin{\left(\mathbf{q}\cdot\dfrac{\mathbf{r}_j-\mathbf{r}_i}{2}\right)}\Big]\\
 \lambda^\text{eff}_{ij}=\lambda_{ij}\times\mathcal{J}_0\Big[\dfrac{2K}{\Omega}\sin{\left(\mathbf{q}\cdot\dfrac{\mathbf{r}_j-\mathbf{r}_{i}}{2}\right)} \Big],
 \end{cases}
 \end{equation}
where $\mathcal{J}_n(x)$ denote Bessel functions of first kind. 

The Hamiltonian in Eq.~\eqref{eq:eqeff} can be gauge transformed so as to recover translational symmetry~\cite{goldman_2013}. In this frame, the momentum-representation Hamiltonian yields
 \begin{align}
&h_\text{eff}(\mathbf{k})=-
\begin{pmatrix}
\,f\left(\mathbf{k}-\mathbf{q}/2 \right)&g(\mathbf{k})\\
g^*(\mathbf{k})&\,f\left(\mathbf{k}+\mathbf{q}/2 \right)-\delta
\end{pmatrix},
\label{eq:heff}\\
&f(\mathbf{k})=2\sum_{i}\lambda^\text{eff}_{\mathbf{a}_i}\cos{(\mathbf{k} \cdot \mathbf{a}_i)}\notag \\
&g(\mathbf{k})=-i\sum_{i}J^\text{eff}_{\mathbf{b}_i}\exp{(-i\mathbf{k}\cdot \mathbf{b}_i)},\notag
\end{align}
where the hopping parameters $\lambda^\text{eff}_{\mathbf{a}_i}$ and $J^\text{eff}_{\mathbf{b}_i}$ are associated with the different links of the brickwall lattice, as defined above.

\begin{figure}[h]
\center
\includegraphics[scale=1]{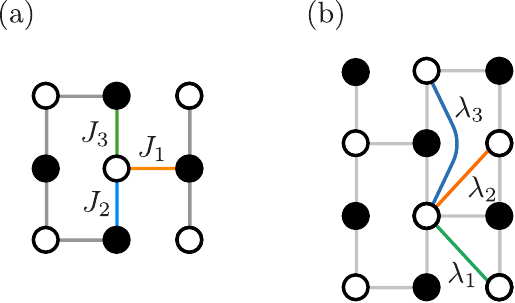}
\caption{Hopping parameters associated with the static Hamiltonian in Eq.~\eqref{t_ham}: (a) NN hopping amplitudes. \text{(b)} Initial NNN hopping amplitudes.}
\label{fig:hopping_bis}
\end{figure}

In the main text, we choose the values for the initial hoppings such that $|J^{\text{eff}}_{ij}|\!=\!J$ and $|\lambda^\text{eff}_{ij}|\!\approx\!0.3J$. Setting $\mathbf{q}\!=\!(6,2)$, which further maximizes the flatness ratio of the lowest topological band of this model~\cite{goldman_2013}, this implies the following values for the hopping parameters associated with the static Hamiltonian $\hat H_0$: $(J_1,J_2,J_3)=(3.69J,5.17J,-5.17J)$ and $(\lambda_1,\lambda_2,\lambda_3)=(-0.76J,-0.76J,-0.76J)$; see Fig.~\ref{fig:hopping_bis} for the definitions of the hopping parameters. Note that these values correspond to a modulation strength  $K\!=\!2 \Omega$.

The calculations shown in Fig.~\ref{fig:figcompa} (main text) include the first-order corrections to the effective Hamiltonian~\eqref{eq:eqeff}. These were  evaluated by numerically calculating the Fourier components in Eq.~\eqref{Fourier_comp} and inserting these into the expression for the effective Hamiltonian in Eq.~\eqref{eq:effectif}. We verified that the sum in Eq.~\eqref{eq:effectif} converged rapidly, namely, that the convergence of the results were reached for $n\!\approx\!10$.

\begin{thebibliography}{103}%
\makeatletter
\providecommand \@ifxundefined [1]{%
 \@ifx{#1\undefined}
}%
\providecommand \@ifnum [1]{%
 \ifnum #1\expandafter \@firstoftwo
 \else \expandafter \@secondoftwo
 \fi
}%
\providecommand \@ifx [1]{%
 \ifx #1\expandafter \@firstoftwo
 \else \expandafter \@secondoftwo
 \fi
}%
\providecommand \natexlab [1]{#1}%
\providecommand \enquote  [1]{``#1''}%
\providecommand \bibnamefont  [1]{#1}%
\providecommand \bibfnamefont [1]{#1}%
\providecommand \citenamefont [1]{#1}%
\providecommand \href@noop [0]{\@secondoftwo}%
\providecommand \href [0]{\begingroup \@sanitize@url \@href}%
\providecommand \@href[1]{\@@startlink{#1}\@@href}%
\providecommand \@@href[1]{\endgroup#1\@@endlink}%
\providecommand \@sanitize@url [0]{\catcode `\\12\catcode `\$12\catcode
  `\&12\catcode `\#12\catcode `\^12\catcode `\_12\catcode `\%12\relax}%
\providecommand \@@startlink[1]{}%
\providecommand \@@endlink[0]{}%
\providecommand \url  [0]{\begingroup\@sanitize@url \@url }%
\providecommand \@url [1]{\endgroup\@href {#1}{\urlprefix }}%
\providecommand \urlprefix  [0]{URL }%
\providecommand \Eprint [0]{\href }%
\providecommand \doibase [0]{http://dx.doi.org/}%
\providecommand \selectlanguage [0]{\@gobble}%
\providecommand \bibinfo  [0]{\@secondoftwo}%
\providecommand \bibfield  [0]{\@secondoftwo}%
\providecommand \translation [1]{[#1]}%
\providecommand \BibitemOpen [0]{}%
\providecommand \bibitemStop [0]{}%
\providecommand \bibitemNoStop [0]{.\EOS\space}%
\providecommand \EOS [0]{\spacefactor3000\relax}%
\providecommand \BibitemShut  [1]{\csname bibitem#1\endcsname}%
\let\auto@bib@innerbib\@empty
\bibitem [{\citenamefont {Grossmann}\ \emph {et~al.}(1991)\citenamefont
  {Grossmann}, \citenamefont {Dittrich}, \citenamefont {Jung},\ and\
  \citenamefont {H{\"a}nggi}}]{Grossmann:1991}%
  \BibitemOpen
  \bibfield  {author} {\bibinfo {author} {\bibfnamefont {F.}~\bibnamefont
  {Grossmann}}, \bibinfo {author} {\bibfnamefont {T.}~\bibnamefont {Dittrich}},
  \bibinfo {author} {\bibfnamefont {P.}~\bibnamefont {Jung}}, \ and\ \bibinfo
  {author} {\bibfnamefont {P.}~\bibnamefont {H{\"a}nggi}},\ }\href
  {https://doi.org/10.1103/PhysRevLett.67.516} {\bibfield  {journal} {\bibinfo
  {journal} {Phys. Rev. Lett.}\ }\textbf {\bibinfo {volume} {67}},\ \bibinfo
  {pages} {516} (\bibinfo {year} {1991})}\BibitemShut {NoStop}%
\bibitem [{\citenamefont {Eckardt}\ \emph {et~al.}(2005)\citenamefont
  {Eckardt}, \citenamefont {Weiss},\ and\ \citenamefont
  {Holthaus}}]{eckardt2005a}%
  \BibitemOpen
  \bibfield  {author} {\bibinfo {author} {\bibfnamefont {A.}~\bibnamefont
  {Eckardt}}, \bibinfo {author} {\bibfnamefont {C.}~\bibnamefont {Weiss}}, \
  and\ \bibinfo {author} {\bibfnamefont {M.}~\bibnamefont {Holthaus}},\ }\href
  {http://link.aps.org/doi/10.1103/PhysRevLett.95.260404} {\bibfield  {journal}
  {\bibinfo  {journal} {Phys. Rev. Lett.}\ }\textbf {\bibinfo {volume} {95}},\
  \bibinfo {pages} {260404} (\bibinfo {year} {2005})}\BibitemShut {NoStop}%
\bibitem [{\citenamefont {Eckardt}\ and\ \citenamefont
  {Holthaus}(2007)}]{eckardt2007}%
  \BibitemOpen
  \bibfield  {author} {\bibinfo {author} {\bibfnamefont {A.}~\bibnamefont
  {Eckardt}}\ and\ \bibinfo {author} {\bibfnamefont {M.}~\bibnamefont
  {Holthaus}},\ }\href {https://doi.org/10.1209/0295-5075/80/50004} {\bibfield
  {journal} {\bibinfo  {journal} {EPL}\ }\textbf {\bibinfo {volume} {80}},\
  \bibinfo {pages} {50004} (\bibinfo {year} {2007})}\BibitemShut {NoStop}%
\bibitem [{\citenamefont {Lignier}\ \emph {et~al.}(2007)\citenamefont
  {Lignier}, \citenamefont {Sias}, \citenamefont {Ciampini}, \citenamefont
  {Singh}, \citenamefont {Zenesini}, \citenamefont {Morsch},\ and\
  \citenamefont {Arimondo}}]{lignier2007}%
  \BibitemOpen
  \bibfield  {author} {\bibinfo {author} {\bibfnamefont {H.}~\bibnamefont
  {Lignier}}, \bibinfo {author} {\bibfnamefont {C.}~\bibnamefont {Sias}},
  \bibinfo {author} {\bibfnamefont {D.}~\bibnamefont {Ciampini}}, \bibinfo
  {author} {\bibfnamefont {Y.}~\bibnamefont {Singh}}, \bibinfo {author}
  {\bibfnamefont {A.}~\bibnamefont {Zenesini}}, \bibinfo {author}
  {\bibfnamefont {O.}~\bibnamefont {Morsch}}, \ and\ \bibinfo {author}
  {\bibfnamefont {E.}~\bibnamefont {Arimondo}},\ }\href
  {http://link.aps.org/doi/10.1103/PhysRevLett.99.220403} {\bibfield  {journal}
  {\bibinfo  {journal} {Phys. Rev. Lett.}\ }\textbf {\bibinfo {volume} {99}},\
  \bibinfo {pages} {220403} (\bibinfo {year} {2007})}\BibitemShut {NoStop}%
\bibitem [{\citenamefont {Kierig}\ \emph {et~al.}(2008)\citenamefont {Kierig},
  \citenamefont {Schnorrberger}, \citenamefont {Schietinger}, \citenamefont
  {Tomkovic},\ and\ \citenamefont {Oberthaler}}]{Kierig:2008}%
  \BibitemOpen
  \bibfield  {author} {\bibinfo {author} {\bibfnamefont {E.}~\bibnamefont
  {Kierig}}, \bibinfo {author} {\bibfnamefont {U.}~\bibnamefont
  {Schnorrberger}}, \bibinfo {author} {\bibfnamefont {A.}~\bibnamefont
  {Schietinger}}, \bibinfo {author} {\bibfnamefont {J.}~\bibnamefont
  {Tomkovic}}, \ and\ \bibinfo {author} {\bibfnamefont {M.~K.}\ \bibnamefont
  {Oberthaler}},\ }\href {https://doi.org/10.1103/PhysRevLett.100.190405}
  {\bibfield  {journal} {\bibinfo  {journal} {Phys. Rev. Lett.}\ }\textbf
  {\bibinfo {volume} {100}},\ \bibinfo {pages} {190405} (\bibinfo {year}
  {2008})}\BibitemShut {NoStop}%
\bibitem [{\citenamefont {Sias}\ \emph {et~al.}(2008)\citenamefont {Sias},
  \citenamefont {Lignier}, \citenamefont {Singh}, \citenamefont {Zenesini},
  \citenamefont {Ciampini}, \citenamefont {Morsch},\ and\ \citenamefont
  {Arimondo}}]{sias2008}%
  \BibitemOpen
  \bibfield  {author} {\bibinfo {author} {\bibfnamefont {C.}~\bibnamefont
  {Sias}}, \bibinfo {author} {\bibfnamefont {H.}~\bibnamefont {Lignier}},
  \bibinfo {author} {\bibfnamefont {Y.~P.}\ \bibnamefont {Singh}}, \bibinfo
  {author} {\bibfnamefont {A.}~\bibnamefont {Zenesini}}, \bibinfo {author}
  {\bibfnamefont {D.}~\bibnamefont {Ciampini}}, \bibinfo {author}
  {\bibfnamefont {O.}~\bibnamefont {Morsch}}, \ and\ \bibinfo {author}
  {\bibfnamefont {E.}~\bibnamefont {Arimondo}},\ }\href
  {http://link.aps.org/doi/10.1103/PhysRevLett.100.040404} {\bibfield
  {journal} {\bibinfo  {journal} {Phys. Rev. Lett.}\ }\textbf {\bibinfo
  {volume} {100}},\ \bibinfo {pages} {040404} (\bibinfo {year}
  {2008})}\BibitemShut {NoStop}%
\bibitem [{\citenamefont {Zenesini}\ \emph {et~al.}(2009)\citenamefont
  {Zenesini}, \citenamefont {Lignier}, \citenamefont {Ciampini}, \citenamefont
  {Morsch},\ and\ \citenamefont {Arimondo}}]{zenesini2009}%
  \BibitemOpen
  \bibfield  {author} {\bibinfo {author} {\bibfnamefont {A.}~\bibnamefont
  {Zenesini}}, \bibinfo {author} {\bibfnamefont {H.}~\bibnamefont {Lignier}},
  \bibinfo {author} {\bibfnamefont {D.}~\bibnamefont {Ciampini}}, \bibinfo
  {author} {\bibfnamefont {O.}~\bibnamefont {Morsch}}, \ and\ \bibinfo {author}
  {\bibfnamefont {E.}~\bibnamefont {Arimondo}},\ }\href
  {http://link.aps.org/doi/10.1103/PhysRevLett.102.100403} {\bibfield
  {journal} {\bibinfo  {journal} {Phys. Rev. Lett.}\ }\textbf {\bibinfo
  {volume} {102}},\ \bibinfo {pages} {100403} (\bibinfo {year}
  {2009})}\BibitemShut {NoStop}%
\bibitem [{\citenamefont {Morsch}\ \emph {et~al.}(2010)\citenamefont {Morsch},
  \citenamefont {Ciampini},\ and\ \citenamefont {Arimondo}}]{Morsch_2010}%
  \BibitemOpen
  \bibfield  {author} {\bibinfo {author} {\bibfnamefont {O.}~\bibnamefont
  {Morsch}}, \bibinfo {author} {\bibfnamefont {D.}~\bibnamefont {Ciampini}}, \
  and\ \bibinfo {author} {\bibfnamefont {E.}~\bibnamefont {Arimondo}},\ }\href
  {http://dx.doi.org/10.1051/epn/2010303} {\bibfield  {journal} {\bibinfo
  {journal} {Europhysics News}\ }\textbf {\bibinfo {volume} {41}},\ \bibinfo
  {pages} {21} (\bibinfo {year} {2010})}\BibitemShut {NoStop}%
\bibitem [{\citenamefont {Eckardt}(2016)}]{Eckardt:2016Review}%
  \BibitemOpen
  \bibfield  {author} {\bibinfo {author} {\bibfnamefont {A.}~\bibnamefont
  {Eckardt}},\ }\href {https://arxiv.org/abs/1606.08041} {\bibfield  {journal}
  {\bibinfo  {journal} {arXiv:1606.08041}\ } (\bibinfo {year}
  {2016})}\BibitemShut {NoStop}%
\bibitem [{\citenamefont {Eckardt}\ \emph {et~al.}(2010)\citenamefont
  {Eckardt}, \citenamefont {Hauke}, \citenamefont {Soltan-Panahi},
  \citenamefont {Becker}, \citenamefont {Sengstock},\ and\ \citenamefont
  {Lewenstein}}]{Eckardt_2010}%
  \BibitemOpen
  \bibfield  {author} {\bibinfo {author} {\bibfnamefont {A.}~\bibnamefont
  {Eckardt}}, \bibinfo {author} {\bibfnamefont {P.}~\bibnamefont {Hauke}},
  \bibinfo {author} {\bibfnamefont {P.}~\bibnamefont {Soltan-Panahi}}, \bibinfo
  {author} {\bibfnamefont {C.}~\bibnamefont {Becker}}, \bibinfo {author}
  {\bibfnamefont {K.}~\bibnamefont {Sengstock}}, \ and\ \bibinfo {author}
  {\bibfnamefont {M.}~\bibnamefont {Lewenstein}},\ }\href
  {http://dx.doi.org/10.1209/0295-5075/89/10010} {\bibfield  {journal}
  {\bibinfo  {journal} {EPL}\ }\textbf {\bibinfo {volume} {89}},\ \bibinfo
  {pages} {10010} (\bibinfo {year} {2010})}\BibitemShut {NoStop}%
\bibitem [{\citenamefont {Struck}\ \emph {et~al.}(2011)\citenamefont {Struck},
  \citenamefont {Olschlager}, \citenamefont {Targat}, \citenamefont
  {Soltan-Panahi}, \citenamefont {Eckardt}, \citenamefont {Lewenstein},
  \citenamefont {Windpassinger},\ and\ \citenamefont
  {Sengstock}}]{struck_2011}%
  \BibitemOpen
  \bibfield  {author} {\bibinfo {author} {\bibfnamefont {J.}~\bibnamefont
  {Struck}}, \bibinfo {author} {\bibfnamefont {C.}~\bibnamefont {Olschlager}},
  \bibinfo {author} {\bibfnamefont {R.~L.}\ \bibnamefont {Targat}}, \bibinfo
  {author} {\bibfnamefont {P.}~\bibnamefont {Soltan-Panahi}}, \bibinfo {author}
  {\bibfnamefont {A.}~\bibnamefont {Eckardt}}, \bibinfo {author} {\bibfnamefont
  {M.}~\bibnamefont {Lewenstein}}, \bibinfo {author} {\bibfnamefont
  {P.}~\bibnamefont {Windpassinger}}, \ and\ \bibinfo {author} {\bibfnamefont
  {K.}~\bibnamefont {Sengstock}},\ }\href
  {http://dx.doi.org/10.1126/science.1207239} {\bibfield  {journal} {\bibinfo
  {journal} {Science}\ }\textbf {\bibinfo {volume} {333}},\ \bibinfo {pages}
  {996} (\bibinfo {year} {2011})}\BibitemShut {NoStop}%
\bibitem [{\citenamefont {Struck}\ \emph {et~al.}(2012)\citenamefont {Struck},
  \citenamefont {\"Olschl\"ager}, \citenamefont {Weinberg}, \citenamefont
  {Hauke}, \citenamefont {Simonet}, \citenamefont {Eckardt}, \citenamefont
  {Lewenstein}, \citenamefont {Sengstock},\ and\ \citenamefont
  {Windpassinger}}]{struck_2012}%
  \BibitemOpen
  \bibfield  {author} {\bibinfo {author} {\bibfnamefont {J.}~\bibnamefont
  {Struck}}, \bibinfo {author} {\bibfnamefont {C.}~\bibnamefont
  {\"Olschl\"ager}}, \bibinfo {author} {\bibfnamefont {M.}~\bibnamefont
  {Weinberg}}, \bibinfo {author} {\bibfnamefont {P.}~\bibnamefont {Hauke}},
  \bibinfo {author} {\bibfnamefont {J.}~\bibnamefont {Simonet}}, \bibinfo
  {author} {\bibfnamefont {A.}~\bibnamefont {Eckardt}}, \bibinfo {author}
  {\bibfnamefont {M.}~\bibnamefont {Lewenstein}}, \bibinfo {author}
  {\bibfnamefont {K.}~\bibnamefont {Sengstock}}, \ and\ \bibinfo {author}
  {\bibfnamefont {P.}~\bibnamefont {Windpassinger}},\ }\href
  {http://link.aps.org/doi/10.1103/PhysRevLett.108.225304} {\bibfield
  {journal} {\bibinfo  {journal} {Phys. Rev. Lett.}\ }\textbf {\bibinfo
  {volume} {108}},\ \bibinfo {pages} {225304} (\bibinfo {year}
  {2012})}\BibitemShut {NoStop}%
\bibitem [{\citenamefont {Struck}\ \emph {et~al.}(2013)\citenamefont {Struck},
  \citenamefont {Weinberg}, \citenamefont {{\"O}lschl{\"a}ger}, \citenamefont
  {Windpassinger}, \citenamefont {Simonet}, \citenamefont {Sengstock},
  \citenamefont {H{\"o}ppner}, \citenamefont {Hauke}, \citenamefont {Eckardt},
  \citenamefont {Lewenstein},\ and\ \citenamefont {Mathey}}]{struck_2013}%
  \BibitemOpen
  \bibfield  {author} {\bibinfo {author} {\bibfnamefont {J.}~\bibnamefont
  {Struck}}, \bibinfo {author} {\bibfnamefont {M.}~\bibnamefont {Weinberg}},
  \bibinfo {author} {\bibfnamefont {C.}~\bibnamefont {{\"O}lschl{\"a}ger}},
  \bibinfo {author} {\bibfnamefont {P.}~\bibnamefont {Windpassinger}}, \bibinfo
  {author} {\bibfnamefont {J.}~\bibnamefont {Simonet}}, \bibinfo {author}
  {\bibfnamefont {K.}~\bibnamefont {Sengstock}}, \bibinfo {author}
  {\bibfnamefont {R.}~\bibnamefont {H{\"o}ppner}}, \bibinfo {author}
  {\bibfnamefont {P.}~\bibnamefont {Hauke}}, \bibinfo {author} {\bibfnamefont
  {A.}~\bibnamefont {Eckardt}}, \bibinfo {author} {\bibfnamefont
  {M.}~\bibnamefont {Lewenstein}}, \ and\ \bibinfo {author} {\bibfnamefont
  {L.}~\bibnamefont {Mathey}},\ }\href {http://dx.doi.org/10.1038/nphys2750}
  {\bibfield  {journal} {\bibinfo  {journal} {Nat. Phys.}\ }\textbf {\bibinfo
  {volume} {9}},\ \bibinfo {pages} {738} (\bibinfo {year} {2013})}\BibitemShut
  {NoStop}%
\bibitem [{\citenamefont {Lim}\ \emph {et~al.}(2008)\citenamefont {Lim},
  \citenamefont {Smith},\ and\ \citenamefont {Hemmerich}}]{Lim:2008}%
  \BibitemOpen
  \bibfield  {author} {\bibinfo {author} {\bibfnamefont {L.-K.}\ \bibnamefont
  {Lim}}, \bibinfo {author} {\bibfnamefont {C.~M.}\ \bibnamefont {Smith}}, \
  and\ \bibinfo {author} {\bibfnamefont {A.}~\bibnamefont {Hemmerich}},\ }\href
  {https://doi.org/10.1103/PhysRevLett.100.130402} {\bibfield  {journal}
  {\bibinfo  {journal} {Phys. Rev. Lett.}\ }\textbf {\bibinfo {volume} {100}},\
  \bibinfo {pages} {130402} (\bibinfo {year} {2008})}\BibitemShut {NoStop}%
\bibitem [{\citenamefont {Kolovsky}(2011)}]{Kolovsky:2011}%
  \BibitemOpen
  \bibfield  {author} {\bibinfo {author} {\bibfnamefont {A.~R.}\ \bibnamefont
  {Kolovsky}},\ }\href {https://doi.org/10.1209/0295-5075/93/20003} {\bibfield
  {journal} {\bibinfo  {journal} {EPL}\ }\textbf {\bibinfo {volume} {93}},\
  \bibinfo {pages} {20003} (\bibinfo {year} {2011})}\BibitemShut {NoStop}%
\bibitem [{\citenamefont {Bermudez}\ \emph {et~al.}(2011)\citenamefont
  {Bermudez}, \citenamefont {Schaetz},\ and\ \citenamefont
  {Porras}}]{Bermudez:2011}%
  \BibitemOpen
  \bibfield  {author} {\bibinfo {author} {\bibfnamefont {A.}~\bibnamefont
  {Bermudez}}, \bibinfo {author} {\bibfnamefont {T.}~\bibnamefont {Schaetz}}, \
  and\ \bibinfo {author} {\bibfnamefont {D.}~\bibnamefont {Porras}},\ }\href
  {https://doi.org/10.1103/PhysRevLett.107.150501} {\bibfield  {journal}
  {\bibinfo  {journal} {Phys. Rev. Lett.}\ }\textbf {\bibinfo {volume} {107}},\
  \bibinfo {pages} {150501} (\bibinfo {year} {2011})}\BibitemShut {NoStop}%
\bibitem [{\citenamefont {Creffield}\ and\ \citenamefont
  {Sols}(2014)}]{creffield_2014}%
  \BibitemOpen
  \bibfield  {author} {\bibinfo {author} {\bibfnamefont {C.~E.}\ \bibnamefont
  {Creffield}}\ and\ \bibinfo {author} {\bibfnamefont {F.}~\bibnamefont
  {Sols}},\ }\href {http://link.aps.org/doi/10.1103/PhysRevA.90.023636}
  {\bibfield  {journal} {\bibinfo  {journal} {Phys. Rev. A}\ }\textbf {\bibinfo
  {volume} {90}},\ \bibinfo {pages} {023636} (\bibinfo {year}
  {2014})}\BibitemShut {NoStop}%
\bibitem [{\citenamefont {Creffield}\ \emph {et~al.}(2016)\citenamefont
  {Creffield}, \citenamefont {Pieplow}, \citenamefont {Sols},\ and\
  \citenamefont {Goldman}}]{creffield_2016}%
  \BibitemOpen
  \bibfield  {author} {\bibinfo {author} {\bibfnamefont {C.~E.}\ \bibnamefont
  {Creffield}}, \bibinfo {author} {\bibfnamefont {G.}~\bibnamefont {Pieplow}},
  \bibinfo {author} {\bibfnamefont {F.}~\bibnamefont {Sols}}, \ and\ \bibinfo
  {author} {\bibfnamefont {N.}~\bibnamefont {Goldman}},\ }\href
  {http://dx.doi.org/10.1088/1367-2630/18/9/093013} {\bibfield  {journal}
  {\bibinfo  {journal} {New Journal of Physics}\ }\textbf {\bibinfo {volume}
  {18}},\ \bibinfo {pages} {093013} (\bibinfo {year} {2016})}\BibitemShut
  {NoStop}%
\bibitem [{\citenamefont {Hauke}\ \emph {et~al.}(2012)\citenamefont {Hauke},
  \citenamefont {Tieleman}, \citenamefont {Celi}, \citenamefont
  {\"Olschl\"ager}, \citenamefont {Simonet}, \citenamefont {Struck},
  \citenamefont {Weinberg}, \citenamefont {Windpassinger}, \citenamefont
  {Sengstock}, \citenamefont {Lewenstein},\ and\ \citenamefont
  {Eckardt}}]{hauke_2012}%
  \BibitemOpen
  \bibfield  {author} {\bibinfo {author} {\bibfnamefont {P.}~\bibnamefont
  {Hauke}}, \bibinfo {author} {\bibfnamefont {O.}~\bibnamefont {Tieleman}},
  \bibinfo {author} {\bibfnamefont {A.}~\bibnamefont {Celi}}, \bibinfo {author}
  {\bibfnamefont {C.}~\bibnamefont {\"Olschl\"ager}}, \bibinfo {author}
  {\bibfnamefont {J.}~\bibnamefont {Simonet}}, \bibinfo {author} {\bibfnamefont
  {J.}~\bibnamefont {Struck}}, \bibinfo {author} {\bibfnamefont
  {M.}~\bibnamefont {Weinberg}}, \bibinfo {author} {\bibfnamefont
  {P.}~\bibnamefont {Windpassinger}}, \bibinfo {author} {\bibfnamefont
  {K.}~\bibnamefont {Sengstock}}, \bibinfo {author} {\bibfnamefont
  {M.}~\bibnamefont {Lewenstein}}, \ and\ \bibinfo {author} {\bibfnamefont
  {A.}~\bibnamefont {Eckardt}},\ }\href {\doibase
  10.1103/PhysRevLett.109.145301} {\bibfield  {journal} {\bibinfo  {journal}
  {Phys. Rev. Lett.}\ }\textbf {\bibinfo {volume} {109}},\ \bibinfo {pages}
  {145301} (\bibinfo {year} {2012})}\BibitemShut {NoStop}%
\bibitem [{\citenamefont {Kosior}\ and\ \citenamefont
  {Sacha}(2014)}]{kosior_2014}%
  \BibitemOpen
  \bibfield  {author} {\bibinfo {author} {\bibfnamefont {A.}~\bibnamefont
  {Kosior}}\ and\ \bibinfo {author} {\bibfnamefont {K.}~\bibnamefont {Sacha}},\
  }\href {http://dx.doi.org/10.1209/0295-5075/107/26006} {\bibfield  {journal}
  {\bibinfo  {journal} {EPL}\ }\textbf {\bibinfo {volume} {107}},\ \bibinfo
  {pages} {26006} (\bibinfo {year} {2014})}\BibitemShut {NoStop}%
\bibitem [{\citenamefont {Struck}\ \emph {et~al.}(2014)\citenamefont {Struck},
  \citenamefont {Simonet},\ and\ \citenamefont {Sengstock}}]{struck_2014}%
  \BibitemOpen
  \bibfield  {author} {\bibinfo {author} {\bibfnamefont {J.}~\bibnamefont
  {Struck}}, \bibinfo {author} {\bibfnamefont {J.}~\bibnamefont {Simonet}}, \
  and\ \bibinfo {author} {\bibfnamefont {K.}~\bibnamefont {Sengstock}},\ }\href
  {http://link.aps.org/doi/10.1103/PhysRevA.90.031601} {\bibfield  {journal}
  {\bibinfo  {journal} {Phys. Rev. A}\ }\textbf {\bibinfo {volume} {90}},\
  \bibinfo {pages} {031601} (\bibinfo {year} {2014})}\BibitemShut {NoStop}%
\bibitem [{\citenamefont {Baur}\ \emph {et~al.}(2014)\citenamefont {Baur},
  \citenamefont {Schleier-Smith},\ and\ \citenamefont {Cooper}}]{baur_2014}%
  \BibitemOpen
  \bibfield  {author} {\bibinfo {author} {\bibfnamefont {S.~K.}\ \bibnamefont
  {Baur}}, \bibinfo {author} {\bibfnamefont {M.~H.}\ \bibnamefont
  {Schleier-Smith}}, \ and\ \bibinfo {author} {\bibfnamefont {N.~R.}\
  \bibnamefont {Cooper}},\ }\href {\doibase 10.1103/PhysRevA.89.051605}
  {\bibfield  {journal} {\bibinfo  {journal} {Phys. Rev. A}\ }\textbf {\bibinfo
  {volume} {89}},\ \bibinfo {pages} {051605} (\bibinfo {year}
  {2014})}\BibitemShut {NoStop}%
\bibitem [{\citenamefont {Zheng}\ and\ \citenamefont
  {Zhai}(2014)}]{Zheng:2014}%
  \BibitemOpen
  \bibfield  {author} {\bibinfo {author} {\bibfnamefont {W.}~\bibnamefont
  {Zheng}}\ and\ \bibinfo {author} {\bibfnamefont {H.}~\bibnamefont {Zhai}},\
  }\href {https://doi.org/10.1103/PhysRevA.89.061603} {\bibfield  {journal}
  {\bibinfo  {journal} {Phys. Rev. A}\ }\textbf {\bibinfo {volume} {89}},\
  \bibinfo {pages} {061603} (\bibinfo {year} {2014})}\BibitemShut {NoStop}%
\bibitem [{\citenamefont {Nascimb{\`e}ne}\ \emph {et~al.}(2015)\citenamefont
  {Nascimb{\`e}ne}, \citenamefont {Goldman}, \citenamefont {Cooper},\ and\
  \citenamefont {Dalibard}}]{nascimbene2015}%
  \BibitemOpen
  \bibfield  {author} {\bibinfo {author} {\bibfnamefont {S.}~\bibnamefont
  {Nascimb{\`e}ne}}, \bibinfo {author} {\bibfnamefont {N.}~\bibnamefont
  {Goldman}}, \bibinfo {author} {\bibfnamefont {N.~R.}\ \bibnamefont {Cooper}},
  \ and\ \bibinfo {author} {\bibfnamefont {J.}~\bibnamefont {Dalibard}},\
  }\href {https://doi.org/10.1103/PhysRevLett.115.140401} {\bibfield  {journal}
  {\bibinfo  {journal} {Phys. Rev. Lett.}\ }\textbf {\bibinfo {volume} {115}},\
  \bibinfo {pages} {140} (\bibinfo {year} {2015})}\BibitemShut {NoStop}%
\bibitem [{\citenamefont {Sacha}\ \emph {et~al.}(2012)\citenamefont {Sacha},
  \citenamefont {Targo\ifmmode~\acute{n}\else \'{n}\fi{}ska},\ and\
  \citenamefont {Zakrzewski}}]{sacha_2012}%
  \BibitemOpen
  \bibfield  {author} {\bibinfo {author} {\bibfnamefont {K.}~\bibnamefont
  {Sacha}}, \bibinfo {author} {\bibfnamefont {K.}~\bibnamefont
  {Targo\ifmmode~\acute{n}\else \'{n}\fi{}ska}}, \ and\ \bibinfo {author}
  {\bibfnamefont {J.}~\bibnamefont {Zakrzewski}},\ }\href
  {http://link.aps.org/doi/10.1103/PhysRevA.85.053613} {\bibfield  {journal}
  {\bibinfo  {journal} {Phys. Rev. A}\ }\textbf {\bibinfo {volume} {85}},\
  \bibinfo {pages} {053613} (\bibinfo {year} {2012})}\BibitemShut {NoStop}%
\bibitem [{\citenamefont {Budich}\ \emph {et~al.}(2016)\citenamefont {Budich},
  \citenamefont {Hu},\ and\ \citenamefont {Zoller}}]{Budich:2016}%
  \BibitemOpen
  \bibfield  {author} {\bibinfo {author} {\bibfnamefont {J.~C.}\ \bibnamefont
  {Budich}}, \bibinfo {author} {\bibfnamefont {Y.}~\bibnamefont {Hu}}, \ and\
  \bibinfo {author} {\bibfnamefont {P.}~\bibnamefont {Zoller}},\ }\href
  {https://arxiv.org/abs/1608.05096} {\bibfield  {journal} {\bibinfo  {journal}
  {arxiv:1608.05096}\ } (\bibinfo {year} {2016})}\BibitemShut {NoStop}%
\bibitem [{\citenamefont {Price}\ \emph {et~al.}(2016)\citenamefont {Price},
  \citenamefont {Ozawa},\ and\ \citenamefont {Goldman}}]{Price:2016}%
  \BibitemOpen
  \bibfield  {author} {\bibinfo {author} {\bibfnamefont {H.~M.}\ \bibnamefont
  {Price}}, \bibinfo {author} {\bibfnamefont {T.}~\bibnamefont {Ozawa}}, \ and\
  \bibinfo {author} {\bibfnamefont {N.}~\bibnamefont {Goldman}},\ }\href
  {https://arxiv.org/abs/1605.09310} {\bibfield  {journal} {\bibinfo  {journal}
  {arxiv:1605.09310}\ } (\bibinfo {year} {2016})}\BibitemShut {NoStop}%
\bibitem [{\citenamefont {Sorensen}\ \emph {et~al.}(2005)\citenamefont
  {Sorensen}, \citenamefont {Demler},\ and\ \citenamefont
  {Lukin}}]{Sorensen:2005}%
  \BibitemOpen
  \bibfield  {author} {\bibinfo {author} {\bibfnamefont {A.~S.}\ \bibnamefont
  {Sorensen}}, \bibinfo {author} {\bibfnamefont {E.}~\bibnamefont {Demler}}, \
  and\ \bibinfo {author} {\bibfnamefont {M.~D.}\ \bibnamefont {Lukin}},\ }\href
  {https://doi.org/10.1103/PhysRevLett.94.086803} {\bibfield  {journal}
  {\bibinfo  {journal} {Phys. Rev. Lett.}\ }\textbf {\bibinfo {volume} {94}},\
  \bibinfo {pages} {086803} (\bibinfo {year} {2005})}\BibitemShut {NoStop}%
\bibitem [{\citenamefont {Xu}\ \emph {et~al.}(2013)\citenamefont {Xu},
  \citenamefont {You},\ and\ \citenamefont {Ueda}}]{Xu:2013}%
  \BibitemOpen
  \bibfield  {author} {\bibinfo {author} {\bibfnamefont {Z.-F.}\ \bibnamefont
  {Xu}}, \bibinfo {author} {\bibfnamefont {L.}~\bibnamefont {You}}, \ and\
  \bibinfo {author} {\bibfnamefont {M.}~\bibnamefont {Ueda}},\ }\href
  {https://doi.org/10.1103/PhysRevA.87.063634} {\bibfield  {journal} {\bibinfo
  {journal} {Phys. Rev. A}\ }\textbf {\bibinfo {volume} {87}},\ \bibinfo
  {pages} {063634} (\bibinfo {year} {2013})}\BibitemShut {NoStop}%
\bibitem [{\citenamefont {Anderson}\ \emph {et~al.}(2013)\citenamefont
  {Anderson}, \citenamefont {Spielman},\ and\ \citenamefont
  {Juzeli{\=u}nas}}]{Anderson:2013}%
  \BibitemOpen
  \bibfield  {author} {\bibinfo {author} {\bibfnamefont {B.~M.}\ \bibnamefont
  {Anderson}}, \bibinfo {author} {\bibfnamefont {I.~B.}\ \bibnamefont
  {Spielman}}, \ and\ \bibinfo {author} {\bibfnamefont {G.}~\bibnamefont
  {Juzeli{\=u}nas}},\ }\href {https://doi.org/10.1103/PhysRevLett.111.125301}
  {\bibfield  {journal} {\bibinfo  {journal} {Phys. Rev. Lett.}\ }\textbf
  {\bibinfo {volume} {111}},\ \bibinfo {pages} {125301} (\bibinfo {year}
  {2013})}\BibitemShut {NoStop}%
\bibitem [{\citenamefont {Goldman}\ and\ \citenamefont
  {Dalibard}(2014)}]{goldman_2014}%
  \BibitemOpen
  \bibfield  {author} {\bibinfo {author} {\bibfnamefont {N.}~\bibnamefont
  {Goldman}}\ and\ \bibinfo {author} {\bibfnamefont {J.}~\bibnamefont
  {Dalibard}},\ }\href {\doibase 10.1103/PhysRevX.4.031027} {\bibfield
  {journal} {\bibinfo  {journal} {Phys. Rev. X}\ }\textbf {\bibinfo {volume}
  {4}},\ \bibinfo {pages} {031027} (\bibinfo {year} {2014})}\BibitemShut
  {NoStop}%
\bibitem [{\citenamefont {Goldman}\ \emph {et~al.}(2016)\citenamefont
  {Goldman}, \citenamefont {Budich},\ and\ \citenamefont
  {Zoller}}]{Goldman:2016Review}%
  \BibitemOpen
  \bibfield  {author} {\bibinfo {author} {\bibfnamefont {N.}~\bibnamefont
  {Goldman}}, \bibinfo {author} {\bibfnamefont {J.~C.}\ \bibnamefont {Budich}},
  \ and\ \bibinfo {author} {\bibfnamefont {P.}~\bibnamefont {Zoller}},\ }\href
  {https://doi.org/10.1038/nphys3803} {\bibfield  {journal} {\bibinfo
  {journal} {Nat. Phys.}\ }\textbf {\bibinfo {volume} {12}},\ \bibinfo {pages}
  {639} (\bibinfo {year} {2016})}\BibitemShut {NoStop}%
\bibitem [{\citenamefont {Miyake}\ \emph {et~al.}(2013)\citenamefont {Miyake},
  \citenamefont {Siviloglou}, \citenamefont {Kennedy}, \citenamefont {Burton},\
  and\ \citenamefont {Ketterle}}]{Miyake_2013}%
  \BibitemOpen
  \bibfield  {author} {\bibinfo {author} {\bibfnamefont {H.}~\bibnamefont
  {Miyake}}, \bibinfo {author} {\bibfnamefont {G.~A.}\ \bibnamefont
  {Siviloglou}}, \bibinfo {author} {\bibfnamefont {C.~J.}\ \bibnamefont
  {Kennedy}}, \bibinfo {author} {\bibfnamefont {W.~C.}\ \bibnamefont {Burton}},
  \ and\ \bibinfo {author} {\bibfnamefont {W.}~\bibnamefont {Ketterle}},\
  }\href {\doibase 10.1103/physrevlett.111.185302} {\bibfield  {journal}
  {\bibinfo  {journal} {Phys. Rev. Lett.}\ }\textbf {\bibinfo {volume} {111}},\
  \bibinfo {pages} {185302} (\bibinfo {year} {2013})}\BibitemShut {NoStop}%
\bibitem [{\citenamefont {Aidelsburger}\ \emph {et~al.}(2013)\citenamefont
  {Aidelsburger}, \citenamefont {Atala}, \citenamefont {Lohse}, \citenamefont
  {Barreiro}, \citenamefont {Paredes},\ and\ \citenamefont
  {Bloch}}]{Aidelsburger_2013}%
  \BibitemOpen
  \bibfield  {author} {\bibinfo {author} {\bibfnamefont {M.}~\bibnamefont
  {Aidelsburger}}, \bibinfo {author} {\bibfnamefont {M.}~\bibnamefont {Atala}},
  \bibinfo {author} {\bibfnamefont {M.}~\bibnamefont {Lohse}}, \bibinfo
  {author} {\bibfnamefont {J.~T.}\ \bibnamefont {Barreiro}}, \bibinfo {author}
  {\bibfnamefont {B.}~\bibnamefont {Paredes}}, \ and\ \bibinfo {author}
  {\bibfnamefont {I.}~\bibnamefont {Bloch}},\ }\href {\doibase
  10.1103/PhysRevLett.111.185301} {\bibfield  {journal} {\bibinfo  {journal}
  {Phys. Rev. Lett.}\ }\textbf {\bibinfo {volume} {111}},\ \bibinfo {pages}
  {185301} (\bibinfo {year} {2013})}\BibitemShut {NoStop}%
\bibitem [{\citenamefont {Aidelsburger}\ \emph {et~al.}(2014)\citenamefont
  {Aidelsburger}, \citenamefont {Lohse}, \citenamefont {Schweizer},
  \citenamefont {Atala}, \citenamefont {Barreiro}, \citenamefont
  {Nascimb{\`{e}}ne}, \citenamefont {Cooper}, \citenamefont {Bloch},\ and\
  \citenamefont {Goldman}}]{Aidelsburger_2014}%
  \BibitemOpen
  \bibfield  {author} {\bibinfo {author} {\bibfnamefont {M.}~\bibnamefont
  {Aidelsburger}}, \bibinfo {author} {\bibfnamefont {M.}~\bibnamefont {Lohse}},
  \bibinfo {author} {\bibfnamefont {C.}~\bibnamefont {Schweizer}}, \bibinfo
  {author} {\bibfnamefont {M.}~\bibnamefont {Atala}}, \bibinfo {author}
  {\bibfnamefont {J.~T.}\ \bibnamefont {Barreiro}}, \bibinfo {author}
  {\bibfnamefont {S.}~\bibnamefont {Nascimb{\`{e}}ne}}, \bibinfo {author}
  {\bibfnamefont {N.~R.}\ \bibnamefont {Cooper}}, \bibinfo {author}
  {\bibfnamefont {I.}~\bibnamefont {Bloch}}, \ and\ \bibinfo {author}
  {\bibfnamefont {N.}~\bibnamefont {Goldman}},\ }\href {\doibase
  10.1038/nphys3171} {\bibfield  {journal} {\bibinfo  {journal} {Nat. Phys.}\
  }\textbf {\bibinfo {volume} {11}},\ \bibinfo {pages} {3171} (\bibinfo {year}
  {2014})}\BibitemShut {NoStop}%
\bibitem [{\citenamefont {Kennedy}\ \emph {et~al.}(2015)\citenamefont
  {Kennedy}, \citenamefont {Burton}, \citenamefont {Chung},\ and\ \citenamefont
  {Ketterle}}]{Kennedy:2015NP}%
  \BibitemOpen
  \bibfield  {author} {\bibinfo {author} {\bibfnamefont {C.~J.}\ \bibnamefont
  {Kennedy}}, \bibinfo {author} {\bibfnamefont {W.~C.}\ \bibnamefont {Burton}},
  \bibinfo {author} {\bibfnamefont {W.~C.}\ \bibnamefont {Chung}}, \ and\
  \bibinfo {author} {\bibfnamefont {W.}~\bibnamefont {Ketterle}},\ }\href
  {https://doi.org/10.1038/nphys3421} {\bibfield  {journal} {\bibinfo
  {journal} {Nat. Phys.}\ }\textbf {\bibinfo {volume} {11}},\ \bibinfo {pages}
  {859} (\bibinfo {year} {2015})}\BibitemShut {NoStop}%
\bibitem [{\citenamefont {Tai}\ \emph {et~al.}(2016)\citenamefont {Tai},
  \citenamefont {Lukin}, \citenamefont {Rispoli}, \citenamefont {Schittko},
  \citenamefont {Menke}, \citenamefont {Borgnia}, \citenamefont {Preiss},
  \citenamefont {Grusdt}, \citenamefont {Kaufman},\ and\ \citenamefont
  {Greiner}}]{Tai:2016}%
  \BibitemOpen
  \bibfield  {author} {\bibinfo {author} {\bibfnamefont {M.~E.}\ \bibnamefont
  {Tai}}, \bibinfo {author} {\bibfnamefont {A.}~\bibnamefont {Lukin}}, \bibinfo
  {author} {\bibfnamefont {M.}~\bibnamefont {Rispoli}}, \bibinfo {author}
  {\bibfnamefont {R.}~\bibnamefont {Schittko}}, \bibinfo {author}
  {\bibfnamefont {T.}~\bibnamefont {Menke}}, \bibinfo {author} {\bibfnamefont
  {D.}~\bibnamefont {Borgnia}}, \bibinfo {author} {\bibfnamefont {P.~M.}\
  \bibnamefont {Preiss}}, \bibinfo {author} {\bibfnamefont {F.}~\bibnamefont
  {Grusdt}}, \bibinfo {author} {\bibfnamefont {A.~M.}\ \bibnamefont {Kaufman}},
  \ and\ \bibinfo {author} {\bibfnamefont {M.}~\bibnamefont {Greiner}},\ }\href
  {https://arxiv.org/abs/1612.05631v1} {\bibfield  {journal} {\bibinfo
  {journal} {arxiv:1612.05631}\ } (\bibinfo {year} {2016})}\BibitemShut
  {NoStop}%
\bibitem [{\citenamefont {Jotzu}\ \emph {et~al.}(2014)\citenamefont {Jotzu},
  \citenamefont {Messer}, \citenamefont {Desbuquois}, \citenamefont {Lebrat},
  \citenamefont {Uehlinger}, \citenamefont {Greif},\ and\ \citenamefont
  {Esslinger}}]{Jotzu_2014}%
  \BibitemOpen
  \bibfield  {author} {\bibinfo {author} {\bibfnamefont {G.}~\bibnamefont
  {Jotzu}}, \bibinfo {author} {\bibfnamefont {M.}~\bibnamefont {Messer}},
  \bibinfo {author} {\bibfnamefont {R.}~\bibnamefont {Desbuquois}}, \bibinfo
  {author} {\bibfnamefont {M.}~\bibnamefont {Lebrat}}, \bibinfo {author}
  {\bibfnamefont {T.}~\bibnamefont {Uehlinger}}, \bibinfo {author}
  {\bibfnamefont {D.}~\bibnamefont {Greif}}, \ and\ \bibinfo {author}
  {\bibfnamefont {T.}~\bibnamefont {Esslinger}},\ }\href {\doibase
  10.1038/nature13915} {\bibfield  {journal} {\bibinfo  {journal} {Nature}\
  }\textbf {\bibinfo {volume} {515}},\ \bibinfo {pages} {237} (\bibinfo {year}
  {2014})}\BibitemShut {NoStop}%
\bibitem [{\citenamefont {Oka}\ and\ \citenamefont {Aoki}(2009)}]{oka_09}%
  \BibitemOpen
  \bibfield  {author} {\bibinfo {author} {\bibfnamefont {T.}~\bibnamefont
  {Oka}}\ and\ \bibinfo {author} {\bibfnamefont {H.}~\bibnamefont {Aoki}},\
  }\href {https://doi.org/10.1103/PhysRevB.79.081406} {\bibfield  {journal}
  {\bibinfo  {journal} {Phys. Rev. B}\ }\textbf {\bibinfo {volume} {79}},\
  \bibinfo {pages} {081406} (\bibinfo {year} {2009})}\BibitemShut {NoStop}%
\bibitem [{\citenamefont {Kitagawa}\ \emph {et~al.}(2010)\citenamefont
  {Kitagawa}, \citenamefont {Berg}, \citenamefont {Rudner},\ and\ \citenamefont
  {Demler}}]{kitagawa2010}%
  \BibitemOpen
  \bibfield  {author} {\bibinfo {author} {\bibfnamefont {T.}~\bibnamefont
  {Kitagawa}}, \bibinfo {author} {\bibfnamefont {E.}~\bibnamefont {Berg}},
  \bibinfo {author} {\bibfnamefont {M.}~\bibnamefont {Rudner}}, \ and\ \bibinfo
  {author} {\bibfnamefont {E.}~\bibnamefont {Demler}},\ }\href
  {https://doi.org/10.1103/PhysRevB.82.235114} {\bibfield  {journal} {\bibinfo
  {journal} {Phys. Rev. B}\ }\textbf {\bibinfo {volume} {82}},\ \bibinfo
  {pages} {235114} (\bibinfo {year} {2010})}\BibitemShut {NoStop}%
\bibitem [{\citenamefont {Calvo}\ \emph {et~al.}(2011)\citenamefont {Calvo},
  \citenamefont {Pastawski}, \citenamefont {Roche},\ and\ \citenamefont
  {Foa~Torres}}]{Calvo:2011}%
  \BibitemOpen
  \bibfield  {author} {\bibinfo {author} {\bibfnamefont {H.~L.}\ \bibnamefont
  {Calvo}}, \bibinfo {author} {\bibfnamefont {H.~M.}\ \bibnamefont
  {Pastawski}}, \bibinfo {author} {\bibfnamefont {S.}~\bibnamefont {Roche}}, \
  and\ \bibinfo {author} {\bibfnamefont {L.~E.~F.}\ \bibnamefont
  {Foa~Torres}},\ }\href {http://dx.doi.org/10.1063/1.3597412} {\bibfield
  {journal} {\bibinfo  {journal} {Appl. Phys. Lett.}\ }\textbf {\bibinfo
  {volume} {98}},\ \bibinfo {pages} {232103} (\bibinfo {year}
  {2011})}\BibitemShut {NoStop}%
\bibitem [{\citenamefont {Kitagawa}\ \emph {et~al.}(2011)\citenamefont
  {Kitagawa}, \citenamefont {Oka}, \citenamefont {Brataas}, \citenamefont
  {Fu},\ and\ \citenamefont {Demler}}]{kitagawa11}%
  \BibitemOpen
  \bibfield  {author} {\bibinfo {author} {\bibfnamefont {T.}~\bibnamefont
  {Kitagawa}}, \bibinfo {author} {\bibfnamefont {T.}~\bibnamefont {Oka}},
  \bibinfo {author} {\bibfnamefont {A.}~\bibnamefont {Brataas}}, \bibinfo
  {author} {\bibfnamefont {L.}~\bibnamefont {Fu}}, \ and\ \bibinfo {author}
  {\bibfnamefont {E.}~\bibnamefont {Demler}},\ }\href
  {https://doi.org/10.1103/PhysRevB.84.235108} {\bibfield  {journal} {\bibinfo
  {journal} {Phys. Rev. B}\ }\textbf {\bibinfo {volume} {84}},\ \bibinfo
  {pages} {235108} (\bibinfo {year} {2011})}\BibitemShut {NoStop}%
\bibitem [{\citenamefont {Lindner}\ \emph {et~al.}(2011)\citenamefont
  {Lindner}, \citenamefont {Refael},\ and\ \citenamefont
  {Galitski}}]{lindner2011}%
  \BibitemOpen
  \bibfield  {author} {\bibinfo {author} {\bibfnamefont {N.~H.}\ \bibnamefont
  {Lindner}}, \bibinfo {author} {\bibfnamefont {G.}~\bibnamefont {Refael}}, \
  and\ \bibinfo {author} {\bibfnamefont {V.}~\bibnamefont {Galitski}},\ }\href
  {https://doi.org/10.1038/nphys1926} {\bibfield  {journal} {\bibinfo
  {journal} {Nat. Phys.}\ }\textbf {\bibinfo {volume} {7}},\ \bibinfo {pages}
  {490} (\bibinfo {year} {2011})}\BibitemShut {NoStop}%
\bibitem [{\citenamefont {Cayssol}\ \emph {et~al.}(2013)\citenamefont
  {Cayssol}, \citenamefont {Dora}, \citenamefont {Simon},\ and\ \citenamefont
  {Moessner}}]{cayssol2013}%
  \BibitemOpen
  \bibfield  {author} {\bibinfo {author} {\bibfnamefont {J.}~\bibnamefont
  {Cayssol}}, \bibinfo {author} {\bibfnamefont {B.}~\bibnamefont {Dora}},
  \bibinfo {author} {\bibfnamefont {F.}~\bibnamefont {Simon}}, \ and\ \bibinfo
  {author} {\bibfnamefont {R.}~\bibnamefont {Moessner}},\ }\href
  {https://doi.org/10.1002/pssr.201206451} {\bibfield  {journal} {\bibinfo
  {journal} {Phys. Status Solidi Rapid Res. Lett.}\ }\textbf {\bibinfo {volume}
  {7}},\ \bibinfo {pages} {101} (\bibinfo {year} {2013})}\BibitemShut {NoStop}%
\bibitem [{\citenamefont {Bukov}\ \emph
  {et~al.}(2015{\natexlab{a}})\citenamefont {Bukov}, \citenamefont
  {D'Alessio},\ and\ \citenamefont {Polkovnikov}}]{bukov_2015}%
  \BibitemOpen
  \bibfield  {author} {\bibinfo {author} {\bibfnamefont {M.}~\bibnamefont
  {Bukov}}, \bibinfo {author} {\bibfnamefont {L.}~\bibnamefont {D'Alessio}}, \
  and\ \bibinfo {author} {\bibfnamefont {A.}~\bibnamefont {Polkovnikov}},\
  }\href {https://doi.org/10.1080/00018732.2015.1055918} {\bibfield  {journal}
  {\bibinfo  {journal} {Adv. Phys.}\ }\textbf {\bibinfo {volume} {64}},\
  \bibinfo {pages} {139} (\bibinfo {year} {2015}{\natexlab{a}})}\BibitemShut
  {NoStop}%
\bibitem [{\citenamefont {Lu}\ \emph {et~al.}(2014)\citenamefont {Lu},
  \citenamefont {Joannopoulos},\ and\ \citenamefont
  {Solja\v{c}i{\'c}}}]{Lu:2014Review}%
  \BibitemOpen
  \bibfield  {author} {\bibinfo {author} {\bibfnamefont {L.}~\bibnamefont
  {Lu}}, \bibinfo {author} {\bibfnamefont {J.~D.}\ \bibnamefont
  {Joannopoulos}}, \ and\ \bibinfo {author} {\bibfnamefont {M.}~\bibnamefont
  {Solja\v{c}i{\'c}}},\ }\href {https://doi.org/10.1038/nphoton.2014.248}
  {\bibfield  {journal} {\bibinfo  {journal} {Nat. Photon.}\ }\textbf {\bibinfo
  {volume} {8}},\ \bibinfo {pages} {821} (\bibinfo {year} {2014})}\BibitemShut
  {NoStop}%
\bibitem [{\citenamefont {Rudner}\ \emph {et~al.}(2013)\citenamefont {Rudner},
  \citenamefont {Lindner}, \citenamefont {Berg},\ and\ \citenamefont
  {Levin}}]{Rudner:2013}%
  \BibitemOpen
  \bibfield  {author} {\bibinfo {author} {\bibfnamefont {M.~S.}\ \bibnamefont
  {Rudner}}, \bibinfo {author} {\bibfnamefont {N.~H.}\ \bibnamefont {Lindner}},
  \bibinfo {author} {\bibfnamefont {E.}~\bibnamefont {Berg}}, \ and\ \bibinfo
  {author} {\bibfnamefont {M.}~\bibnamefont {Levin}},\ }\href
  {https://doi.org/10.1103/physrevx.3.031005} {\bibfield  {journal} {\bibinfo
  {journal} {Phys. Rev. X}\ }\textbf {\bibinfo {volume} {3}},\ \bibinfo {pages}
  {031005} (\bibinfo {year} {2013})}\BibitemShut {NoStop}%
\bibitem [{\citenamefont {Carpentier}\ \emph {et~al.}(2015)\citenamefont
  {Carpentier}, \citenamefont {Delplace}, \citenamefont {Fruchart},\ and\
  \citenamefont {Gawedzki}}]{Carpentier:2015}%
  \BibitemOpen
  \bibfield  {author} {\bibinfo {author} {\bibfnamefont {D.}~\bibnamefont
  {Carpentier}}, \bibinfo {author} {\bibfnamefont {P.}~\bibnamefont
  {Delplace}}, \bibinfo {author} {\bibfnamefont {M.}~\bibnamefont {Fruchart}},
  \ and\ \bibinfo {author} {\bibfnamefont {K.}~\bibnamefont {Gawedzki}},\
  }\href {http://journals.aps.org/prl/abstract/10.1103/PhysRevLett.114.106806}
  {\bibfield  {journal} {\bibinfo  {journal} {Phys. Rev. Lett.}\ }\textbf
  {\bibinfo {volume} {114}},\ \bibinfo {pages} {106806} (\bibinfo {year}
  {2015})}\BibitemShut {NoStop}%
\bibitem [{\citenamefont {Nathan}\ and\ \citenamefont
  {Rudner}(2015)}]{Nathan:2015}%
  \BibitemOpen
  \bibfield  {author} {\bibinfo {author} {\bibfnamefont {F.}~\bibnamefont
  {Nathan}}\ and\ \bibinfo {author} {\bibfnamefont {M.~S.}\ \bibnamefont
  {Rudner}},\ }\href
  {http://iopscience.iop.org/article/10.1088/1367-2630/17/12/125014/meta}
  {\bibfield  {journal} {\bibinfo  {journal} {New J. Phys.}\ }\textbf {\bibinfo
  {volume} {17}},\ \bibinfo {pages} {1} (\bibinfo {year} {2015})}\BibitemShut
  {NoStop}%
\bibitem [{\citenamefont {Foa~Torres}\ \emph {et~al.}(2014)\citenamefont
  {Foa~Torres}, \citenamefont {Perez-Piskunow}, \citenamefont {Balseiro},\ and\
  \citenamefont {Usaj}}]{torres_2014}%
  \BibitemOpen
  \bibfield  {author} {\bibinfo {author} {\bibfnamefont {L.~E.~F.}\
  \bibnamefont {Foa~Torres}}, \bibinfo {author} {\bibfnamefont {P.~M.}\
  \bibnamefont {Perez-Piskunow}}, \bibinfo {author} {\bibfnamefont {C.~A.}\
  \bibnamefont {Balseiro}}, \ and\ \bibinfo {author} {\bibfnamefont
  {G.}~\bibnamefont {Usaj}},\ }\href
  {http://link.aps.org/doi/10.1103/PhysRevLett.113.266801} {\bibfield
  {journal} {\bibinfo  {journal} {Phys. Rev. Lett.}\ }\textbf {\bibinfo
  {volume} {113}},\ \bibinfo {pages} {266801} (\bibinfo {year}
  {2014})}\BibitemShut {NoStop}%
\bibitem [{\citenamefont {Fl{\"a}schner}\ \emph {et~al.}(2016)\citenamefont
  {Fl{\"a}schner}, \citenamefont {Rem}, \citenamefont {Tarnowski},
  \citenamefont {Vogel}, \citenamefont {L{\"u}hmann}, \citenamefont
  {Sengstock},\ and\ \citenamefont {Weitenberg}}]{Flaschner_2016}%
  \BibitemOpen
  \bibfield  {author} {\bibinfo {author} {\bibfnamefont {N.}~\bibnamefont
  {Fl{\"a}schner}}, \bibinfo {author} {\bibfnamefont {B.~S.}\ \bibnamefont
  {Rem}}, \bibinfo {author} {\bibfnamefont {M.}~\bibnamefont {Tarnowski}},
  \bibinfo {author} {\bibfnamefont {D.}~\bibnamefont {Vogel}}, \bibinfo
  {author} {\bibfnamefont {D.-S.}\ \bibnamefont {L{\"u}hmann}}, \bibinfo
  {author} {\bibfnamefont {K.}~\bibnamefont {Sengstock}}, \ and\ \bibinfo
  {author} {\bibfnamefont {C.}~\bibnamefont {Weitenberg}},\ }\href {\doibase
  10.1126/science.aad4568} {\bibfield  {journal} {\bibinfo  {journal}
  {Science}\ }\textbf {\bibinfo {volume} {352}},\ \bibinfo {pages} {1091}
  (\bibinfo {year} {2016})}\BibitemShut {NoStop}%
\bibitem [{\citenamefont {Kitagawa}\ \emph {et~al.}(2012)\citenamefont
  {Kitagawa}, \citenamefont {Broome}, \citenamefont {Fedrizzi}, \citenamefont
  {Rudner}, \citenamefont {Berg}, \citenamefont {Kassal}, \citenamefont
  {Aspuru-Guzik}, \citenamefont {Demler},\ and\ \citenamefont
  {White}}]{Kitagawa_2012}%
  \BibitemOpen
  \bibfield  {author} {\bibinfo {author} {\bibfnamefont {T.}~\bibnamefont
  {Kitagawa}}, \bibinfo {author} {\bibfnamefont {M.~A.}\ \bibnamefont
  {Broome}}, \bibinfo {author} {\bibfnamefont {A.}~\bibnamefont {Fedrizzi}},
  \bibinfo {author} {\bibfnamefont {M.~S.}\ \bibnamefont {Rudner}}, \bibinfo
  {author} {\bibfnamefont {E.}~\bibnamefont {Berg}}, \bibinfo {author}
  {\bibfnamefont {I.}~\bibnamefont {Kassal}}, \bibinfo {author} {\bibfnamefont
  {A.}~\bibnamefont {Aspuru-Guzik}}, \bibinfo {author} {\bibfnamefont
  {E.}~\bibnamefont {Demler}}, \ and\ \bibinfo {author} {\bibfnamefont {A.~G.}\
  \bibnamefont {White}},\ }\href {\doibase 10.1038/ncomms1872} {\bibfield
  {journal} {\bibinfo  {journal} {Nat. Commun.}\ }\textbf {\bibinfo {volume}
  {3}},\ \bibinfo {pages} {882} (\bibinfo {year} {2012})}\BibitemShut {NoStop}%
\bibitem [{\citenamefont {Cardano}\ \emph
  {et~al.}(2016{\natexlab{a}})\citenamefont {Cardano}, \citenamefont {Maffei},
  \citenamefont {Massa}, \citenamefont {Piccirillo}, \citenamefont {de~Lisio},
  \citenamefont {De~Filippis}, \citenamefont {Cataudella}, \citenamefont
  {Santamato},\ and\ \citenamefont {Marrucci}}]{Cardano_2016}%
  \BibitemOpen
  \bibfield  {author} {\bibinfo {author} {\bibfnamefont {F.}~\bibnamefont
  {Cardano}}, \bibinfo {author} {\bibfnamefont {M.}~\bibnamefont {Maffei}},
  \bibinfo {author} {\bibfnamefont {F.}~\bibnamefont {Massa}}, \bibinfo
  {author} {\bibfnamefont {B.}~\bibnamefont {Piccirillo}}, \bibinfo {author}
  {\bibfnamefont {C.}~\bibnamefont {de~Lisio}}, \bibinfo {author}
  {\bibfnamefont {G.}~\bibnamefont {De~Filippis}}, \bibinfo {author}
  {\bibfnamefont {V.}~\bibnamefont {Cataudella}}, \bibinfo {author}
  {\bibfnamefont {E.}~\bibnamefont {Santamato}}, \ and\ \bibinfo {author}
  {\bibfnamefont {L.}~\bibnamefont {Marrucci}},\ }\href
  {http://dx.doi.org/10.1038/ncomms11439} {\bibfield  {journal} {\bibinfo
  {journal} {Nat. Commun.}\ }\textbf {\bibinfo {volume} {7}},\ \bibinfo {pages}
  {11439} (\bibinfo {year} {2016}{\natexlab{a}})}\BibitemShut {NoStop}%
\bibitem [{\citenamefont {Cardano}\ \emph
  {et~al.}(2016{\natexlab{b}})\citenamefont {Cardano}, \citenamefont
  {D'Errico}, \citenamefont {Dauphin}, \citenamefont {Maffei}, \citenamefont
  {Piccirillo}, \citenamefont {de~Lisio}, \citenamefont {Filippis},
  \citenamefont {Cataudella}, \citenamefont {Santamato}, \citenamefont
  {Marrucci}, \citenamefont {Lewenstein},\ and\ \citenamefont
  {Massignan}}]{Cardano2016a}%
  \BibitemOpen
  \bibfield  {author} {\bibinfo {author} {\bibfnamefont {F.}~\bibnamefont
  {Cardano}}, \bibinfo {author} {\bibfnamefont {A.}~\bibnamefont {D'Errico}},
  \bibinfo {author} {\bibfnamefont {A.}~\bibnamefont {Dauphin}}, \bibinfo
  {author} {\bibfnamefont {M.}~\bibnamefont {Maffei}}, \bibinfo {author}
  {\bibfnamefont {B.}~\bibnamefont {Piccirillo}}, \bibinfo {author}
  {\bibfnamefont {C.}~\bibnamefont {de~Lisio}}, \bibinfo {author}
  {\bibfnamefont {G.~D.}\ \bibnamefont {Filippis}}, \bibinfo {author}
  {\bibfnamefont {V.}~\bibnamefont {Cataudella}}, \bibinfo {author}
  {\bibfnamefont {E.}~\bibnamefont {Santamato}}, \bibinfo {author}
  {\bibfnamefont {L.}~\bibnamefont {Marrucci}}, \bibinfo {author}
  {\bibfnamefont {M.}~\bibnamefont {Lewenstein}}, \ and\ \bibinfo {author}
  {\bibfnamefont {P.}~\bibnamefont {Massignan}},\ }\href
  {https://arxiv.org/abs/1610.06322} {\bibfield  {journal} {\bibinfo  {journal}
  {arXiv:1610.06322}\ } (\bibinfo {year} {2016}{\natexlab{b}})}\BibitemShut
  {NoStop}%
\bibitem [{\citenamefont {Groh}\ \emph {et~al.}(2016)\citenamefont {Groh},
  \citenamefont {Brakhane}, \citenamefont {Alt}, \citenamefont {Meschede},
  \citenamefont {Asb\'oth},\ and\ \citenamefont {Alberti}}]{groh_2016}%
  \BibitemOpen
  \bibfield  {author} {\bibinfo {author} {\bibfnamefont {T.}~\bibnamefont
  {Groh}}, \bibinfo {author} {\bibfnamefont {S.}~\bibnamefont {Brakhane}},
  \bibinfo {author} {\bibfnamefont {W.}~\bibnamefont {Alt}}, \bibinfo {author}
  {\bibfnamefont {D.}~\bibnamefont {Meschede}}, \bibinfo {author}
  {\bibfnamefont {J.~K.}\ \bibnamefont {Asb\'oth}}, \ and\ \bibinfo {author}
  {\bibfnamefont {A.}~\bibnamefont {Alberti}},\ }\href
  {http://link.aps.org/doi/10.1103/PhysRevA.94.013620} {\bibfield  {journal}
  {\bibinfo  {journal} {Phys. Rev. A}\ }\textbf {\bibinfo {volume} {94}},\
  \bibinfo {pages} {013620} (\bibinfo {year} {2016})}\BibitemShut {NoStop}%
\bibitem [{\citenamefont {Rechtsman}\ \emph {et~al.}(2013)\citenamefont
  {Rechtsman}, \citenamefont {Zeuner}, \citenamefont {Plotnik}, \citenamefont
  {Lumer}, \citenamefont {Podolsky}, \citenamefont {Dreisow}, \citenamefont
  {Nolte}, \citenamefont {Segev},\ and\ \citenamefont
  {Szameit}}]{Rechtsman_2013}%
  \BibitemOpen
  \bibfield  {author} {\bibinfo {author} {\bibfnamefont {M.~C.}\ \bibnamefont
  {Rechtsman}}, \bibinfo {author} {\bibfnamefont {J.~M.}\ \bibnamefont
  {Zeuner}}, \bibinfo {author} {\bibfnamefont {Y.}~\bibnamefont {Plotnik}},
  \bibinfo {author} {\bibfnamefont {Y.}~\bibnamefont {Lumer}}, \bibinfo
  {author} {\bibfnamefont {D.}~\bibnamefont {Podolsky}}, \bibinfo {author}
  {\bibfnamefont {F.}~\bibnamefont {Dreisow}}, \bibinfo {author} {\bibfnamefont
  {S.}~\bibnamefont {Nolte}}, \bibinfo {author} {\bibfnamefont
  {M.}~\bibnamefont {Segev}}, \ and\ \bibinfo {author} {\bibfnamefont
  {A.}~\bibnamefont {Szameit}},\ }\href {http://dx.doi.org/10.1038/nature12066}
  {\bibfield  {journal} {\bibinfo  {journal} {Nature}\ }\textbf {\bibinfo
  {volume} {496}},\ \bibinfo {pages} {196} (\bibinfo {year}
  {2013})}\BibitemShut {NoStop}%
\bibitem [{\citenamefont {{Mukherjee}}\ \emph {et~al.}(2016)\citenamefont
  {{Mukherjee}}, \citenamefont {{Spracklen}}, \citenamefont {{Valiente}},
  \citenamefont {{Andersson}}, \citenamefont {{{\"O}hberg}}, \citenamefont
  {{Goldman}},\ and\ \citenamefont {{Thomson}}}]{mukherjee_2016}%
  \BibitemOpen
  \bibfield  {author} {\bibinfo {author} {\bibfnamefont {S.}~\bibnamefont
  {{Mukherjee}}}, \bibinfo {author} {\bibfnamefont {A.}~\bibnamefont
  {{Spracklen}}}, \bibinfo {author} {\bibfnamefont {M.}~\bibnamefont
  {{Valiente}}}, \bibinfo {author} {\bibfnamefont {E.}~\bibnamefont
  {{Andersson}}}, \bibinfo {author} {\bibfnamefont {P.}~\bibnamefont
  {{{\"O}hberg}}}, \bibinfo {author} {\bibfnamefont {N.}~\bibnamefont
  {{Goldman}}}, \ and\ \bibinfo {author} {\bibfnamefont {R.~R.}\ \bibnamefont
  {{Thomson}}},\ }\href {https://arxiv.org/abs/1604.05612} {\bibfield
  {journal} {\bibinfo  {journal} {arXiv:1604.05612}\ } (\bibinfo {year}
  {2016})}\BibitemShut {NoStop}%
\bibitem [{\citenamefont {Maczewsky}\ \emph {et~al.}(2016)\citenamefont
  {Maczewsky}, \citenamefont {Zeuner}, \citenamefont {Nolte},\ and\
  \citenamefont {Szameit}}]{Maczewsky:2016}%
  \BibitemOpen
  \bibfield  {author} {\bibinfo {author} {\bibfnamefont {L.~J.}\ \bibnamefont
  {Maczewsky}}, \bibinfo {author} {\bibfnamefont {J.~M.}\ \bibnamefont
  {Zeuner}}, \bibinfo {author} {\bibfnamefont {S.}~\bibnamefont {Nolte}}, \
  and\ \bibinfo {author} {\bibfnamefont {A.}~\bibnamefont {Szameit}},\ }\href
  {https://arxiv.org/abs/1605.03877} {\bibfield  {journal} {\bibinfo  {journal}
  {arXiv:1605.03877}\ } (\bibinfo {year} {2016})}\BibitemShut {NoStop}%
\bibitem [{\citenamefont {Salerno}\ \emph {et~al.}(2016)\citenamefont
  {Salerno}, \citenamefont {Ozawa}, \citenamefont {Price},\ and\ \citenamefont
  {Carusotto}}]{Salerno:2016A}%
  \BibitemOpen
  \bibfield  {author} {\bibinfo {author} {\bibfnamefont {G.}~\bibnamefont
  {Salerno}}, \bibinfo {author} {\bibfnamefont {T.}~\bibnamefont {Ozawa}},
  \bibinfo {author} {\bibfnamefont {H.~M.}\ \bibnamefont {Price}}, \ and\
  \bibinfo {author} {\bibfnamefont {I.}~\bibnamefont {Carusotto}},\ }\href
  {https://doi.org/10.1103/PhysRevB.93.085105} {\bibfield  {journal} {\bibinfo
  {journal} {Phys. Rev. B}\ }\textbf {\bibinfo {volume} {93}},\ \bibinfo
  {pages} {085105} (\bibinfo {year} {2016})}\BibitemShut {NoStop}%
\bibitem [{\citenamefont {Huber}(2016)}]{Huber:2016}%
  \BibitemOpen
  \bibfield  {author} {\bibinfo {author} {\bibfnamefont {S.~D.}\ \bibnamefont
  {Huber}},\ }\href {https://doi.org/10.1038/nphys3801} {\bibfield  {journal}
  {\bibinfo  {journal} {Nat. Phys.}\ }\textbf {\bibinfo {volume} {12}},\
  \bibinfo {pages} {621} (\bibinfo {year} {2016})}\BibitemShut {NoStop}%
\bibitem [{\citenamefont {Eckardt}\ \emph {et~al.}(2009)\citenamefont
  {Eckardt}, \citenamefont {Holthaus}, \citenamefont {Lignier}, \citenamefont
  {Zenesini}, \citenamefont {Ciampini}, \citenamefont {Morsch},\ and\
  \citenamefont {Arimondo}}]{eckardt2009}%
  \BibitemOpen
  \bibfield  {author} {\bibinfo {author} {\bibfnamefont {A.}~\bibnamefont
  {Eckardt}}, \bibinfo {author} {\bibfnamefont {M.}~\bibnamefont {Holthaus}},
  \bibinfo {author} {\bibfnamefont {H.}~\bibnamefont {Lignier}}, \bibinfo
  {author} {\bibfnamefont {A.}~\bibnamefont {Zenesini}}, \bibinfo {author}
  {\bibfnamefont {D.}~\bibnamefont {Ciampini}}, \bibinfo {author}
  {\bibfnamefont {O.}~\bibnamefont {Morsch}}, \ and\ \bibinfo {author}
  {\bibfnamefont {E.}~\bibnamefont {Arimondo}},\ }\href
  {https://doi.org/10.1103/physreva.79.013611} {\bibfield  {journal} {\bibinfo
  {journal} {Phys. Rev. A}\ }\textbf {\bibinfo {volume} {79}},\ \bibinfo
  {pages} {013611} (\bibinfo {year} {2009})}\BibitemShut {NoStop}%
\bibitem [{\citenamefont {Baur}\ and\ \citenamefont
  {Cooper}(2013)}]{Baur:2013}%
  \BibitemOpen
  \bibfield  {author} {\bibinfo {author} {\bibfnamefont {S.~K.}\ \bibnamefont
  {Baur}}\ and\ \bibinfo {author} {\bibfnamefont {N.~R.}\ \bibnamefont
  {Cooper}},\ }\href {https://doi.org/10.1103/physreva.88.033603} {\bibfield
  {journal} {\bibinfo  {journal} {Phys. Rev. A}\ }\textbf {\bibinfo {volume}
  {88}},\ \bibinfo {pages} {033603} (\bibinfo {year} {2013})}\BibitemShut
  {NoStop}%
\bibitem [{\citenamefont {Seetharam}\ \emph {et~al.}(2015)\citenamefont
  {Seetharam}, \citenamefont {Bardyn}, \citenamefont {Lindner}, \citenamefont
  {Rudner},\ and\ \citenamefont {Refael}}]{Seetharam:2015}%
  \BibitemOpen
  \bibfield  {author} {\bibinfo {author} {\bibfnamefont {K.~I.}\ \bibnamefont
  {Seetharam}}, \bibinfo {author} {\bibfnamefont {C.-E.}\ \bibnamefont
  {Bardyn}}, \bibinfo {author} {\bibfnamefont {N.~H.}\ \bibnamefont {Lindner}},
  \bibinfo {author} {\bibfnamefont {M.~S.}\ \bibnamefont {Rudner}}, \ and\
  \bibinfo {author} {\bibfnamefont {G.}~\bibnamefont {Refael}},\ }\href
  {https://doi.org/10.1103/PhysRevX.5.041050} {\bibfield  {journal} {\bibinfo
  {journal} {Phys. Rev. X}\ }\textbf {\bibinfo {volume} {5}},\ \bibinfo {pages}
  {041050} (\bibinfo {year} {2015})}\BibitemShut {NoStop}%
\bibitem [{\citenamefont {{Ho}}\ and\ \citenamefont
  {{Abanin}}(2016)}]{Ho_2016}%
  \BibitemOpen
  \bibfield  {author} {\bibinfo {author} {\bibfnamefont {W.~W.}\ \bibnamefont
  {{Ho}}}\ and\ \bibinfo {author} {\bibfnamefont {D.~A.}\ \bibnamefont
  {{Abanin}}},\ }\href {https://arxiv.org/abs/1611.05024} {\bibfield  {journal}
  {\bibinfo  {journal} {arXiv:1611.05024}\ } (\bibinfo {year}
  {2016})}\BibitemShut {NoStop}%
\bibitem [{\citenamefont {{Novi{\v c}enko}}\ \emph {et~al.}(2016)\citenamefont
  {{Novi{\v c}enko}}, \citenamefont {{Anisimovas}},\ and\ \citenamefont
  {{Juzeli{\=u}nas}}}]{novicenko_2016}%
  \BibitemOpen
  \bibfield  {author} {\bibinfo {author} {\bibfnamefont {V.}~\bibnamefont
  {{Novi{\v c}enko}}}, \bibinfo {author} {\bibfnamefont {E.}~\bibnamefont
  {{Anisimovas}}}, \ and\ \bibinfo {author} {\bibfnamefont {G.}~\bibnamefont
  {{Juzeli{\=u}nas}}},\ }\href {https://arxiv.org/abs/1608.08420} {\bibfield
  {journal} {\bibinfo  {journal} {arXiv:1608.08420}\ } (\bibinfo {year}
  {2016})}\BibitemShut {NoStop}%
\bibitem [{\citenamefont {Hasan}\ and\ \citenamefont
  {Kane}(2010)}]{Hasan_2010}%
  \BibitemOpen
  \bibfield  {author} {\bibinfo {author} {\bibfnamefont {M.~Z.}\ \bibnamefont
  {Hasan}}\ and\ \bibinfo {author} {\bibfnamefont {C.~L.}\ \bibnamefont
  {Kane}},\ }\href {\doibase 10.1103/revmodphys.82.3045} {\bibfield  {journal}
  {\bibinfo  {journal} {Rev. Mod. Phys.}\ }\textbf {\bibinfo {volume} {82}},\
  \bibinfo {pages} {3045} (\bibinfo {year} {2010})}\BibitemShut {NoStop}%
\bibitem [{\citenamefont {Lacki}\ and\ \citenamefont
  {Zakrzewski}(2013)}]{lacki_2012}%
  \BibitemOpen
  \bibfield  {author} {\bibinfo {author} {\bibfnamefont {M.}~\bibnamefont
  {Lacki}}\ and\ \bibinfo {author} {\bibfnamefont {J.}~\bibnamefont
  {Zakrzewski}},\ }\href
  {http://link.aps.org/doi/10.1103/PhysRevLett.110.065301} {\bibfield
  {journal} {\bibinfo  {journal} {Phys. Rev. Lett.}\ }\textbf {\bibinfo
  {volume} {110}},\ \bibinfo {pages} {065301} (\bibinfo {year}
  {2013})}\BibitemShut {NoStop}%
\bibitem [{\citenamefont {Lacki}\ \emph {et~al.}(2013)\citenamefont {Lacki},
  \citenamefont {Delande},\ and\ \citenamefont {Zakrzewski}}]{lacki_2013}%
  \BibitemOpen
  \bibfield  {author} {\bibinfo {author} {\bibfnamefont {M.}~\bibnamefont
  {Lacki}}, \bibinfo {author} {\bibfnamefont {D.}~\bibnamefont {Delande}}, \
  and\ \bibinfo {author} {\bibfnamefont {J.}~\bibnamefont {Zakrzewski}},\
  }\href {http://dx.doi.org/10.1088/1367-2630/15/1/013062} {\bibfield
  {journal} {\bibinfo  {journal} {New J. Phys.}\ }\textbf {\bibinfo {volume}
  {15}},\ \bibinfo {pages} {013062} (\bibinfo {year} {2013})}\BibitemShut
  {NoStop}%
\bibitem [{\citenamefont {Lazarides}\ \emph {et~al.}(2014)\citenamefont
  {Lazarides}, \citenamefont {Das},\ and\ \citenamefont
  {Moessner}}]{lazarides_2014}%
  \BibitemOpen
  \bibfield  {author} {\bibinfo {author} {\bibfnamefont {A.}~\bibnamefont
  {Lazarides}}, \bibinfo {author} {\bibfnamefont {A.}~\bibnamefont {Das}}, \
  and\ \bibinfo {author} {\bibfnamefont {R.}~\bibnamefont {Moessner}},\ }\href
  {http://link.aps.org/doi/10.1103/PhysRevE.90.012110} {\bibfield  {journal}
  {\bibinfo  {journal} {Phys. Rev. E}\ }\textbf {\bibinfo {volume} {90}},\
  \bibinfo {pages} {012110} (\bibinfo {year} {2014})}\BibitemShut {NoStop}%
\bibitem [{\citenamefont {D'Alessio}\ and\ \citenamefont
  {Rigol}(2014)}]{dalessio_2014}%
  \BibitemOpen
  \bibfield  {author} {\bibinfo {author} {\bibfnamefont {L.}~\bibnamefont
  {D'Alessio}}\ and\ \bibinfo {author} {\bibfnamefont {M.}~\bibnamefont
  {Rigol}},\ }\href {http://link.aps.org/doi/10.1103/PhysRevX.4.041048}
  {\bibfield  {journal} {\bibinfo  {journal} {Phys. Rev. X}\ }\textbf {\bibinfo
  {volume} {4}},\ \bibinfo {pages} {041048} (\bibinfo {year}
  {2014})}\BibitemShut {NoStop}%
\bibitem [{\citenamefont {Bukov}\ \emph
  {et~al.}(2015{\natexlab{b}})\citenamefont {Bukov}, \citenamefont
  {Gopalakrishnan}, \citenamefont {Knap},\ and\ \citenamefont
  {Demler}}]{Bukov:2015}%
  \BibitemOpen
  \bibfield  {author} {\bibinfo {author} {\bibfnamefont {M.}~\bibnamefont
  {Bukov}}, \bibinfo {author} {\bibfnamefont {S.}~\bibnamefont
  {Gopalakrishnan}}, \bibinfo {author} {\bibfnamefont {M.}~\bibnamefont
  {Knap}}, \ and\ \bibinfo {author} {\bibfnamefont {E.}~\bibnamefont
  {Demler}},\ }\href {https://doi.org/10.1103/PhysRevLett.115.205301}
  {\bibfield  {journal} {\bibinfo  {journal} {Phys. Rev. Lett.}\ }\textbf
  {\bibinfo {volume} {115}},\ \bibinfo {pages} {205301} (\bibinfo {year}
  {2015}{\natexlab{b}})}\BibitemShut {NoStop}%
\bibitem [{\citenamefont {Bilitewski}\ and\ \citenamefont
  {Cooper}(2015)}]{bilitewski_2015}%
  \BibitemOpen
  \bibfield  {author} {\bibinfo {author} {\bibfnamefont {T.}~\bibnamefont
  {Bilitewski}}\ and\ \bibinfo {author} {\bibfnamefont {N.~R.}\ \bibnamefont
  {Cooper}},\ }\href {http://link.aps.org/doi/10.1103/PhysRevA.91.033601}
  {\bibfield  {journal} {\bibinfo  {journal} {Phys. Rev. A}\ }\textbf {\bibinfo
  {volume} {91}},\ \bibinfo {pages} {033601} (\bibinfo {year}
  {2015})}\BibitemShut {NoStop}%
\bibitem [{\citenamefont {Choudhury}\ and\ \citenamefont
  {Mueller}(2015)}]{Choudhury:2015}%
  \BibitemOpen
  \bibfield  {author} {\bibinfo {author} {\bibfnamefont {S.}~\bibnamefont
  {Choudhury}}\ and\ \bibinfo {author} {\bibfnamefont {E.~J.}\ \bibnamefont
  {Mueller}},\ }\href {https://doi.org/10.1103/PhysRevA.91.023624} {\bibfield
  {journal} {\bibinfo  {journal} {Phys. Rev. A}\ }\textbf {\bibinfo {volume}
  {91}},\ \bibinfo {pages} {023624} (\bibinfo {year} {2015})}\BibitemShut
  {NoStop}%
\bibitem [{\citenamefont {Str{\"a}ter}\ and\ \citenamefont
  {Eckardt}(2016)}]{strater_2016}%
  \BibitemOpen
  \bibfield  {author} {\bibinfo {author} {\bibfnamefont {C.}~\bibnamefont
  {Str{\"a}ter}}\ and\ \bibinfo {author} {\bibfnamefont {A.}~\bibnamefont
  {Eckardt}},\ }\href {http://dx.doi.org/10.1515/zna-2016-0129} {\bibfield
  {journal} {\bibinfo  {journal} {Z. Naturforsch. A}\ }\textbf {\bibinfo
  {volume} {71}} (\bibinfo {year} {2016})}\BibitemShut {NoStop}%
\bibitem [{\citenamefont {Bukov}\ \emph {et~al.}(2016)\citenamefont {Bukov},
  \citenamefont {Heyl}, \citenamefont {Huse},\ and\ \citenamefont
  {Polkovnikov}}]{Bukov:2016}%
  \BibitemOpen
  \bibfield  {author} {\bibinfo {author} {\bibfnamefont {M.}~\bibnamefont
  {Bukov}}, \bibinfo {author} {\bibfnamefont {M.}~\bibnamefont {Heyl}},
  \bibinfo {author} {\bibfnamefont {D.~A.}\ \bibnamefont {Huse}}, \ and\
  \bibinfo {author} {\bibfnamefont {A.}~\bibnamefont {Polkovnikov}},\ }\href
  {https://doi.org/10.1103/PhysRevB.93.155132} {\bibfield  {journal} {\bibinfo
  {journal} {Phys. Rev. B}\ }\textbf {\bibinfo {volume} {93}},\ \bibinfo
  {pages} {155132} (\bibinfo {year} {2016})}\BibitemShut {NoStop}%
\bibitem [{\citenamefont {Celi}\ \emph {et~al.}(2016)\citenamefont {Celi},
  \citenamefont {Grass}, \citenamefont {Ferris}, \citenamefont {Padhi},
  \citenamefont {Ravent\'os}, \citenamefont {Simonet}, \citenamefont
  {Sengstock},\ and\ \citenamefont {Lewenstein}}]{celi_2016}%
  \BibitemOpen
  \bibfield  {author} {\bibinfo {author} {\bibfnamefont {A.}~\bibnamefont
  {Celi}}, \bibinfo {author} {\bibfnamefont {T.}~\bibnamefont {Grass}},
  \bibinfo {author} {\bibfnamefont {A.~J.}\ \bibnamefont {Ferris}}, \bibinfo
  {author} {\bibfnamefont {B.}~\bibnamefont {Padhi}}, \bibinfo {author}
  {\bibfnamefont {D.}~\bibnamefont {Ravent\'os}}, \bibinfo {author}
  {\bibfnamefont {J.}~\bibnamefont {Simonet}}, \bibinfo {author} {\bibfnamefont
  {K.}~\bibnamefont {Sengstock}}, \ and\ \bibinfo {author} {\bibfnamefont
  {M.}~\bibnamefont {Lewenstein}},\ }\href
  {http://link.aps.org/doi/10.1103/PhysRevB.94.075110} {\bibfield  {journal}
  {\bibinfo  {journal} {Phys. Rev. B}\ }\textbf {\bibinfo {volume} {94}},\
  \bibinfo {pages} {075110} (\bibinfo {year} {2016})}\BibitemShut {NoStop}%
\bibitem [{\citenamefont {Lellouch}\ \emph {et~al.}(2016)\citenamefont
  {Lellouch}, \citenamefont {Bukov}, \citenamefont {Demler},\ and\
  \citenamefont {Goldman}}]{lellouch_2016}%
  \BibitemOpen
  \bibfield  {author} {\bibinfo {author} {\bibfnamefont {S.}~\bibnamefont
  {Lellouch}}, \bibinfo {author} {\bibfnamefont {M.}~\bibnamefont {Bukov}},
  \bibinfo {author} {\bibfnamefont {E.}~\bibnamefont {Demler}}, \ and\ \bibinfo
  {author} {\bibfnamefont {N.}~\bibnamefont {Goldman}},\ }\href
  {https://arxiv.org/abs/1610.02972} {\bibfield  {journal} {\bibinfo  {journal}
  {arXiv:1610.02972}\ } (\bibinfo {year} {2016})}\BibitemShut {NoStop}%
\bibitem [{\citenamefont {Caio}\ \emph {et~al.}(2015)\citenamefont {Caio},
  \citenamefont {Cooper},\ and\ \citenamefont {Bhaseen}}]{caio_2015}%
  \BibitemOpen
  \bibfield  {author} {\bibinfo {author} {\bibfnamefont {M.~D.}\ \bibnamefont
  {Caio}}, \bibinfo {author} {\bibfnamefont {N.~R.}\ \bibnamefont {Cooper}}, \
  and\ \bibinfo {author} {\bibfnamefont {M.~J.}\ \bibnamefont {Bhaseen}},\
  }\href {\doibase 10.1103/PhysRevLett.115.236403} {\bibfield  {journal}
  {\bibinfo  {journal} {Phys. Rev. Lett.}\ }\textbf {\bibinfo {volume} {115}},\
  \bibinfo {pages} {236403} (\bibinfo {year} {2015})}\BibitemShut {NoStop}%
\bibitem [{\citenamefont {D'Alessio}\ and\ \citenamefont
  {Rigol}(2015)}]{dalessio_2015}%
  \BibitemOpen
  \bibfield  {author} {\bibinfo {author} {\bibfnamefont {L.}~\bibnamefont
  {D'Alessio}}\ and\ \bibinfo {author} {\bibfnamefont {M.}~\bibnamefont
  {Rigol}},\ }\href {http://dx.doi.org/10.1038/ncomms9336} {\bibfield
  {journal} {\bibinfo  {journal} {Nat. Commun.}\ }\textbf {\bibinfo {volume}
  {6}},\ \bibinfo {pages} {8336} (\bibinfo {year} {2015})}\BibitemShut
  {NoStop}%
\bibitem [{\citenamefont {{Caio}}\ \emph {et~al.}(2016)\citenamefont {{Caio}},
  \citenamefont {{Cooper}},\ and\ \citenamefont {{Bhaseen}}}]{caio_2016}%
  \BibitemOpen
  \bibfield  {author} {\bibinfo {author} {\bibfnamefont {M.~D.}\ \bibnamefont
  {{Caio}}}, \bibinfo {author} {\bibfnamefont {N.~R.}\ \bibnamefont
  {{Cooper}}}, \ and\ \bibinfo {author} {\bibfnamefont {M.~J.}\ \bibnamefont
  {{Bhaseen}}},\ }\href {https://doi.org/10.1103/PhysRevB.94.155104} {\bibfield
   {journal} {\bibinfo  {journal} {Phys. Rev. B}\ }\textbf {\bibinfo {volume}
  {94}},\ \bibinfo {pages} {155104} (\bibinfo {year} {2016})}\BibitemShut
  {NoStop}%
\bibitem [{\citenamefont {{Hu}}\ \emph {et~al.}(2016)\citenamefont {{Hu}},
  \citenamefont {{Zoller}},\ and\ \citenamefont {{Budich}}}]{hu_2016}%
  \BibitemOpen
  \bibfield  {author} {\bibinfo {author} {\bibfnamefont {Y.}~\bibnamefont
  {{Hu}}}, \bibinfo {author} {\bibfnamefont {P.}~\bibnamefont {{Zoller}}}, \
  and\ \bibinfo {author} {\bibfnamefont {J.~C.}\ \bibnamefont {{Budich}}},\
  }\href {https://doi.org/10.1103/PhysRevLett.117.126803} {\bibfield  {journal}
  {\bibinfo  {journal} {Phys. Rev. Lett.}\ }\textbf {\bibinfo {volume} {117}},\
  \bibinfo {pages} {126803} (\bibinfo {year} {2016})}\BibitemShut {NoStop}%
\bibitem [{\citenamefont {{Wilson}}\ \emph {et~al.}(2016)\citenamefont
  {{Wilson}}, \citenamefont {{Song}},\ and\ \citenamefont
  {{Refael}}}]{wilson_2016}%
  \BibitemOpen
  \bibfield  {author} {\bibinfo {author} {\bibfnamefont {J.~H.}\ \bibnamefont
  {{Wilson}}}, \bibinfo {author} {\bibfnamefont {J.~C.~W.}\ \bibnamefont
  {{Song}}}, \ and\ \bibinfo {author} {\bibfnamefont {G.}~\bibnamefont
  {{Refael}}},\ }\href {https://doi.org/10.1103/PhysRevLett.117.235302}
  {\bibfield  {journal} {\bibinfo  {journal} {Phys. Rev. Lett.}\ }\textbf
  {\bibinfo {volume} {117}},\ \bibinfo {pages} {235302} (\bibinfo {year}
  {2016})}\BibitemShut {NoStop}%
\bibitem [{\citenamefont {Dehghani}\ and\ \citenamefont
  {Mitra}(2016)}]{dehghani_2016}%
  \BibitemOpen
  \bibfield  {author} {\bibinfo {author} {\bibfnamefont {H.}~\bibnamefont
  {Dehghani}}\ and\ \bibinfo {author} {\bibfnamefont {A.}~\bibnamefont
  {Mitra}},\ }\href {\doibase 10.1103/PhysRevB.93.205437} {\bibfield  {journal}
  {\bibinfo  {journal} {Phys. Rev. B}\ }\textbf {\bibinfo {volume} {93}},\
  \bibinfo {pages} {205437} (\bibinfo {year} {2016})}\BibitemShut {NoStop}%
\bibitem [{\citenamefont {{Nur {\"U}nal}}\ \emph {et~al.}(2016)\citenamefont
  {{Nur {\"U}nal}}, \citenamefont {{Mueller}},\ and\ \citenamefont
  {{Oktel}}}]{nurunal_2016}%
  \BibitemOpen
  \bibfield  {author} {\bibinfo {author} {\bibfnamefont {F.}~\bibnamefont {{Nur
  {\"U}nal}}}, \bibinfo {author} {\bibfnamefont {E.~J.}\ \bibnamefont
  {{Mueller}}}, \ and\ \bibinfo {author} {\bibfnamefont {M.~{\"O}.}\
  \bibnamefont {{Oktel}}},\ }\href {\doibase
  https://doi.org/10.1103/PhysRevA.94.053604} {\bibfield  {journal} {\bibinfo
  {journal} {Phys. Rev. A}\ }\textbf {\bibinfo {volume} {94}},\ \bibinfo
  {pages} {053604} (\bibinfo {year} {2016})}\BibitemShut {NoStop}%
\bibitem [{\citenamefont {{Wang}}\ \emph {et~al.}(2016)\citenamefont {{Wang}},
  \citenamefont {{Zhang}}, \citenamefont {{Chen}}, \citenamefont {{Yu}},\ and\
  \citenamefont {{Zhai}}}]{wangc_2016}%
  \BibitemOpen
  \bibfield  {author} {\bibinfo {author} {\bibfnamefont {C.}~\bibnamefont
  {{Wang}}}, \bibinfo {author} {\bibfnamefont {P.}~\bibnamefont {{Zhang}}},
  \bibinfo {author} {\bibfnamefont {X.}~\bibnamefont {{Chen}}}, \bibinfo
  {author} {\bibfnamefont {J.}~\bibnamefont {{Yu}}}, \ and\ \bibinfo {author}
  {\bibfnamefont {H.}~\bibnamefont {{Zhai}}},\ }\href
  {https://arxiv.org/abs/1611.03304} {\bibfield  {journal} {\bibinfo  {journal}
  {arXiv:1611.03304}\ } (\bibinfo {year} {2016})}\BibitemShut {NoStop}%
\bibitem [{\citenamefont {Goldman}\ \emph {et~al.}(2015)\citenamefont
  {Goldman}, \citenamefont {Dalibard}, \citenamefont {Aidelsburger},\ and\
  \citenamefont {Cooper}}]{goldman_2015}%
  \BibitemOpen
  \bibfield  {author} {\bibinfo {author} {\bibfnamefont {N.}~\bibnamefont
  {Goldman}}, \bibinfo {author} {\bibfnamefont {J.}~\bibnamefont {Dalibard}},
  \bibinfo {author} {\bibfnamefont {M.}~\bibnamefont {Aidelsburger}}, \ and\
  \bibinfo {author} {\bibfnamefont {N.~R.}\ \bibnamefont {Cooper}},\ }\href
  {\doibase 10.1103/PhysRevA.91.033632} {\bibfield  {journal} {\bibinfo
  {journal} {Phys. Rev. A}\ }\textbf {\bibinfo {volume} {91}},\ \bibinfo
  {pages} {033632} (\bibinfo {year} {2015})}\BibitemShut {NoStop}%
\bibitem [{\citenamefont {Dauphin}\ and\ \citenamefont
  {Goldman}(2013)}]{dauphin_2013}%
  \BibitemOpen
  \bibfield  {author} {\bibinfo {author} {\bibfnamefont {A.}~\bibnamefont
  {Dauphin}}\ and\ \bibinfo {author} {\bibfnamefont {N.}~\bibnamefont
  {Goldman}},\ }\href {\doibase 10.1103/PhysRevLett.111.135302} {\bibfield
  {journal} {\bibinfo  {journal} {Phys. Rev. Lett.}\ }\textbf {\bibinfo
  {volume} {111}},\ \bibinfo {pages} {135302} (\bibinfo {year}
  {2013})}\BibitemShut {NoStop}%
\bibitem [{\citenamefont {Rahav}\ \emph {et~al.}(2003)\citenamefont {Rahav},
  \citenamefont {Gilary},\ and\ \citenamefont {Fishman}}]{rahav_03_pra}%
  \BibitemOpen
  \bibfield  {author} {\bibinfo {author} {\bibfnamefont {S.}~\bibnamefont
  {Rahav}}, \bibinfo {author} {\bibfnamefont {I.}~\bibnamefont {Gilary}}, \
  and\ \bibinfo {author} {\bibfnamefont {S.}~\bibnamefont {Fishman}},\ }\href
  {https://doi.org/10.1103/PhysRevA.68.013820} {\bibfield  {journal} {\bibinfo
  {journal} {Phys. Rev. A}\ }\textbf {\bibinfo {volume} {68}},\ \bibinfo
  {pages} {013820} (\bibinfo {year} {2003})}\BibitemShut {NoStop}%
\bibitem [{\citenamefont {Eckardt}\ and\ \citenamefont
  {Anisimovas}(2015)}]{eckardt_15}%
  \BibitemOpen
  \bibfield  {author} {\bibinfo {author} {\bibfnamefont {A.}~\bibnamefont
  {Eckardt}}\ and\ \bibinfo {author} {\bibfnamefont {E.}~\bibnamefont
  {Anisimovas}},\ }\href {https://doi.org/10.1088/1367-2630/17/9/093039}
  {\bibfield  {journal} {\bibinfo  {journal} {New J. Phys.}\ }\textbf {\bibinfo
  {volume} {17}},\ \bibinfo {pages} {093039} (\bibinfo {year}
  {2015})}\BibitemShut {NoStop}%
\bibitem [{\citenamefont {Mikami}\ \emph {et~al.}(2016)\citenamefont {Mikami},
  \citenamefont {Kitamura}, \citenamefont {Yasuda}, \citenamefont {Tsuji},
  \citenamefont {Oka},\ and\ \citenamefont {Aoki}}]{mikami_15}%
  \BibitemOpen
  \bibfield  {author} {\bibinfo {author} {\bibfnamefont {T.}~\bibnamefont
  {Mikami}}, \bibinfo {author} {\bibfnamefont {S.}~\bibnamefont {Kitamura}},
  \bibinfo {author} {\bibfnamefont {K.}~\bibnamefont {Yasuda}}, \bibinfo
  {author} {\bibfnamefont {N.}~\bibnamefont {Tsuji}}, \bibinfo {author}
  {\bibfnamefont {T.}~\bibnamefont {Oka}}, \ and\ \bibinfo {author}
  {\bibfnamefont {H.}~\bibnamefont {Aoki}},\ }\href
  {https://doi.org/10.1103/PhysRevB.93.144307} {\bibfield  {journal} {\bibinfo
  {journal} {Phys. Rev. B}\ }\textbf {\bibinfo {volume} {93}},\ \bibinfo
  {pages} {144307} (\bibinfo {year} {2016})}\BibitemShut {NoStop}%
\bibitem [{\citenamefont {Xiao}\ \emph {et~al.}(2010)\citenamefont {Xiao},
  \citenamefont {Chang},\ and\ \citenamefont {Niu}}]{xiao_2010}%
  \BibitemOpen
  \bibfield  {author} {\bibinfo {author} {\bibfnamefont {D.}~\bibnamefont
  {Xiao}}, \bibinfo {author} {\bibfnamefont {M.-C.}\ \bibnamefont {Chang}}, \
  and\ \bibinfo {author} {\bibfnamefont {Q.}~\bibnamefont {Niu}},\ }\href
  {\doibase 10.1103/RevModPhys.82.1959} {\bibfield  {journal} {\bibinfo
  {journal} {Rev. Mod. Phys.}\ }\textbf {\bibinfo {volume} {82}},\ \bibinfo
  {pages} {1959} (\bibinfo {year} {2010})}\BibitemShut {NoStop}%
\bibitem [{\citenamefont {{Price}}\ \emph {et~al.}(2016)\citenamefont
  {{Price}}, \citenamefont {{Zilberberg}}, \citenamefont {{Ozawa}},
  \citenamefont {{Carusotto}},\ and\ \citenamefont {{Goldman}}}]{price_2016}%
  \BibitemOpen
  \bibfield  {author} {\bibinfo {author} {\bibfnamefont {H.~M.}\ \bibnamefont
  {{Price}}}, \bibinfo {author} {\bibfnamefont {O.}~\bibnamefont
  {{Zilberberg}}}, \bibinfo {author} {\bibfnamefont {T.}~\bibnamefont
  {{Ozawa}}}, \bibinfo {author} {\bibfnamefont {I.}~\bibnamefont
  {{Carusotto}}}, \ and\ \bibinfo {author} {\bibfnamefont {N.}~\bibnamefont
  {{Goldman}}},\ }\href {https://doi.org/10.1103/PhysRevB.93.245113} {\bibfield
   {journal} {\bibinfo  {journal} {Phys. Rev. B}\ }\textbf {\bibinfo {volume}
  {93}},\ \bibinfo {pages} {245113} (\bibinfo {year} {2016})}\BibitemShut
  {NoStop}%
\bibitem [{\citenamefont {Price}\ and\ \citenamefont
  {Cooper}(2012)}]{Price:2012}%
  \BibitemOpen
  \bibfield  {author} {\bibinfo {author} {\bibfnamefont {H.~M.}\ \bibnamefont
  {Price}}\ and\ \bibinfo {author} {\bibfnamefont {N.~R.}\ \bibnamefont
  {Cooper}},\ }\href {https://doi.org/10.1103/physreva.85.033620} {\bibfield
  {journal} {\bibinfo  {journal} {Phys. Rev. A}\ }\textbf {\bibinfo {volume}
  {85}},\ \bibinfo {pages} {033620} (\bibinfo {year} {2012})}\BibitemShut
  {NoStop}%
\bibitem [{\citenamefont {Alba}\ \emph {et~al.}(2011)\citenamefont {Alba},
  \citenamefont {Fernandez-Gonzalvo}, \citenamefont {Mur-Petit}, \citenamefont
  {Pachos},\ and\ \citenamefont {Garcia-Ripoll}}]{Alba_2011}%
  \BibitemOpen
  \bibfield  {author} {\bibinfo {author} {\bibfnamefont {E.}~\bibnamefont
  {Alba}}, \bibinfo {author} {\bibfnamefont {X.}~\bibnamefont
  {Fernandez-Gonzalvo}}, \bibinfo {author} {\bibfnamefont {J.}~\bibnamefont
  {Mur-Petit}}, \bibinfo {author} {\bibfnamefont {J.~K.}\ \bibnamefont
  {Pachos}}, \ and\ \bibinfo {author} {\bibfnamefont {J.~J.}\ \bibnamefont
  {Garcia-Ripoll}},\ }\href {\doibase 10.1103/physrevlett.107.235301}
  {\bibfield  {journal} {\bibinfo  {journal} {Phys. Rev. Lett.}\ }\textbf
  {\bibinfo {volume} {107}},\ \bibinfo {pages} {235301} (\bibinfo {year}
  {2011})}\BibitemShut {NoStop}%
\bibitem [{\citenamefont {Goldman}\ \emph {et~al.}(2013)\citenamefont
  {Goldman}, \citenamefont {Anisimovas}, \citenamefont {Gerbier}, \citenamefont
  {{\"O}hberg}, \citenamefont {Spielman},\ and\ \citenamefont
  {Juzeli{\=u}nas}}]{goldman_2013}%
  \BibitemOpen
  \bibfield  {author} {\bibinfo {author} {\bibfnamefont {N.}~\bibnamefont
  {Goldman}}, \bibinfo {author} {\bibfnamefont {E.}~\bibnamefont {Anisimovas}},
  \bibinfo {author} {\bibfnamefont {F.}~\bibnamefont {Gerbier}}, \bibinfo
  {author} {\bibfnamefont {P.}~\bibnamefont {{\"O}hberg}}, \bibinfo {author}
  {\bibfnamefont {I.~B.}\ \bibnamefont {Spielman}}, \ and\ \bibinfo {author}
  {\bibfnamefont {G.}~\bibnamefont {Juzeli{\=u}nas}},\ }\href
  {http://stacks.iop.org/1367-2630/15/i=1/a=013025} {\bibfield  {journal}
  {\bibinfo  {journal} {New J. Phys.}\ }\textbf {\bibinfo {volume} {15}},\
  \bibinfo {pages} {013025} (\bibinfo {year} {2013})}\BibitemShut {NoStop}%
\bibitem [{\citenamefont {Anisimovas}\ \emph {et~al.}(2014)\citenamefont
  {Anisimovas}, \citenamefont {Gerbier}, \citenamefont {Andrijauskas},\ and\
  \citenamefont {Goldman}}]{Anisimovas:2014}%
  \BibitemOpen
  \bibfield  {author} {\bibinfo {author} {\bibfnamefont {E.}~\bibnamefont
  {Anisimovas}}, \bibinfo {author} {\bibfnamefont {F.}~\bibnamefont {Gerbier}},
  \bibinfo {author} {\bibfnamefont {T.}~\bibnamefont {Andrijauskas}}, \ and\
  \bibinfo {author} {\bibfnamefont {N.}~\bibnamefont {Goldman}},\ }\href
  {https://doi.org/10.1103/PhysRevA.89.013632} {\bibfield  {journal} {\bibinfo
  {journal} {Phys. Rev. A}\ }\textbf {\bibinfo {volume} {89}},\ \bibinfo
  {pages} {013632} (\bibinfo {year} {2014})}\BibitemShut {NoStop}%
\bibitem [{\citenamefont {{Dziarmaga}}(2010)}]{Dziarmaga_2010}%
  \BibitemOpen
  \bibfield  {author} {\bibinfo {author} {\bibfnamefont {J.}~\bibnamefont
  {{Dziarmaga}}},\ }\href {https://doi.org/10.1080/00018732.2010.514702}
  {\bibfield  {journal} {\bibinfo  {journal} {Adv. Phys.}\ }\textbf {\bibinfo
  {volume} {59}},\ \bibinfo {pages} {1063} (\bibinfo {year}
  {2010})}\BibitemShut {NoStop}%
\bibitem [{\citenamefont {Thouless}\ \emph {et~al.}(1982)\citenamefont
  {Thouless}, \citenamefont {Kohmoto}, \citenamefont {Nightingale},\ and\
  \citenamefont {den Nijs}}]{thouless_1982}%
  \BibitemOpen
  \bibfield  {author} {\bibinfo {author} {\bibfnamefont {D.~J.}\ \bibnamefont
  {Thouless}}, \bibinfo {author} {\bibfnamefont {M.}~\bibnamefont {Kohmoto}},
  \bibinfo {author} {\bibfnamefont {M.~P.}\ \bibnamefont {Nightingale}}, \ and\
  \bibinfo {author} {\bibfnamefont {M.}~\bibnamefont {den Nijs}},\ }\href
  {\doibase 10.1103/PhysRevLett.49.405} {\bibfield  {journal} {\bibinfo
  {journal} {Phys. Rev. Lett.}\ }\textbf {\bibinfo {volume} {49}},\ \bibinfo
  {pages} {405} (\bibinfo {year} {1982})}\BibitemShut {NoStop}%
\bibitem [{\citenamefont {Aidelsburger}(2016)}]{Monika_thesis}%
  \BibitemOpen
  \bibfield  {author} {\bibinfo {author} {\bibfnamefont {M.}~\bibnamefont
  {Aidelsburger}},\ }\href@noop {} {\emph {\bibinfo {title} {Artificial Gauge
  Fields with Ultracold Atoms in Optical Lattices}}},\ \bibinfo {edition}
  {1st}\ ed.\ (\bibinfo  {publisher} {Springer},\ \bibinfo {year}
  {2016})\BibitemShut {NoStop}%
\bibitem [{\citenamefont {Bukov}\ and\ \citenamefont
  {Polkovnikov}(2014)}]{bukov_2014}%
  \BibitemOpen
  \bibfield  {author} {\bibinfo {author} {\bibfnamefont {M.}~\bibnamefont
  {Bukov}}\ and\ \bibinfo {author} {\bibfnamefont {A.}~\bibnamefont
  {Polkovnikov}},\ }\href {\doibase 10.1103/PhysRevA.90.043613} {\bibfield
  {journal} {\bibinfo  {journal} {Phys. Rev. A}\ }\textbf {\bibinfo {volume}
  {90}},\ \bibinfo {pages} {043613} (\bibinfo {year} {2014})}\BibitemShut
  {NoStop}%
\bibitem [{\citenamefont {Price}\ \emph {et~al.}(2015)\citenamefont {Price},
  \citenamefont {Zilberberg}, \citenamefont {Ozawa}, \citenamefont
  {Carusotto},\ and\ \citenamefont {Goldman}}]{Price:20154D}%
  \BibitemOpen
  \bibfield  {author} {\bibinfo {author} {\bibfnamefont {H.~M.}\ \bibnamefont
  {Price}}, \bibinfo {author} {\bibfnamefont {O.}~\bibnamefont {Zilberberg}},
  \bibinfo {author} {\bibfnamefont {T.}~\bibnamefont {Ozawa}}, \bibinfo
  {author} {\bibfnamefont {I.}~\bibnamefont {Carusotto}}, \ and\ \bibinfo
  {author} {\bibfnamefont {N.}~\bibnamefont {Goldman}},\ }\href
  {https://doi.org/10.1103/PhysRevLett.115.195303} {\bibfield  {journal}
  {\bibinfo  {journal} {Phys. Rev. Lett.}\ }\textbf {\bibinfo {volume} {115}},\
  \bibinfo {pages} {195303} (\bibinfo {year} {2015})}\BibitemShut {NoStop}%
\bibitem [{\citenamefont {Grushin}\ \emph {et~al.}(2014)\citenamefont
  {Grushin}, \citenamefont {Gomez-Leon},\ and\ \citenamefont
  {Neupert}}]{Grushin:2014}%
  \BibitemOpen
  \bibfield  {author} {\bibinfo {author} {\bibfnamefont {A.~G.}\ \bibnamefont
  {Grushin}}, \bibinfo {author} {\bibfnamefont {A.}~\bibnamefont {Gomez-Leon}},
  \ and\ \bibinfo {author} {\bibfnamefont {T.}~\bibnamefont {Neupert}},\ }\href
  {https://doi.org/10.1103/PhysRevLett.112.156801} {\bibfield  {journal}
  {\bibinfo  {journal} {Phys. Rev. Lett.}\ }\textbf {\bibinfo {volume} {112}},\
  \bibinfo {pages} {156801} (\bibinfo {year} {2014})}\BibitemShut {NoStop}%
\bibitem [{\citenamefont {Anisimovas}\ \emph {et~al.}(2015)\citenamefont
  {Anisimovas}, \citenamefont {Zlabys}, \citenamefont {Anderson}, \citenamefont
  {Juzeliunas},\ and\ \citenamefont {Eckardt}}]{anisimovas2015}%
  \BibitemOpen
  \bibfield  {author} {\bibinfo {author} {\bibfnamefont {E.}~\bibnamefont
  {Anisimovas}}, \bibinfo {author} {\bibfnamefont {G.}~\bibnamefont {Zlabys}},
  \bibinfo {author} {\bibfnamefont {B.~M.}\ \bibnamefont {Anderson}}, \bibinfo
  {author} {\bibfnamefont {G.}~\bibnamefont {Juzeliunas}}, \ and\ \bibinfo
  {author} {\bibfnamefont {A.}~\bibnamefont {Eckardt}},\ }\href
  {https://doi.org/10.1103/PhysRevB.91.245135} {\bibfield  {journal} {\bibinfo
  {journal} {Phys. Rev. B}\ }\textbf {\bibinfo {volume} {91}},\ \bibinfo
  {pages} {245135} (\bibinfo {year} {2015})}\BibitemShut {NoStop}%
\end{thebibliography}
\end{document}